\documentclass[a4wide]{elsart}
\usepackage{amsmath}
\usepackage{amssymb}
\usepackage{graphicx,dcolumn}

\newcommand{\sign}[0]{\text{sign}}
\newcommand{\bra}[1]{\langle #1|}
\newcommand{\ket}[1]{|#1\rangle}
\newcommand{\psibeta}[0]{\psi^{l}}
\newcommand{\ketbra}[0]{\ket{\psibeta}\bra{\psibeta}}
\newcommand{\lambdabeta}[0]{\lambda_{l}}

\newcommand{\braket}[2]{\langle #1|#2\rangle}
\newcommand{\dbydt}[0]{\frac{d}{d\tau}}
\newcommand{\Tr}[0]{\text{Tr}}

\begin{document}
\begin{frontmatter}
\title{Numerical Methods for the QCD Overlap Operator IV: Hybrid Monte
  Carlo}
\author[a]{N. Cundy},
\author[d]{S. Krieg$^{\text{b},}$}, 
\author[d]{G. Arnold}, 
\author[c]{A. Frommer},
\author[d]{Th. Lippert$^{\text{b},}$} and
\author[b]{K. Schilling} 
\address[a] {Department of Physics, University of Regensburg, Universit\"atstra\ss e 31, 93040 Regensburg,
  Germany.}
\address[b] {Fachbereich C, Fachgruppe Physik, University of Wuppertal,
Gaussstra\ss e 20,
42119 Wuppertal,
  Germany.}
\address[c]{Fachbereich C, Fachgruppe Mathematik, University of Wuppertal,Gaussstra\ss e 20
42119 Wuppertal,
  Germany.}
\address[d]{J\"ulich Supercomputing Center,
    J\"ulich Research
 Center, 52425 J\"ulich, Germany.}

\begin{abstract}
The computational costs of calculating the matrix sign function of
the overlap operator together with fundamental numerical problems
related to the discontinuity of the sign function in the kernel
eigenvalues are the major obstacles towards simulations with
dynamical overlap fermions using the Hybrid Monte Carlo algorithm.
In a previous paper of the present series we introduced optimal
numerical approximation of the sign function and have developed
highly advanced preconditioning and relaxation techniques which
speed up the the inversion of the overlap operator by nearly an
order of magnitude.

\noindent In this fourth paper of the series we construct an HMC
algorithm for overlap fermions. We approximate the matrix sign
function using the Zolotarev rational approximation, treating the
smallest eigenvalues of the Wilson operator exactly within the
fermionic force.  Based on this we derive the fermionic force for
the overlap operator. We explicitly solve the problem of the Dirac
delta-function terms arising through zero crossings of eigenvalues 
of the Wilson operator. The main advantage of this scheme is that its
energy violations scale better than O($\Delta\tau^2$) and thus are
comparable with the violations of the standard leapfrog algorithm
over the course of a trajectory. We explicitly prove that our
algorithm satisfies reversibility and area conservation. We present
test results from our algorithm on $4^4$, $6^4$, and $8^4$ lattices.
\end{abstract}

\end{frontmatter}

\section{Introduction}


For more than two decades lattice QCD simulations with light
dynamical quarks remained intractable as the chiral symmetry of
the underlying QCD Lagrangian, which holds in the case of zero mass
quarks, could not be embedded properly into flavour conserving
fermion lattice discretization schemes. A standard workaround takes
recourse to fairly heavy quarks instead and extrapolates the results
over a wide range of quark masses to the very light quark mass
regime.  Unfortunately, simulating far beyond the realm of chiral
perturbation theory such extrapolations carry systematic
errors. These errors have to be avoided in order to achieve a high
precision of phenomenological observables.\footnote{Simulations
  with staggered fermions are less prone to fluctuations. However, it
  is yet unclear if staggered fermions have the correct continuum
  limit in case of their one-flavour approximation by the fourth
  root trick~\cite{Creutz:2008hx}.}

During the 90's the first chiral lattice Dirac operators were
written down.  While studying domain wall fermions~\cite{Kaplan:1992bt}, an early attempt to formulate chiral fermions
on the lattice, Neuberger and Narayanan realised that the
Nielsen-Ninomiya theorem~\cite{Nishy-Ninny} can be circumvented by
placing an infinite number of fermions on the lattice. This insight
led them to the overlap lattice Dirac operator~\cite{Narayanan:1993ss,Neuberger:1998fp}. Afterwards, Hasenfratz,
while working with classically perfect fermions, realised that the
small mass bottleneck could be overcome by using a discretization
scheme that obeys a lattice variant of chiral symmetry~\cite{Hasenfratz}, as expressed by the Ginsparg-Wilson relation for
the quark propagator~\cite{Ginsparg:1982bj}, which in turn implies a
novel version of chiral symmetry on the lattice~\cite{Luscher:1998du}. Neuberger showed that the overlap operator
obeys the Ginsparg-Wilson relation. Theoretically, such a scheme
induces a dramatic reduction in fluctuations in the smallest Dirac operator eigenvalue compared to naive Wilson fermions in the vicinity of zero
quark mass.  However, there is a numerical obstacle: the
implementation of the overlap operator requires the very frequent
solution of linear systems involving the inverse matrix square root
or, equivalently, the matrix sign function (of the Hermitian
Wilson-Dirac operator $Q$).

The problem of approximating the application of $\sign(Q)$ on a
vector has been discussed in a number of papers, using polynomial
approximations~\cite{HeJaLe99,Hernandez:1999gu,HeJaLe00,Bu98,Hernandez:2000iw},
Lanczos based methods~\cite{Bor99c,Bor99b,Bor99a,Vor00} and
multi-shift CG combined with a partial fraction expansion~\cite{Neuberger:1998my,Neu00,EHKN00,EHN98}.  The Zolotarev partial
fraction approximation (ZPFE), first applied to lattice QCD in~\cite{EFL02} -- the first paper in the present series -- was shown
to be the optimal approximation to the matrix sign function. ZPFE
has led to an improvement of over a factor of 3 compared to the
Chebyshev polynomial approach~\cite{HeJaLe00}. This technique to
compute the sign function has meanwhile been established as the
method of choice~\cite{OVERLAP1,Dong:2003im,LIUOVERLAP}. Moreover,
it is the natural starting point for both the treatment of dynamical
overlap fermions~\cite{FODOR} and optimised domain wall fermions~\cite{LIU1,LIU2,LIULATTICE}.\footnote{An alternative scheme is to use
  a 5-dimensional representation of the overlap operator -- see
  references~\cite{Borici:2001ua,Neuberger:1999re,Wenger:2004gd}.  An
  HMC algorithm using a polynomial approximation to the sign
  function was tested in~\cite{Liu:1998hj}, and found to be
  inferior to the ZPFE.}

Until recently, simulations with overlap fermions were restricted to
the quenched model where fermion loops are neglected~\cite{Jansen1,Jansen2,Giusti1,Giusti2}, or hybrid calculations using
staggered or Wilson sea quarks, because of two reasons: (i) the
sheer costs of the evaluation of sign functions of matrices with
extremely high dimensions and (ii) the lack of exact algorithms
including dynamical overlap fermions.

In this paper, we construct a Hybrid Monte Carlo
algorithm~\cite{KennedyDuane} for QCD with overlap fermions.\footnote
{Earlier work done concerning dynamical overlap fermions in the
  Schwinger model can be found in
  ~\cite{BODE,Schwinger1,Schwinger2,Durr:2004ta}.
 More recently,
   some exploratory studies in full QCD
  ~\cite{FODOR,FODOR2,DeGrand:2004nq} have been presented.}  Mostly, we just
need to adapt the algorithm as existing for example for Wilson
fermions to the overlap operator.  However, the discontinuity in the
matrix sign function used in the overlap construction causes
additional complications: as soon as one of the eigenvalues of $Q$
crosses zero, the change in the sign of the eigenvalue leads to a
discontinuity in the fermionic part of the action, which introduces
the occurrence of a Dirac delta function in the HMC fermionic force,
as first discussed in ~\cite{FODOR,FODOR2}. This effect is
predominantly seen in strong coupling: eigenvalue crossings are rare
in the weak coupling limit~\cite{Hernandez:1998et}.

The major part of this paper will deal with the proper integration
of this singular force. Important new contributions in this paper
are: (1) We calculate the energy violation at the crossing exactly;
(2) We explicitly treat the implementation of the eigenvalue
projection technique in the HMC algorithm; (3) Most importantly, we
introduce a momentum update when there is an eigenvalue crossing
with improved energy conservation properties, with the leading
errors being of order O($\Delta\tau^2$).  This is based on our
generalized scheme to construct higher-order exact treatment of the
eigenvalue crossing. We are going to demonstrate that this feature,
along with the dependency of the small eigenvalue density leads to
an HMC algorithm with computational complexity of O($V^{5/4}$). (4)
We discuss what happens when two eigenvalues cross within the same
molecular dynamics step.\footnote{In a forthcoming paper we will
  present an advanced scheme for the treatment of two small
  eigenvalues.}

The main practical difficulty in running dynamical overlap
simulations is the cost in computer time. Overlap fermions are at
least O(100) times more expensive to compute than Wilson
fermions. In other papers of this
series~\cite{vdEetal:03a,cundy2,Krieg:2004xg}, we discussed how the
inversion of the overlap operator, a very time consuming part of the
HMC algorithm, can be accelerated. Here we will make use of all
these methods. We estimate the time required with our current
algorithms for a full HMC simulation on moderate lattices at
realistic masses. Early results were outlined in
\cite{Cundy:2004xf}.

Section \ref{sec:HMC2} gives a brief introduction to the Hybrid
Monte Carlo method for generating dynamical configurations. Sections
\ref{sec:overlap} to \ref{sec:overlap2} outline our method for
adapting HMC to the overlap operator.  Section \ref{sec:numerical}
gives first numerical results, demonstrating that the algorithm
works in practice. After a brief conclusion, we include two
appendices, proving that the proposed algorithm satisfies detailed
balance and describe the additional implications of a combined
transmission/reflection update within a normal leapfrog integration.
Furthermore we describe the correction update when two eigenvalues
cross zero during the same time step.

\section{Hybrid Monte Carlo}\label{sec:HMC}
\subsection{HMC for Wilson fermions}\label{sec:HMC2}

In order to make the paper self-contained, we give a short review of
the HMC algorithm for the case of Wilson
fermions~\cite{KennedyDuane}.

The standard Wilson Dirac operator on the lattice, with a mass $-m_0$, is
\begin{align}
  M_{xy} = & 1_{xy} -
  \kappa \sum_{\mu}\Big[\left(1-\gamma_{\mu}\right)U_{\mu}(x)\delta_{x+e_{\mu},y} \nonumber\\
  &\phantom{1_{xy}}+\left(1+\gamma_{\mu}\right){U^{\dagger}}_{\mu}(x-e_{\mu})\delta_{x
    -e_{\mu},y}
  \Big],\nonumber\\
  \kappa =& \frac{1}{8-2m_0}.
\end{align}
The Wilson operator is $\gamma_5$-Hermitian, implying that one can
construct a Hermitian Wilson operator
\begin{gather}
Q=\gamma_5 M.
\end{gather}
The HMC method updates the gauge field in two steps:
(1) a molecular dynamics evolution of the gauge field and
(2) a Metropolis step which renders the algorithm exact. In the molecular
dynamics step, a momentum $\Pi$ is introduced, which is conjugate to
the gauge fields $U$. Using a fictitious computer time, $\tau$, the momentum field is defined such that 
\begin{gather}
dU/d\tau = \dot{U} = i\Pi{U}.\label{eq:Ueq}
\end{gather}
 Since $U\in$ SU($N_C$), $\Pi$ must be a Hermitian traceless matrix. Following the
classical equations of motion of this system will allow to generate the
correct ensemble. Using the Wilson gauge action, the total energy of this system is 
\begin{align}
  E(\tau) =& \beta\sum_x\left[1-\frac{1}{2 N_C} Tr
  \big(U(x)_{\mu\nu}+U(x)_{\mu\nu}^{\dagger}\big)\right] + X_W^{\dagger}\Phi +
  \frac{1}{2}\sum\Tr \Pi^2,\label{eq:energyclassical}\\
  X_W =& Q^{-2}\Phi.
\end{align}
$U(x)_{\mu\nu}$ is called the plaquette, $\Phi$ is a Gaussian random spinor
field, and $N_C$ is the number of colours (for QCD $N_C = 3$). The second
equation of motion can be inferred from the condition
\begin{align}
\dot{E} = & \,\, 0\nonumber\\
 = & \sum_{x,\mu}\Tr\left[ \frac{1}{2}\frac{d}{d\tau}(\Pi_{\mu}(x)^2) -i
 \left(\frac{\beta}{6}\Pi_{\mu}(x)U_{\mu}(x)V_{\mu}(x)+\Pi_{\mu}(x)F_{\mu}(x) -
 h.c.\right)\right].
\end{align}
This leads to
\begin{gather}
\dot{\Pi}_{\mu}(x) =
 i\left[\left(\frac{\beta}{6}U_{\mu}(x)V_{\mu}(x) +
 F_{\mu}(x)- h.c.\right)\right]_{\text{Traceless}}. 
\end{gather}
Here $V_{\mu}(x)$ is the staple, the sum over all the remaining parts of those
plaquettes which contain the specific gauge link $U_{\mu}(x)$. $F_{\mu}(x)$ is
the fermionic force, which can be found by differentiating $X_W^{\dagger}\Phi$
with respect to $U_{\mu}(x)$. For Wilson fermions, the fermionic force is
given by
\begin{gather}
  F_{\mu}(x) = \kappa\left[\Big(MX_W\Big)_{x+e_{\mu}}
  X_W^{\dagger}(x)(1+\gamma_{\mu}) + X_W(x+e_{\mu})
  \Big(MX_W\Big)^{\dagger}(x)(1-\gamma_{\mu})\right].
\end{gather}
These classical equations of motion have to be solved
numerically. An important requirement for the Markov process is to
maintain \emph{detailed balance}, in order to guarantee the
generation of the correct canonical distribution. To maintain
detailed balance, each molecular dynamics update from computer time
$\tau$ to computer time $\tau+\Delta\tau$ needs to be both area
conserving, {\it i.e.} the Jacobian for the update is 1, and
reversible.\footnote{Starting with a configuration $\{U(0),\Pi(0)\}$,
  we perform a molecular dynamics step to reach a configuration
  $\{U(\Delta\tau),\Pi(\Delta\tau)\}$. If the update is reversible,
  carrying out the molecular dynamics step backwards, with
  $\Delta\tau \rightarrow -\Delta\tau$ and $\Pi \rightarrow -\Pi$,
  will return to the original configuration.} The leapfrog
algorithm fulfils both these requirements, and, over an entire
trajectory, conserves energy up to order $\Delta\tau^2$.\footnote{There are superior integrators to the simple leapfrog, such as the Omelyan integrator~\cite{Omelyan,Takaishi:2005tz}, and the Sexton-Weingarten Integrator~\cite{SW}, but the methods described here can easily be adapted to these integrators; indeed we are using the Omelyan integrator in our current production algorithm.}  For later
convenience, we will write it in terms of a four-step procedure
updating the momentum fields and gauge fields in turn (see algorithm
\ref{alg:standard}). The HMC update consists of $n_{md}$ molecular
dynamics steps followed by a Metropolis step, which corrects for the
small violations in energy conservation due to the numerical
integration.  At the end of the trajectory, the configuration is
either accepted or rejected, with the probability of acceptance
being $P_{\text{acc}}=\text{min}(1,\exp(E-E'))$.  Here $E$ is the
initial energy at the start of the trajectory, and $E'$ the final
energy. To generate a sufficiently large number of new
configurations, a high rate of acceptance is required. Thus, the
molecular dynamics procedure has to conserve energy as well as
possible. Any sizable violation of energy conservation will lower
the acceptance rate, and thus the efficiency of the algorithm.
\begin{algorithm}
\begin{enumerate}
\item $\Pi(\tau+\Delta\tau/2) = \Pi(\tau) + \Delta\tau \dot{\Pi}(\tau)/2$.
\item $U(\tau+\Delta\tau/2) = e^{i(\Delta\tau/2) \Pi(\tau+\Delta\tau/2)}U(\tau)$.\item $U(\tau+\Delta\tau) = e^{i(\Delta\tau/2)
  \Pi(\tau+\Delta\tau/2)}U(\tau+\Delta\tau/2)$.
\item $\Pi(\tau+\Delta\tau) = \Pi(\tau+\Delta\tau/2) + \Delta\tau
  \dot{\Pi}(\tau+\Delta\tau)/2$.
\end{enumerate}
\caption{The standard leapfrog update.}\label{alg:standard}
\end{algorithm}

\subsection{Naive HMC with overlap fermions}\label{sec:overlap}

The overlap operator is given by~\cite{Neuberger:1998my}:
\begin{gather}
D = (1+\mu) + \gamma_5(1-\mu)\;\text{sign}(Q),
\end{gather}
where $\mu$ is a mass term. The bare fermion mass is 
\begin{gather}m_b = 2\mu
m_0/(1-\mu),\label{eq:barefermionmass}
\end{gather}
where $m_0$ is the quark mass of the Wilson operator $D_W = \gamma_5Q$.\\ The
Hermitian overlap operator reads
\begin{gather}
H = \gamma_5 D.
\end{gather}
Thus, the pseudo-fermion action is given by $S_{pf} = \phi^{\dagger} H^{-2} \phi$. 

\subsubsection{Eigenvalues outside Zolotarev range.} 
We approximate the matrix sign function using a rational
approximation. It is advantageous while calculating the sign function
to treat the smallest eigenvalues of $Q$ explicitly in a spectral
representation. If we project out the lowest $n_e$ eigenvectors of
$Q$, then the rational fraction expansion, which is used to
approximate the sign function for the eigenvalues of $Q^2$ within
the fixed range $[\alpha^2,\beta^2]$, is modified to
\begin{align}
  \sign (Q) =& aQ
  \sum_{k}A^{k}\omega_{k}(\alpha,\beta)\left(1-\sum_{l=1}^{n_e}\ket{\psibeta}\bra{\psibeta}\right)+
  \sum_{l=1}^{n_e}\ket{\psibeta}\bra{\psibeta}\epsilon(\lambdabeta),\label{eq:5}\\
  A^{k} =& \frac{1}{a^2 Q^2 +\zeta_{k}(\alpha,\beta)}.
\end{align}
Here $a = 1/\alpha$, $\ket{\psibeta}$ are the eigenvectors of $Q$
with eigenvalue $\lambdabeta$, and $\epsilon(\lambda)$ denotes the
sign function.  We shall assume that $\omega$ and $\zeta$, the
coefficients of the rational fraction, are known (here we used the
Zolotarev coefficients~\cite{Zolotarev,zolotarevpaper}). During the
course of the Hybrid Monte Carlo, we keep the coefficients $\alpha$
and $\beta$ fixed to maintain the acceptance rate, which requires
the projection of all eigenvectors of $Q$ with eigenvalues $|\lambda| < \alpha$.

\subsubsection{Eigenvalues inside Zolotarev range.} 
Of course, one is free to project out eigenvectors within the
Zolotarev range as well, either exactly, as in (\ref{eq:5}), or from
the rational approximation itself:
\begin{align}
  \sum_{k}\frac{\omega_{k}}{a^2Q^2+\zeta_{k}} =& \sum_{k}\frac{\omega_{k}}{a^2Q^2+\zeta_{k}}\left(1-\sum_{l=n_e+1}^{n_p}\ket{\psibeta}\bra{\psibeta}\right)+\nonumber\\
  &\sum_{l=n_e+1}^{n_p}\ket{\psibeta}\bra{\psibeta}\sum_{k}\frac{\omega_{k}}{a^2\lambdabeta^2+\zeta_{k}}.\label{eq:6}
\end{align}
This accelerates the calculation of the multi-mass solver.
Our preferred method that allows to simply fulfil the requirements
for detailed balance, is to project out a fixed number, $n_p$, of
eigenvectors, treating the $n_e$ eigenvectors below $\alpha$
explicitly according to (\ref{eq:5}), and using (\ref{eq:6}) for any
eigenvectors that lie within the range of the rational fraction
approximation.

\subsubsection{Differentiating Eigenvectors.}
In order to calculate the fermionic force one needs to differentiate
the sign function with respect to fictitious time $\tau$, which
means that one must differentiate both the rational fraction and the
small eigenvalues and eigenvectors. To differentiate the
eigenvalues, we start with the eigenvalue equation
\begin{gather}
Q\ket{\psibeta} = \lambdabeta\ket{\psibeta}.
\end{gather}
We now perform an infinitesimal change in the matrix $Q$,
$Q\rightarrow Q+\delta Q$. The new eigenvalue equation reads
\begin{gather}
\left(Q+\delta Q\right)\left(\ket{\psibeta} + \ket{\delta}\right)=\left(\lambdabeta + \delta{\lambda}\right)\left(\ket{\psibeta} + \ket{\delta}\right).
\end{gather}
Since we are free to define $\ket{\delta}$ so that
$\braket{\psibeta}{\delta} = 0$, we immediately have:
\begin{align}
\dot{\lambdabeta} =& \bra{\psibeta}\dot{Q}\ket{\psibeta}\label{eq:lambdadot}\\
\dbydt\ket{\psibeta} =& - P_{l}\dot{Q}\ket{\psibeta}\label{eq:psidot}\\
P_{l} =& (1-\ketbra)(Q-\lambdabeta)^{-1}(1-\ketbra).\label{eq:Pbeta} 
\end{align}
We have added a second eigenvector projector to (\ref{eq:Pbeta}) so
that both $\bra{\phi}P_l$ and $P_l\ket{\phi}$ can be calculated
numerically.\footnote{These equations are just used to calculate the
  fermionic force. We cannot use the updates to provide a time
  evolution of the eigenvectors, for reasons of reversibility, computer time and
  accuracy.} We note that this expansion is only valid if the
separation between the eigenvalues is sufficiently large (see
appendix \ref{sec:updatetwice}). We will treat the case when the two eigenvalues are near degenerate in a forthcoming paper~\cite{Cundy:2007df,Cundy:2007dp}.

We use a CG multi-mass solver to perform the inversion of
$Q-\lambda$ which is required in (\ref{eq:psidot}). We exploit the
normal-equation trick:
\begin{align}
  \frac{1}{Q-\lambda_i}(1-\ket{\psi_i}\bra{\psi_i}) =
  (Q+\lambda_i)\frac{1}{Q^2-\lambda^2_i}&(1-\sum_{j=1}^{n_p}\ket{\psi_j}\bra{\psi_j})+\phantom{a}\nonumber\\
  &\sum_{j=1;j\neq i}^{n_p}\ket{\psi_j}\bra{\psi_j}\frac{\lambda_j +
    \lambda_i}{\lambda_j^2-\lambda_i^2},\nonumber
\end{align}
to convert the inversion of $Q-\lambda_i$ required in the calculation of
$\dbydt\ket{\psibeta}$ as given in (\ref{eq:psidot}) into a positive definite
form---we are working in a subspace orthonormal to all the eigenvectors with
eigenvalues lower than $\lambda_i$. We can now use a CG multi-mass solver to
evaluate the inversion. With a sufficiently large number of kernel eigenvalues projected out of the Zolotarev approximation, this procedure proved to be faster and more stable than using a
Minimal Residual inversion or a Chebyshev approximation of the
inverse~\cite{Degrandschaefer2} for each eigenvector. The multi-mass inversion
converges well, as long as we project into the subspace frequently enough
during the inversion, and $n_p$ is sufficiently larger than $n_e$ to ensure
that the condition number of the multi-mass inversion remains under control.
Tuning $\alpha$ is therefore a balancing act between different parts of the
algorithm: a large $\alpha$ will correspond to a slower multi-mass inversion
for the eigenvector differentiation (because we will be increasing the
condition number for the inversion), but a faster inversion for the sign
function (since the Zolotarev weights generally increase as the condition
number decreases).

The eigenvectors have to be determined whenever we calculate the fermionic
force. We used the Arnoldi PARPACK package with Chebyshev acceleration
\cite{Neff,PARPACK} to calculate the lowest eigenvalues, and a CG minimisation
of the Ritz functional~\cite{Kalkreuter:1996mm} to ensure that no small
eigenvectors were missed by the Arnoldi, and that all the eigenvalues were
calculated to the required accuracy. This method proved to be considerably
faster than using entirely a CG minimisation, even though we can use the old
eigenvectors as a starting point for the new calculation when using the CG
algorithm.

We are now in a position to calculate the fermionic force in the usual
 manner (sums over $k$, $l$ and repeated
spatial indices $\mu$ and $x$ will be assumed from this point onwards):
\begin{align}
\dot{U}_{\mu}(x)F_{\mu}(x)+F^{\dagger}_{\mu}(x)\dot{U}^{\dagger}_{\mu}(x) 
=&-(1-\mu^2) \bra{X}\left(\gamma_5 \dbydt\sign\, Q
+\dbydt \sign\, Q \gamma_5\right) \ket{X},\\
X = & H^{-2}\phi.
\end{align}
The differential of the sign function is:
\begin{gather}
\left(\dbydt\sign\, Q\right)_{nm} = \dot{U}_{\mu}(x)
\left[\tilde{F}^R_{\mu,nm}(x) + \tilde{F}^P_{\mu,nm}(x) + \tilde{F}^S_{\mu,nm}(x) + \tilde{F}^D_{\mu,nm}(x)\right] + h.c.,
\end{gather}
where $\tilde{F}^R$ is the term generated by differentiating the rational
approximation, $\tilde{F}^P$ is the term generated by differentiating the
eigenvector projector $1-\ketbra$, while $\tilde{F}^S$ and $\tilde{F}^D$ come
from the differential of $\epsilon(\lambda)$ in (\ref{eq:5}). These four terms
are:
\begin{align}
\tilde{F}^R_{\mu,nm}(x) =& \kappa a
\omega_{\eta}A^{k}_{na}\left[a^2Q_{ax}\gamma_5(1-\gamma_{\mu})\delta_{x+e_{\mu},c}Q_{cd}\right.\nonumber\\
-&\left.\zeta_{k}\gamma_5(1-\gamma_{\mu})\delta_{a,x}\delta_{x+e_{\mu},d}\right]A^{k}_{de}\left(1-\ketbra\right)_{em},\nonumber\\
\tilde{F}^P_{\mu,nm}(x) =&-\kappa  QA^k_{na}P_{l,
  ax}\gamma_5(1-\gamma_{\mu})\delta_{x-e_{\mu},c}\left(\ketbra\right)_{cm}\nonumber\\
-&\kappa  QA^k_{na}\left(\ketbra\right)_{ax}\gamma_5(1-\gamma_{\mu})\delta_{x,c+e_{\mu}}P_{l, cm},\nonumber\\
\tilde{F}^S_{\mu,nm}(x) =&\kappa P_{l,
  nx}\gamma_5\epsilon(\lambdabeta)(1-\gamma_{\mu})\delta_{x,c+e_{\mu}}\left(\ketbra\right)_{cm}\nonumber\\
+&\kappa
\left(\ketbra\right)_{nx}\gamma_5(1-\gamma_{\mu})\delta_{x,c+e_{\mu}}\epsilon(\lambdabeta)P_{l,
  cm},\nonumber\\
\tilde{F}^D_{\mu,nm}(x) = &-\left(\ketbra\right)_{nm}
\frac{d\epsilon(\lambdabeta)}{d\lambdabeta}\bra{\psibeta}_x\gamma_5(1-{\gamma_{\mu}})\delta_{x,c+e_{\mu}}\ket{\psibeta}_c
.\label{eq:diffsignfunction}
\end{align}
For later convenience, we define $F^G$ as the gauge field force, and $F^C$ as the continuous
part of the fermionic force, i.e.,
\begin{align}
F^G_{\mu}(x) = & \bra{X}i\frac{\beta}{6}U_{\mu}(x)V_{\mu}(x)\ket{X}+h.c.\;,\label{eq:fg}\\
F^C_{\mu}(x) = &-(1-\mu^2)iU_{\mu}(x) \bra{X}\left\{\gamma_5,\tilde{F}^R_{\mu}(x) +
\tilde{F}^P_{\mu}(x) + \tilde{F}^S_{\mu}(x)  \right\}  \ket{X}+h.c.\; .\label{eq:fc}
\end{align}
In order to calculate the fermionic force, we need to invert the overlap
operator twice, perform two multi-mass inversions of the Zolotarev rational
function, and, as remarked before, calculate four multi-mass inversions of the
Wilson operator to obtain the force from the small eigenvalue projection.
This formula can be inserted into a standard HMC routine for Wilson or staggered
fermions. However, the last term in (\ref{eq:diffsignfunction}) contains a
Dirac delta function (the derivative of the sign function) that will ruin the
performance of standard integrators. Whenever an eigenvalue of the Wilson
operator crosses zero on a HMC trajectory, special attention has to be paid to
this last term. We note in passing that eigenvalue crossings of the Wilson
operator are associated with a change in the topological index $Q_f =
-\frac{1}{2}\Tr (\sign\, Q)$.
\section{Eigenvalue Crossings}\label{sec:eigenvalues}
\subsection{Possible strategies}
There are several possibilities which can be taken to overcome the problem of
the eigenvalue crossing:
\begin{enumerate}
\item Ignore the problem in the hope that it will vanish with increasing
  values of $\beta$. In this spirit, one would use chiral projection
  ~\cite{BODE,Cundy:2005mr,DeGrand2006} to allow for sampling across different
  topological sectors, and a small path length to compensate for the low
  acceptance rate. In principle this recipe might allow the simulation to
  bypass the potential wall. In practice, with the pseudo-fermion estimate of the determinant used here, the height of the action jump at the topological sector boundary is too large and attempted crossings too frequent to allow this method to be practical. In future work, we will show that using another representation of the determinant can reduce the height of the action jump to O(1), which might make this method practical~\cite{Cundy:2008zc,cundyforthcoming08}. 
\item Replace $\epsilon$ with some continuous function for small
  $\lambda$~\cite{BODE}. The substitute of the sign function will have
  to be broad enough so that the low eigenvalues are affected, but not too
  broad as this might lead to a large deviation in the final Monte Carlo
  ensemble.  In principle, it should be possible to narrow the revised
  $\epsilon$ as the time step decreases.  One can reduce these artifacts by
  using a more accurate overlap operator in the accept/reject step than in the
  molecular dynamics~\cite{BODE}, but this will lower the acceptance rate. Our
  experience with this method suggests that it cannot achieve an acceptable
  acceptance rate.
\item\label{item:meth4} Restrict the simulation to one topological sector, either by
  always reflecting (see section \ref{sec:reflection}) when one encounters a
  potential eigenvalue crossing ~\cite{Egri:2005cx}, or by using an action  
  which suppresses small eigenvalues of the kernel operator (both by chosing a topology preserving gauge action and by adding a fermionic term)
  ~\cite{Fukaya:2005cw}
. These methods have the disadvantage that one needs to
  calculate the re-weighting factors to combine the different topological
  sectors, and (more seriously) there are unresolved issues concerning whether
  these methods are ergodic.
\item\label{item:meth3} As soon as one encounters a crossing, one repeats the
  micro-canonical step, with the integration over the delta function treated
  exactly~\cite{FODOR,FODOR2}.
\end{enumerate}  
Albeit being more costly (per HMC trajectory; the situation when the auto-correlation is taken into account is unclear) than the first three
 approaches, method (\ref{item:meth3}) is our strategy of choice as it
 offers best control over systematic errors from the Dirac delta
 function's contribution to the fermionic force, as will be explained
 in the following sections. Moreover, it will provide a systematic way to
 improve the scaling of the HMC with overlap fermions.

\subsection{Computing the discontinuity}
The eigenvalue crossing induces a discontinuity in the fermionic contribution
to the action, the kinetic energy, and the fermionic force. From the Monte
Carlo energy (\ref{eq:energyclassical}) the second equation of motion is
derived imposing energy conservation:
\begin{gather}
\frac{d  E}{d\tau} = 0 = \ldots + \Phi^{\dagger}\frac{d  }{d\tau}\Big(H^{-2}\Big)\Phi.
\end{gather}
Care needs to be taken when differentiating the fermionic contribution to the
action near to a discontinuity in the overlap operator.  We note that for
discontinuous functions $a$ and $b$, with $a=a^c(\tau) +
\Delta_a\theta(\tau-\tau_c)$ and $b=b^c(\tau) + \Delta_b\theta(\tau-\tau_c)$
and $a^c$ and $b^c$ being continuous functions, the differential of the
product of $a$ and $b$ is not the usual formula:
\begin{align}
  \frac{d}{d\tau} \Big(a b\Big) = & \lim_{\delta\tau\rightarrow
    0}\frac{1}{\delta\tau}\Big[\Big(a^c(\tau+\delta\tau) +
  \Delta_a\Big)\Big(b^c(\tau+\delta\tau) +
  \Delta_b\Big)-a^c(\tau)b^c(\tau)\Big]\nonumber\\
  =&\lim_{\delta\tau\rightarrow0}\frac{1}{\delta\tau}\Big[\Delta_a
  b^c(\tau+\delta\tau)+ \Big(a^c(\tau+\delta\tau) +
  \Delta_a\Big)\Delta_b\Big] + \phantom{a}\nonumber\\
&\Big(a^c(\tau)+\Delta_a\Big)\frac{d b^c}{d\tau} + \frac{da^c(\tau)}{d\tau} b^c(\tau).
\end{align}
With $b=a^{-1}$, one can show that $\Delta_a b^c(\tau)+ (a^c(\tau+\delta\tau) +
\Delta_a)\Delta_b=0$, and that
\begin{gather}
\frac{d a^{-1}}{d\tau} = - a^{-1}(\tau+\delta\tau) \frac{d a}{d\tau} a^{-1}(\tau).\nonumber
\end{gather}
Therefore,
\begin{gather}
\frac{d}{d\tau} H^{-2} =  -
  \lim_{\delta\tau\rightarrow0}H^{-2}(\tau+\delta\tau)\left.\frac{d H^2}{d\tau}
  \right|_{\tau} H^{-2}(\tau).
\end{gather}

If there is a discontinuity in $H$, such as when the eigenvalue $\lambda_1$
crosses zero, then we need to take one inversion just before the crossing,
leading to $X_- = H^{-2}(\lambda_1)\phi$, and the other inversion just after
it, with  $X_+ = H^{-2}(-\lambda_1)\phi$:
\begin{align}
\frac{d E}{d\tau} = 0 =&\sum_{x,\mu} \Tr\left[\frac{1}{2}
  \frac{d}{d\tau} \Pi^2_{\mu}(x) - i \frac{\beta}{6} \Big(\Pi_{\mu}(x)
  U_{\mu}(x)V_{\mu}(x) - h.c\Big) +\right. \nonumber\\
&\left.(1-\mu^2)\bra{X_+}\left(\gamma_5 \frac{d}{d\tau}\sign\ Q
  +\frac{d}{d\tau}\sign\ Q \gamma_5 \right)\ket{X_-} \right]. \label{eq:30}
\end{align}
We denote the momenta just before and after the eigenvalue
crossing as $\Pi^-$ and $\Pi^+$ respectively (with the smallest
eigenvalues $\lambda_-$ and $\lambda_+$). We can recast
equation (\ref{eq:30}) into the form
\begin{align}
\frac{1}{2\delta\tau}\big[(\Pi^+)^2& - (\Pi^-)^2\big] = \nonumber\\&- (1-\mu^2)
\bra{X_+}\left(\gamma_5\ketbra + \ketbra\gamma_5\right)\ket{X_-}
\frac{\epsilon(\lambda_-) - \epsilon(\lambda_+)}{\delta\tau}. 
\end{align}
This shows that integrating over the Dirac $\delta$ function in the
fermionic force will produce a discontinuity in the kinetic energy:
\begin{align}
(\Pi^+)^2 - (\Pi^-)^2 = &\, 4d(\tau_c)\label{eq:31b}\\
d(\tau) =&
-(1-\mu^2)\epsilon(\lambda_-)\bra{X_+(\tau)}\gamma_5\ket{\psibeta(\tau)}
\braket{\psibeta(\tau)}{X_-(\tau)}\nonumber\\
&-(1-\mu^2)\epsilon(\lambda_-)\braket{X_+(\tau)}{\psibeta(\tau)}\bra{\psibeta(\tau)}\gamma_5\ket{X_-(\tau)}.
\end{align}
It is straightforward to show that this discontinuity in the kinetic
energy will exactly cancel the discontinuity in the pseudo-fermion
action $\langle X_+|\Phi\rangle - \langle
X_-|\Phi\rangle$. Therefore, energy would be conserved across the
eigenvalue crossing in an exact integration.  Of course, numerical
integration schemes are not exact and will produce a huge error at
the discontinuity.  Our task is to develop an integration algorithm
such as to maintain area conservation, energy conservation and
reversibility in the presence of Dirac delta function forces.

\section{HMC with Overlap Fermions}\label{sec:overlap2}
\label{sec:prescription}
In this section we propose to add a correction step to the
\textit{standard leapfrog} algorithm (algorithm
\ref{alg:standard}). The correction step allows us to handle the above
discontinuities from eigenvalue crossing as desired. In order to
realize a proper HMC scheme for overlap fermions, the correction
step has to be area preserving\footnote{We note in passing that area
  preservation is not a strict requirement. Non-area preserving
  higher-order methods will be introduced in subsequent papers which
  will increase the rate of transmission~\cite{Cundy:2005mr} and  allow a solution to problem of mixing between low lying eigenvectors~\cite{Cundy:2007df}.} and
reversible (ergodicity). In order to preserve the $O(\Delta t^2)$
complexity of the standard HMC, one must satisfy equation (\ref{eq:30}) with
O($\Delta\tau^2$) errors or better. In the first part of this
section, we shall introduce a general method suitable correction
steps can be derived in a systematic manner. Thereafter, we shall
present our integrator of choice, leaving the proof to appendix
\ref{app:detailedbalance} that its correction step is area
conserving, reversible and conserves energy with only
O($\Delta\tau^2$) errors.
 
\subsection{Notation}

The space-time lattice is 4-dimensional with $V$ lattice sites and $4V$ links.
In general, the SU($N_C$) gauge group is used (although in practice, throughout this work, we will be using $N_C=3$), so the gauge field $U_{\tau}$ contains a member of SU($N_C$) on
every link, and the momentum field $\Pi_{\tau}$ is represented by a Hermitian,
traceless $N_C\times N_C$ matrix on each link. The subscript $\tau$ refers to
the computer time at which the gauge or momentum field is calculated. For
convenience, $\Delta\tau$ is set to 1, the start of the algorithm is at time
0, so that $U_0$ is the original gauge field, and $U_1$ is the final gauge
field at the end of the leapfrog correction step. $\tau_c + 1/2$ is the
computer time at which the eigenvalue is 0 --- the calculation of $\tau_c$
is discussed at the end of this section. The superscript ``$-$'' is used
to indicate that the effects of the crossing are not yet included into the
momentum update, and ``$+$'' indicates that the momentum has been updated to
account for the crossing. $\Pi^+ \equiv \Pi^+_{\frac{1}{2}}$, etc., and $U_c \equiv
U_{\frac{1}{2} + \tau_c}^- = U_{\frac{1}{2} + \tau_c}^+$. Finally, the
notation $(A,B)$ shall be used to represent $\sum_{x,\mu} \Tr(A_{\mu}(x)
B_{\mu}(x))$. The continuous part of the (Hermitian) force is $F_{\tau} =
F^G_{\tau} + F^C_{\tau}$, with $F^G$ and $F^C$ defined in equations
(\ref{eq:fg}) and (\ref{eq:fc}).

$U$ contains an element of SU($N_C$) for every link,
while $\Pi$ is a generator of $U$, i.e. it contains a Hermitian,
traceless $N_C\times N_C$ matrix on every link. 

In order to simplify the notation for the correction step in its most general
form in the following, $\Pi$ is expanded in terms of an orthonormal basis.
This basis is defined by a set of orthonormal basis vectors which are divided
into $N_S$ subsets,
$\{\{\eta^1_1,\eta^1_2,\ldots\},\{\eta^2_1,\eta^2_2,\ldots\},\ldots\}$. The
parameter $N_k$ gives the number of $\eta$ vectors in each subset $k$. The
$\eta^k_i$ are $4V(N_C^2-1)$ Hermitian traceless matrices which satisfy
$(\eta^k_i,\eta^m_j) = \delta_{ij}\delta^{km}$. $N_k$ and $N_S$ are defined
such that the subscripts ($i$ and $j$) run from $1$ to $N_k$, where $N_k$ is
not necessarily constant for all the $k$, while the superscripts ($k$ and $m$)
run from $1$ to $N_S$. For the moment, we shall leave the $N_k$s and $N_S$ as
arbitrary parameters, which satisfy the constraint
\begin{gather}
\sum_{k = 1}^{N_S} N_k = 4V(N_C^2-1).\nonumber
\end{gather}
Thus, the basis is complete, so that
\begin{gather}
\Pi = \sum_{k=1}^{N_S}\sum_{i=1}^{N_k} \eta_i^k (\eta_i^k,\Pi).
\end{gather}
The $\eta$ matrices are functions of $U_c$, and are otherwise independent of
the momentum. $\eta^1_1\equiv\eta$ is defined as being normal to the gauge
field surface where the eigenvalue is zero (see appendix
\ref{app:detailedbalance} and equation (\ref{eq:etadef})). $\eta^1_2$ is
proportional to $F - \eta(\eta,F)$, where $F = F^+_{\tau_c + 1/2} -
F^-_{\tau_c + 1/2}$. The other $\eta$ matrices are arbitrary.

\subsection{The classical mechanics case}

Before the situation of the molecular dynamics with the full overlap operator
is considered, the much simpler problem of a classical mechanics particle
approaching a potential wall of height $-2d$ is reviewed.

The potential energy in this case is defined as $V(q) = -2d\theta(q)$, (i.e.
$V=0$ for $q<0$, and $V=-2d$ for $q>0$). Note that $d$ may be positive or
negative. First, consider the one-dimensional case. The kinetic energy of the
particle before it hits the wall is $\frac{1}{2} (\Pi^{-})^2$.  After it has
hit the wall, there are two possibilities: if the initial momentum is large
enough, the particle will continue onwards with a changed momentum $\Pi^+ =
\sqrt({\Pi^-}^2 + 4d)$; this case is called \underline{transmission}.  If
the momentum is too small, the particle will carry insufficient kinetic energy
to cross the wall and will be reflected (elastically), giving a final momentum
$\Pi^+=-\Pi^-$. This case is called \underline{reflection}. The total energy
$\frac{1}{2}\Pi^2 + V$ is of course conserved in both cases.

Next consider the classical mechanics particle in three dimensions. The
coordinate system shall be defined in terms of the orthonormal basis vectors
$\eta, \eta_1^1,$ and $\eta^1_2$ as just given above. It is assumed that the
potential is given by $V(\mathbf{q})=-2d\theta((\mathbf{q},\mathbf{\eta}))$,
such that $\eta$ is the normal vector to the potential wall.

The three components of the momentum are defined as $(\Pi^-,\eta)$,
$(\Pi^-,\eta_1^2)$ and $(\Pi^-,\eta_2^2)$. It is well known what happens to
the particle in classical mechanics after it hits the potential wall; in the
case of transmission, 
\begin{gather}
(\Pi^+,\eta)=\sqrt{(\Pi^-,\eta)^2 + 4d},
\end{gather}
is obtained, and the transverse moments are
\begin{gather}
(\Pi^+,\eta^1_i)=(\Pi^-,\eta^1_i).
\end{gather}
For reflection,  the final  momenta will be  
\begin{align}
(\Pi^+,\eta)=&-(\Pi^-,\eta),\nonumber
\intertext{and}
(\Pi^+,\eta^1_i)=&(\Pi^-,\eta^1_i),
\end{align}
respectively.  Again, both cases conserve energy.

In fact, there is no deeper reason why the molecular dynamics trajectory must
follow the classical mechanics trajectory, as long as it is energy conserving,
area conserving, reversible and ergodic. One can equally well use the scheme
\begin{align}
(\Pi^+,\eta) =& (\Pi^-,\eta)\nonumber\\
(\Pi^+,\eta^1_i) =& (\Pi^-,\eta^1_i)\sqrt{1+4d/((\Pi^-,\eta^1_1)^2 +
(\Pi^-,\eta^1_2)^2)},
\end{align}
as long as it conserves the area. It is evident that this update, as
well as many others which can be chosen, conserves energy. The given
example also conserves the area, as a short calculation
demonstrates:
\begin{align}
  \left|\frac{\partial (\Pi^+,\eta^1_i)}{\partial (\Pi^-,\eta^1_j)}\right| =
  &\left|\begin{array}{c c} J_{11}&
      J_{12}\\
      J_{21}&J_{22}
\end{array}\right|=1,\nonumber\\
J_{11}=&\sqrt{1+\frac{4d}{(\Pi^-,\eta_1^1)^2+(\Pi^-,\eta^1_2)^2}} -
\nonumber\\&\frac{4d
  (\Pi^-,\eta_1^1)^2}{((\Pi^-,\eta_1^1)^2+(\Pi^-,\eta^1_2)^2)^2}\left(\sqrt{1+\frac{4d}{(\Pi^-,\eta_1^1)^2+(\Pi^-,\eta^1_2)^2}}\right)^{-1},
\nonumber\\
J_{12}=&- \frac{4d
  (\Pi^-,\eta_1^1)(\Pi^-,\eta_2^1)}{((\Pi^-,\eta_1^1)^2+(\Pi^-,\eta^1_2)^2)^2}\left(\sqrt{1+\frac{4d}{(\Pi^-,\eta_1^1)^2+(\Pi^-,\eta^1_2)^2}}\right)^{-1},\nonumber\\
J_{21}=&- \frac{4d
  (\Pi^-,\eta_1^1)(\Pi^-,\eta_2^1)}{((\Pi^-,\eta_1^1)^2+(\Pi^-,\eta^1_2)^2)^2}\left(\sqrt{1+\frac{4d}{(\Pi^-,\eta_1^1)^2+(\Pi^-,\eta^1_2)^2}}\right)^{-1},\nonumber\\
J_{22}=&\sqrt{1+\frac{4d}{(\Pi^-,\eta_1^1)^2+(\Pi^-,\eta^1_2)^2}} -
\nonumber\\&\frac{4d
  (\Pi^-,\eta^1_2)^2}{((\Pi^-,\eta^1_1)^2+(\Pi^-,\eta^1_2)^2)^2}\left(\sqrt{1+\frac{4d}{(\Pi^-,\eta_1^1)^2+(\Pi^-,\eta^1_2)^2}}\right)^{-1}.
\end{align} 
It can be shown that this latter update is reversible, since reversing the
time changes $d\rightarrow -d$. Thus, instead of creating a discontinuity in
the momentum {\it normal} to the potential wall, we are introducing the
discontinuity in the directions {\it transverse} to the potential wall.
Furthermore, one can also consider a 5-dimensional case, with two additional
basis vectors $\eta^2_1$ and $\eta^2_2$, which provides the option of changing
momentum in any or all of the three directions defined by $\eta$, $\{\eta^1_1,
\eta^1_2\}$, and $\{\eta^2_1, \eta^2_2\}$. This general scheme will allow to
achieve the $O(\Delta \tau^2)$-scalability of the approach.

\subsection{The QCD situation}
 
The are two differences in lattice QCD molecular dynamics to the
classical mechanics example. Firstly, one has a considerably larger
space, the configuration space of the gauge fields on the
lattice. Assuming that one is working in SU($N_c$), with a
$D$-dimensional lattice, and $V$ lattice sites, then the gauge and
momentum fields exist within a $DV(N_C^2-1)$-dimensional space,
which gives a lot of freedom in how to update the momentum fields.
Secondly, a numerical integration is performed rather than an exact
integration, and care must be taken as to the effects of time
discretization on both the energy conservation and the area
conservation.

\subsubsection*{Simple update algorithm}
Now, the QCD correction update can be constructed, $(\Pi_0,U_0)
\rightarrow (\Pi_1,U_1)$.  The first step is to update the gauge and
momentum to time $\tau+\Delta\tau/2$, as in the leapfrog procedure
(algorithm \ref{alg:standard}), to yield fields $\Pi^-$ and $U^-$:
\begin{align}
  \Pi^{-} =& \Pi_0 + (F_0^-)
  \frac{\Delta\tau}{2}\nonumber\\
  U^{{-}} =& e^{i\Delta\tau\Pi_{-}/2}U_0.
\end{align}
Next, the integration proceeds up to the crossing point itself:
\begin{align}
\Pi^-_{\frac{1}{2} + \tau_c} =& \Pi^{-} + \tau_c F^{-}\label{eq:notusedupdate1}\\
U_c = & e^{i\tau_c\Pi^{-}}U^{{-}}.
\end{align}
We now perform the correction step. For transmission, we use equation (\ref{eq:31b}):
\begin{gather}
\left(\Pi^+_{\frac{1}{2}+\tau_c},\Pi^+_{\frac{1}{2}+\tau_c}\right) = \left(\Pi^-_{\frac{1}{2}+\tau_c},\Pi^-_{\frac{1}{2}+\tau_c}\right)  + 4d.\label{eq:momupdateU1}
\end{gather}
In terms of our basis, a general solution of
(\ref{eq:momupdateU1})\footnote{It should be observed that this is
  not the most general solution: there are other possibilities
  involving error functions.} is
\begin{align}
\Pi^+_{\frac{1}{2}+\tau_c}=&\Pi^-_{\frac{1}{2}+\tau_c}+\sum_{k=1}^{N_S}
\left(\sum_{i = 1}^{N_k} \eta_i^k
(\eta_i^k,\Pi^-_{\frac{1}{2}+\tau_c})\right)\left(\sqrt{1+ \frac{d_k}{\sum_{i}
(\eta_i^k,\Pi^-_{\frac{1}{2}+\tau_c})^2}}-1\right),\nonumber\\
\sum d_k =& 4d\label{eq:pisoln}.
\end{align}
Finally, we move back to computer time $\tau + \Delta\tau/2$, and
complete the rest of the normal leapfrog integration
\begin{align}
\Pi^+ =& \Pi^+_{\frac{1}{2}+\tau_c} - \tau_c\left(F_{+}\right)\label{eq:notusedupdate2}\\
U^+ = &e^{-i\tau_c\Pi^{+}}U_{{c}}\nonumber\\
U_1 =& e^{i\Delta\tau/2\Pi^{+}}U^+\nonumber\\
\Pi_1 = & \Pi^+ + F_1^+ \frac{\Delta\tau}{2}.
\end{align}
This defines the simple update algorithm. Unfortunately we cannot
make use of it because of two reasons: firstly, there is the
possibility that one of the square roots in equation
(\ref{eq:pisoln}) might be imaginary (which would mean that $\Pi$ would no longer be Hermitian); and secondly because the steps
described in equations (\ref{eq:notusedupdate1}) and
(\ref{eq:notusedupdate2}) violate detailed balance. Although
$\tau_c$ is a function of the momentum, this alone is not enough to
violate detailed balance, but, as shown in appendix
\ref{app:detailedbalance}, an update of the momentum parallel to
$\eta$ will violate area conservation. 

\subsubsection*{Maintaining detailed balance}
In order to satisfy detailed balance, we can update the momentum in
directions normal to $\eta$, {\em i.e.} we replace
(\ref{eq:notusedupdate1}) and (\ref{eq:notusedupdate2}) by
$\Pi^{\pm}_{1/2+\tau_c} = \Pi^{\pm} + \tau_c(F^{\pm} -
\eta(\eta,F^{\pm}))$; note however that this replacement comes at
the cost of an $O(\tau_c)$ violation of energy conservation. In
appendix \ref{app:detailedbalance}1.1, we shall prove that a more
flexible general momentum integration step which satisfies detailed
balance is given by
\begin{align}
\Pi^+ =& \Pi^- - \tau_c (F-\eta(\eta,F)) - \frac{\tau_c}{3}\Tr F +
\eta(\eta,\Pi^-)\left(\sqrt{1+\frac{d_1}{(\eta,\Pi^-)^2}}-1\right)+\nonumber\\
& \sum_{k = 2}^{N_S} \bigg[\eta^k_{1}(\eta^k_{1},\Pi^- -
\frac{\tau_c}{2}(F^+  + F^-)) + \eta^k_{2}(\eta^k_{2},\Pi^- -
\frac{\tau_c}{2}(F^+  + F^-))\bigg]\times\nonumber\\
&\left[\sqrt{1 + \frac{d_k}{[\eta^k_{1},\Pi^- -
\frac{\tau_c}{2}(F^+  + F^-)]^2 +[\eta^k_{2},\Pi^- -
\frac{\tau_c}{2}(F^+  + F^-)]^2 }} - 1\right].\label{eq:mostgensolution}
\end{align}
We have inserted the fermionic force dependence here for later
convenience. To satisfy detailed balance, we need $N_k=2 \;\forall
k$, and the $d_k$ should be functions only of the gauge field at the
crossing and (for $k \neq 1$) $(\eta,\Pi^-)$, and odd functions of
$\Delta\tau$.


If we set $d_1 = 4d$, and $d_k=0\;\forall k\neq1$, then we get an
algorithm similar to that proposed in \cite{FODOR,FODOR2}. However,
the significant disadvantage with this method --- as remarked
earlier --- is, that it has an $O(\tau_c)$ energy conservation
violating term (see appendix \ref{app:corrsteperror}): $\Delta
E_{\tau} = \tau_c ((F^+,\eta)(\Pi^+,\eta)-
(F^-,\eta)(\Pi^-,\eta))$. This term would cripple and extremely slow
down the Overlap-HMC algorithm for large lattices. In this case a
much smaller time step should be applied at large volumes (as the
number of correction steps increases) --- rendering the HMC
impractical.

\subsubsection*{Improved error behaviour}
Using the just proposed general update will allow a better, $O(\Delta \tau^2)$, scaling of the energy conservation, as shown in 
\ref{app:corrsteperror}. Instead of adding $2d$ to the kinetic
energy, we add $2d-\Delta E_{\tau}$, by setting $\sum d_k = 4d
-2\Delta E_{\tau}$. We can also remove some (though not all) of the
O$(\Delta\tau^2)$ error in the same way,
leaving us with errors which are close to O$(\Delta\tau^3)$, as demonstrated numerically in section \ref{sec:energycons2}. The presence of $O(\Delta \tau^2)$ errors is not inconsistent with reversibility because $\tau_c$ is an even function of $\Delta \tau$.

These are the most important improvements of our method over that
proposed in~\cite{FODOR,FODOR2}.

\subsubsection*{Transmission and reflection}
A further problem arises when one of the square roots in equation
(\ref{eq:mostgensolution}) becomes imaginary. In analogy to the
classical mechanics picture discussed above, following~\cite{FODOR,FODOR2}, we reflect off the $\lambda=0$ surface. On the
one hand, if the momentum is large enough, then we pass through into
the new topological sector with a changed momentum. As in the classical mechanics example, we call this
case \textit{transmission},\footnote{It is called \textit{refraction}
  in ~\cite{FODOR,FODOR2}} and this is the scenario described so
far in this section. On the other hand, if the momentum is too
small, we bounce off the $\lambda=0$ surface. As before, this is denoted as
\textit{reflection}. Unlike transmission, a reflection update is
accompanied neither with a change in topological index nor with a
discontinuity in the fermionic action.

Still there is a further subtlety which needs to be addressed before
one can formulate the algorithms for transmission and
reflection. With the techniques used in this article, the combined update step should also be area
conserving. 
As an example, suppose that we set
$d_1 = 4d$, and all the other $d_k = 0$, and we reflect whenever
$1+\frac{4d}{(\eta,\Pi^-)^2}<0$. The method by which we choose to
reflect or transmit is denoted as the \textit{reflection condition}. One
can write:
\begin{gather}
\Pi^+ =
\Pi^{+{\text{transmission}}}\theta\left(1+\frac{4d}{(\eta,\Pi^-)^2}\right)
+ \Pi^{+{\text{reflection}}}\left(1-\theta\left(1+\frac{4d}{(\eta,\Pi^-)^2}\right)\right).
\end{gather}
When calculating the Jacobian one must differentiate the $\theta$
(step) functions, which might lead to a Jacobian of the form
$1-\delta(1+4d/(\eta,\Pi^-)^2)$.  This issue is outlined in more
detail in appendices \ref{App:areaconsrefltrans} and \ref{App:A.2.4}. In these appendices we demonstrate that the $\delta$-function is not present if the reflection condition is independent of the momentum (i.e. we
reflect when $|d| > d_{max}$), or if there is no discontinuity between the
transmission and reflection updates. We also argue in the appendices that even if the $\delta-$function is present in the Jacobian, it will not affect the final ensemble. We test this conclusion numerically in section
\ref{sec:tuningdmax}. Both theoretically and in the numerical
experiment, no systematic error is found.

\subsection{Integration over the discontinuity}

A first approximation of $\tau_c$ can be determined by a Taylor
expansion of the eigenvalue around zero (see equation
(\ref{eq:lambdadot})):
\begin{gather}
  \tau_c = -\frac{\bra{\psibeta} Q\ket{\psibeta}}{\bra{\psibeta}
    \dot{Q}\ket{\psibeta}}.\label{eq:firsttauc}
\end{gather}
This gives the correct value of $\tau_c$ up to order $\tau_c^2$.  We
can, in principle, calculate $\tau_c$ using (\ref{eq:firsttauc})
from either the gauge field $U^+$ or the gauge field $U^-$. There
will be a small deviation between the two calculations, and
therefore a small discrepancy between $d$ and $\eta$ calculated from
the two gauge fields. This will lead to a breakdown in reversibility
if we apply algorithm \ref{alg:algmod1} naively. There are two
possible ways in which this problem can be overcome. Firstly, one
can calculate $\tau_c$ from $U^-$ if $\tau_c/\Delta\tau < 0$
(i.e. we have already passed the crossing at molecular dynamics time
$\tau+\Delta\tau/2$), and use $U^+$ to calculate $\tau_c$ otherwise.
Calculating $\tau_c$ using $U^+$ requires an iterative procedure; we
used a combination of simple iteration and the Van
Wijngaarden-Dekker-Brent (VWDB) method~\cite{Brent,Brent2} for root
finding, which was reasonably effective. The second, and our
preferred, possibility is to calculate $\tau_c$ exactly using a root
finding algorithm, such as either Newton-Raphson or VWDB. We used a Newton-Raphson algorithm,
iterating
\begin{gather}
  \tau_c^n = \tau_c^{n-1}-\frac{\bra{\psibeta(\Delta\tau/2 + \tau_c^{n-1})}
    Q(\Delta\tau/2 + \tau_c^{n-1})\ket{\psibeta(\Delta\tau/2 + \tau_c^{n-1})}}{\bra{\psibeta(\Delta\tau/2 +
      \tau_c^{n-1})} \dot{Q}(\Delta\tau/2 + \tau_c^{n-1})\ket{\psibeta(\Delta\tau/2 + \tau_c^{n-1})}},
\end{gather}
so that we could improve the accuracy of the eigenvector as the
Newton-Raphson iteration progressed. The difficultly with bounded methods such as bisection or VWDB is that trying to constrain the
eigenvector between two bounds could lead to a high accuracy eigenvector
escaping bounds calculated with a low accuracy. Using Newton-Raphson allows us to adjust the accuracy of the eigenvector as we proceed with the calculation. This method, of finding the exact point where the eigenvalue crosses, has the advantage that it removes one source of $O(\Delta\tau^2)$ energy conservation violating terms, and it also was faster than the simple iterative method on our small trial lattices.  However, the eigenvector with the smallest eigenvalue
needs to be calculated to a very high accuracy. This method breaks
down if the crossing is close to a minimum of the eigenvalue. In
this case, it is necessary to use a different algorithm, such as
Brent's method, to get sufficiently close enough to the solution for
Newton-Raphson to converge.

\subsubsection{Transmission}

The update scheme proposed for transmission is given by (see
algorithm \ref{alg:algmod1}):
\begin{align}
  \Pi^+ =& \Pi^- + \tau_c(F) - \eta \tau_c(\eta,F)
  - \frac{\tau_c}{3}\Tr(F)+ \nonumber\\
  & \left(\eta^k_1 (\eta^k_1, \Pi^- - \frac{\tau_c}{2}(F^- + F^+)) + \eta_2^k
    (\eta_2^2, \Pi^-
    - \frac{\tau_c}{2}(F^- + F^+))\right)\times\nonumber\\
  &\left(\sqrt{1 + \frac{d_k }{(\eta^k_1, \Pi^- - \frac{\tau_c}{2}(F^- +
        F^+))^2 + (\eta_2^k, \Pi^- - \frac{\tau_c}{2}(F^- + F^+))^2}}-1\right)
\label{eq:40}\\
d_k =& \frac{1}{N_S}(4d - 2\tau_c(F^-,\eta)(\Pi^-,\eta)
+2\tau_c(F^+,\eta)(\Pi^+,\eta) + \tau_c^2(F^- + F^+,F^+ - F^-)).\nonumber
\end{align}

As mentioned earlier, $\eta_1^k$, and $\eta_2^k$ must be
orthonormal and orthogonal to $\eta$ and $F$. We constructed these
vectors using the following procedure:
\begin{itemize}
\item We start by taking three unit vectors, $\alpha$, $\beta$ and
  $\gamma$ corresponding to putting one of the eight Gell-Mann
  matrices on one particular link.
\item We then construct $\eta^k_1= \alpha \cos\theta + \beta
  \sin\theta\cos\phi + \gamma \sin\theta\sin\phi$, choosing the
  angles $\theta $ and $\phi$ so that $\eta^k_1$ is normal to $\eta$
  and $F$.
\item $\eta^k_2$ is then constructed in the same way from three
  different Gell-Mann matrices on the same link.
\end{itemize}
Therefore $N_S$ is the number of links on the lattice. It should be
noted that this procedure does not lead to a higher transmission
rate than other choices.\footnote{Although it is possible to use
  non-area conserving algorithms with a higher transmission
  rate --- see~\cite{Cundy:2005mr}.} The change of momentum for any pair
of $\eta-$vectors is proportional to $1/N_S$, but there are $N_S$
pairs of vectors, so the overall probability of transmission remains
constant. Indeed, we will show in a forthcoming paper~\cite{Cundy:2005mr} that (if we are updating the
momentum normal to $\eta$ and assuming that $\Pi$ is Gaussian
distributed) the probability of transmission is always
$\text{min}(1,\exp(-\frac{1}{2}\sum d_k)) + O(1/V)$, no matter which area-conserving algorithm
is chosen. We investigate below how this choice of transmission
algorithm will affect the numerical stability and the transmission
rate.

\begin{algorithm}
\begin{enumerate}
\item $\Pi^-(\tau+\Delta\tau/2) = \Pi(\tau) + \Delta\tau/2 \dot{\Pi}(\tau)$;
\item $U^-(\tau+\Delta\tau/2) = e^{i(\Delta\tau/2)
  \Pi^-(\tau+\Delta\tau/2)}U(\tau)$;
\item $U_c = e^{i\tau_c\Pi^-} U^-$;
\item The momentum update given in equation \ref{eq:40};
\item $U^+ = e^{-i\tau_c \Pi^+}  U_c$;
\item $U(\tau+\Delta\tau) = e^{i(\Delta\tau/2)
  \Pi^+(\tau+\Delta\tau/2)}U^+(\tau+\Delta\tau/2)$;
\item $\Pi(\tau+\Delta\tau) = \Pi^+(\tau+\Delta\tau/2) +
  (\Delta\tau/2) \dot{\Pi}_+(\tau+\Delta\tau)$.
\end{enumerate}
\caption{The modified leapfrog algorithm for
  transmission.}\label{alg:algmod1}
\end{algorithm}
\begin{algorithm}
\begin{enumerate}
\item $\Pi^-(\tau+\Delta\tau/2) = \Pi(\tau) + (\Delta\tau/2)
  \dot{\Pi}(\tau)$;
\item $U^-(\tau+\Delta\tau/2) = e^{i(\Delta\tau/2)
  \Pi^-(\tau+\Delta\tau/2)}U(\tau)$;
\item $U_c = e^{i\tau_c\Pi^-} U^-$;
\item The momentum update given in equation (\ref{eq:41});
\item $U^+ = e^{i\tau_c \Pi^+} U_c$;
\item $U(\tau+\Delta\tau) = e^{i(\Delta\tau/2)
  \Pi^+(\tau+\Delta\tau/2)}U^+(\tau+\Delta\tau/2)$;
\item $\Pi(\tau+\Delta\tau) = \Pi^+(\tau+\Delta\tau/2) +
  (\Delta\tau/2) \dot{\Pi}_+(\tau+\Delta\tau)$.
\end{enumerate}
\caption{The modified leapfrog algorithm for
  reflection.}\label{alg:algmod2}
\end{algorithm}

\subsubsection{Reflection}\label{sec:reflection}
A reflection takes place whenever one of the momentum updates in the
transmission algorithm would lead to an imaginary momentum. To
correct for the reflection's O($\tau$) errors, one can use a method
similar to the one which we use for transmission.  Indeed, we used
this algorithm for a while, and it worked well. However, there
remains a problem with this method, which is what to do when the
O($\tau_c$) error becomes sufficiently large that the square roots
in the O($\tau_c$) correction become imaginary. Although this scenario did not
occur in practice, it is a worrying possibility with no obvious
answer. We have subsequently switched to a method first proposed by
Fodor \textit{et.\ al.}~\cite{Egri:2005cx}, which is based
around a different idea.  Our current update for reflection (see
algorithm \ref{alg:algmod2}) is
\begin{align}
\Pi^+ =& \Pi^--2 \eta
(\eta,\Pi^-)-2\tau_c(F^-) + 2\eta \tau_c(\eta,F^-)
 + 2\frac{\tau_c}{3}\Tr(F^-). \label{eq:41}
\end{align}

Note the change of sign in the gauge field update in step (5) of
algorithm \ref{alg:algmod2}, as compared to algorithm
\ref{alg:algmod1}.

The sign of the smallest eigenvalue should change for a transmission
step, and remain the same for reflection. However on a few rare
occasions, this did not occur. Such odd phenomena happen if the
$\lambda=0$ surface is not smooth near to the point of crossing, so
that the eigenvalue might try to cross again in the same molecular
dynamics step. We corrected for this by adding a second correction
step (i.e. we repeated steps (3) (4) and (5)) if the sign of the
smallest eigenvalue did not behave as expected.

Two different eigenvalues crossing during the same micro-canonical
step will also cause this algorithm to break down, because of
possible mixing between the two eigenvectors. This issue is
discussed in appendix \ref{sec:updatetwice}. This phenomenon will be discussed in a future paper~\cite{Cundy:2007df}.

\section{Numerical Results}\label{sec:numerical}

In order to test the overlap HMC algorithm numerically on various
small lattices, $4^4$, $N_F=2$ ensembles with $\beta = 5.4$ were
generated, one set of runs with parameter settings $\kappa = 0.225$
and $\mu = 0.05, 0.1, 0.2, 0.3, 0.4$ and $0.5$, the other with $\mu
= 0.5$ and $\kappa = 0.18, 0.19, 0.2, 0.21, 0.22$ and $0.23$. We
have also generated three $6^4$, $\beta=5.6$ ensembles, with $\kappa
= 0.2$ and $\mu = 0.3,0.1,0.05$, and two $8^4$, $\beta = 5.6$
ensembles at $\kappa = 0.2$, and $\mu = 0.1$ or $0.3$, although we
did not analyse the small statistics $8^4$ data sets for most of the
results presented in this section. Phenomenological
calculations using the overlap operator in the quenched
approximation have covered masses in the range $\mu \sim
0.007\rightarrow 0.4$~\cite{Dong:2003im} or $\mu \sim
0.015\rightarrow 0.037$~\cite{Garron:2003cb}, roughly a factor of 10
below our present mass range.\footnote{In order to express $\mu$ as a
  physical mass, we used $r_0=0.49\text{fm}$ to calculate the lattice spacing as
  approximately $a^{-1}\sim 590\text{MeV}$ on our $6^4$ lattices. The
  renormalisation constant was not possible to measure on these
  lattices because the error was too large, so, for this rough
  estimate, we will take $Z_m = 1$. This implies that on these
  ensembles a quark mass of $\mu = 0.05$ corresponds to a physical
  mass of $\sim 93$ MeV, {\em i.e.} around the value of the strange
  quark mass, however with a very large error.} Throughout these
simulations, we took advantage of relaxation and preconditioning
techniques developed in~\cite{vdEetal:03a,cundy2}. The accuracy of
the preconditioner and the number of projected eigenvalues were
optimised for our HMC program, which gave a gain factor of $\sim
30$. We used anti-periodic boundary conditions in the time direction
and periodic boundary conditions in the spatial directions.

\subsection{Energy conservation and topological index}\label{sec:energycons}

\begin{figure}
\begin{center}
\begin{tabular}{c}
\includegraphics[width = 14cm,height =
  9cm]{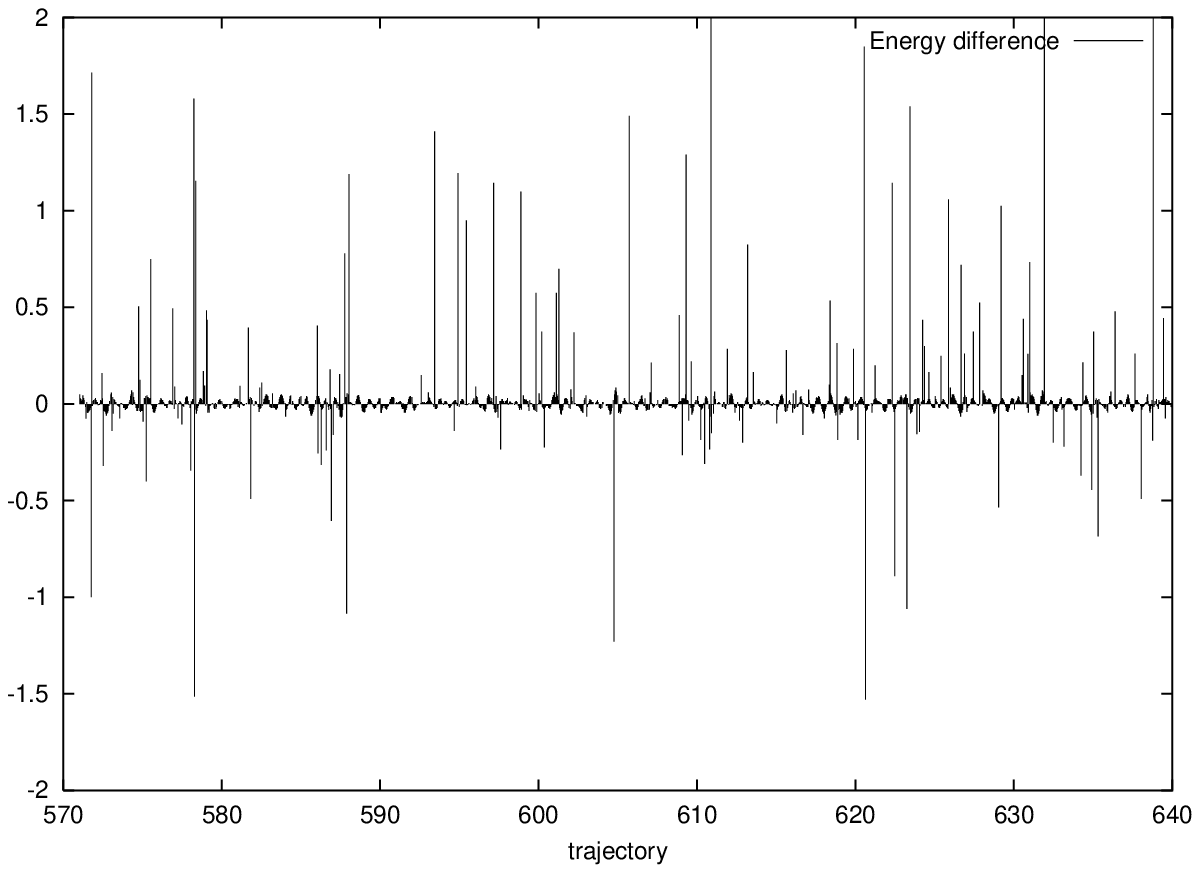}\\\includegraphics[width = 14cm,height =
  9cm]{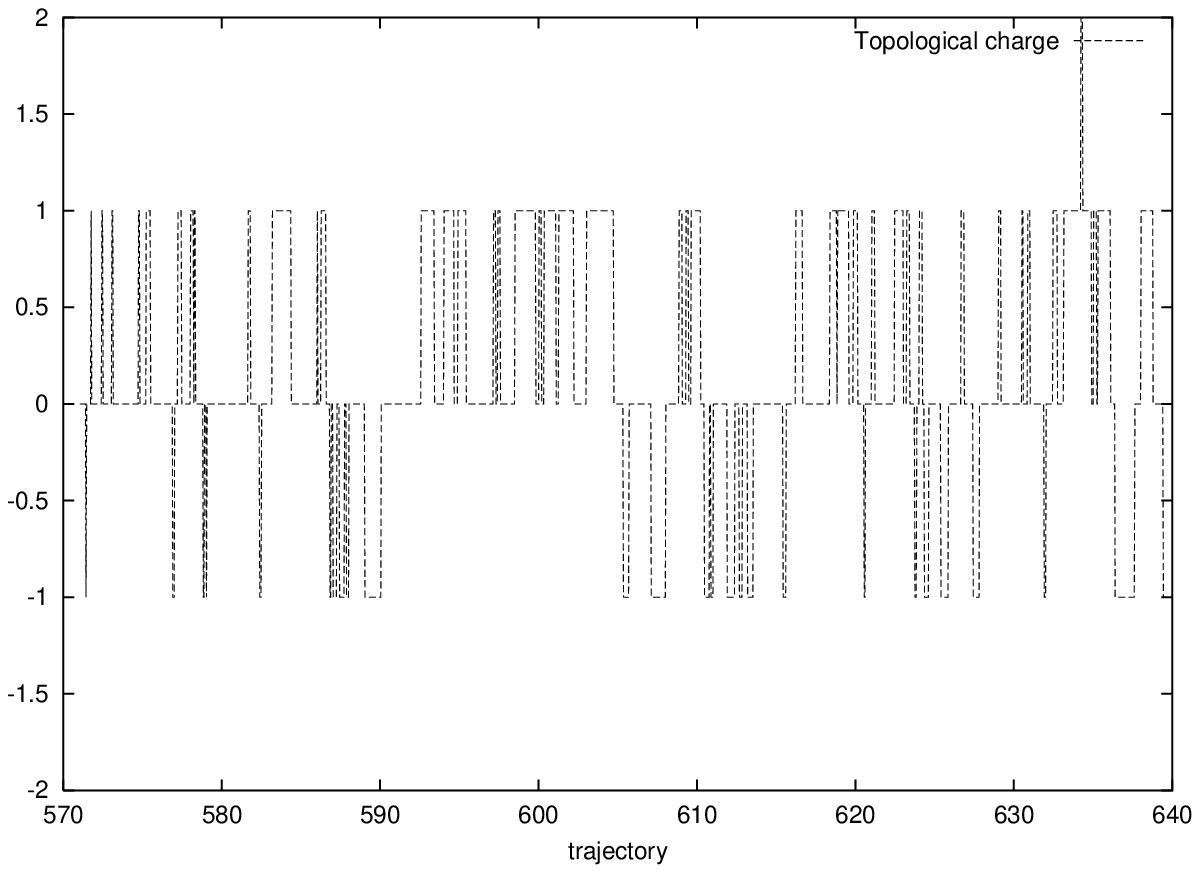}
\end{tabular}
\end{center}
\caption{The energy difference between two micro-canonical steps,
  and the topological index plotted against the trajectory number,
  for a $\beta = 5.4$, $\mu = 0.5$, $\kappa = 0.225$ ensemble
  generated without the correction step.}\label{fig:fig1b}
\end{figure}

\begin{figure}
\begin{center}
\begin{tabular}{c}
\includegraphics[width = 14cm,height =
  9cm]{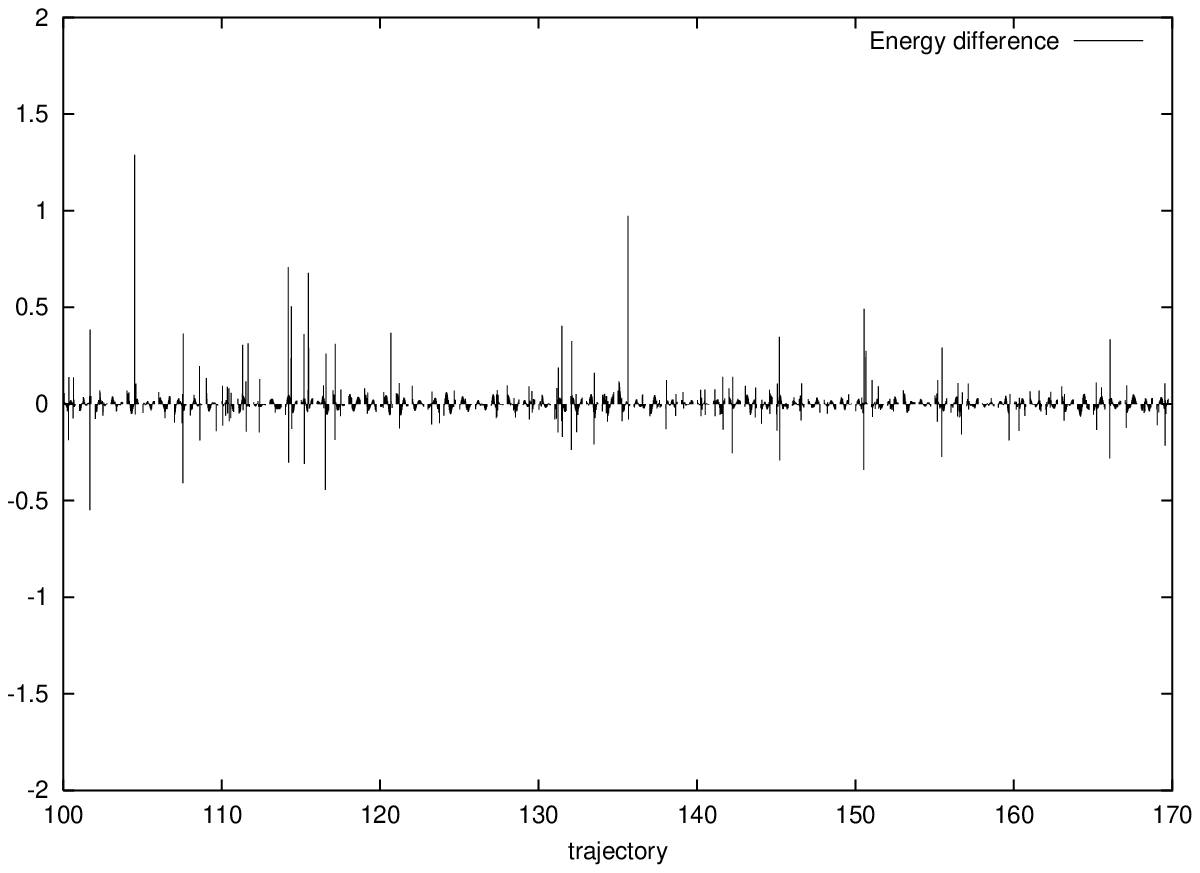}\\\includegraphics[width = 14cm,height =
  9cm]{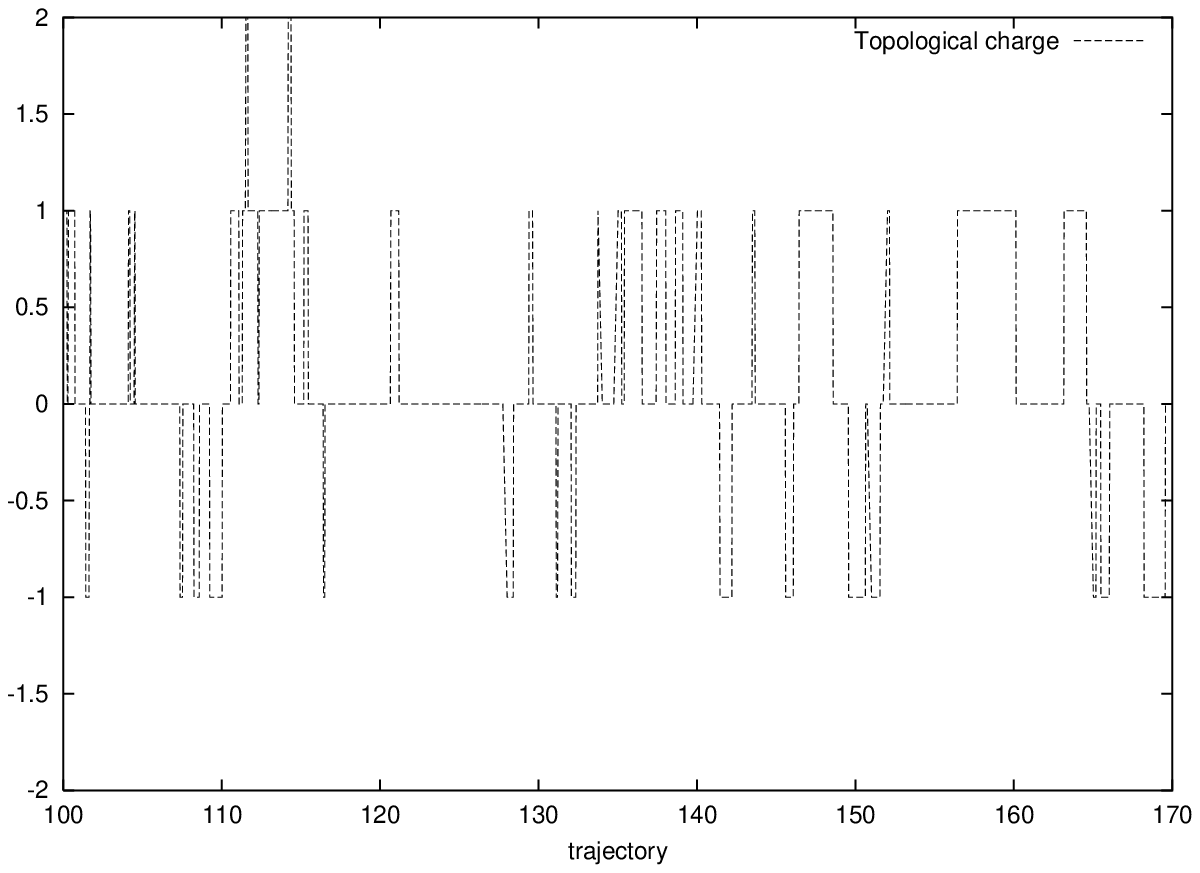}
\end{tabular}
\end{center}
\caption{The energy difference between two micro-canonical steps
  and the topological index plotted against the trajectory number
  for 70 trajectories of the $\beta = 5.4$, $\mu = 0.5$, $\kappa =
  0.225$ ensemble, generated with the correction
  step.}\label{fig:fig1}
\end{figure}

\subsubsection{The correction step and the number of
  topological index changes} 

Our first concern is to verify the impact of the correction step on
energy conservation, {\em i.e.} on the size of energy fluctuations
during a Monte Carlo run. This can be achieved by a comparison of
the upper plots of figures \ref{fig:fig1b} and \ref{fig:fig1}, which
display the energy variations during the individual trajectories for
a typical run. The lower plots in these figures exhibit the
tunnelling histories of the systems through the topological sectors.
A large spike in the energy difference signals a breakdown in energy
conservation. Without the correction step (figure \ref{fig:fig1b})
there are spikes in the energy difference whenever the topological
charge changes --- which occurs when there is an eigenvalue
crossing. This confirms our expectation that energy is not conserved
when the topological index changes if the standard leapfrog
algorithm is used. In all subsequent plots, we shall use \textit{the
  corrected update}. Figure \ref{fig:fig1} is based on the same
parameters as figure \ref{fig:fig1b}; generally we find that most of
the spikes disappear. The remaining spikes are caused when a positive
and a negative eigenvalue of $Q$ both approach, but do not cross, $0$
simultaneously. This can cause large mixing between the two
eigenvectors, and a very large fermionic force. This second effect
will not affect the acceptance rate. It should be noted that
sometimes the topological index bounces back within the same time
step, which can cause a small discontinuity in the energy (unless,
like our code, the HMC algorithm is designed to pick up this
possibility).

\begin{figure}
\begin{center}
\begin{tabular}{c}
\includegraphics[width = 14cm,height =
  9cm]{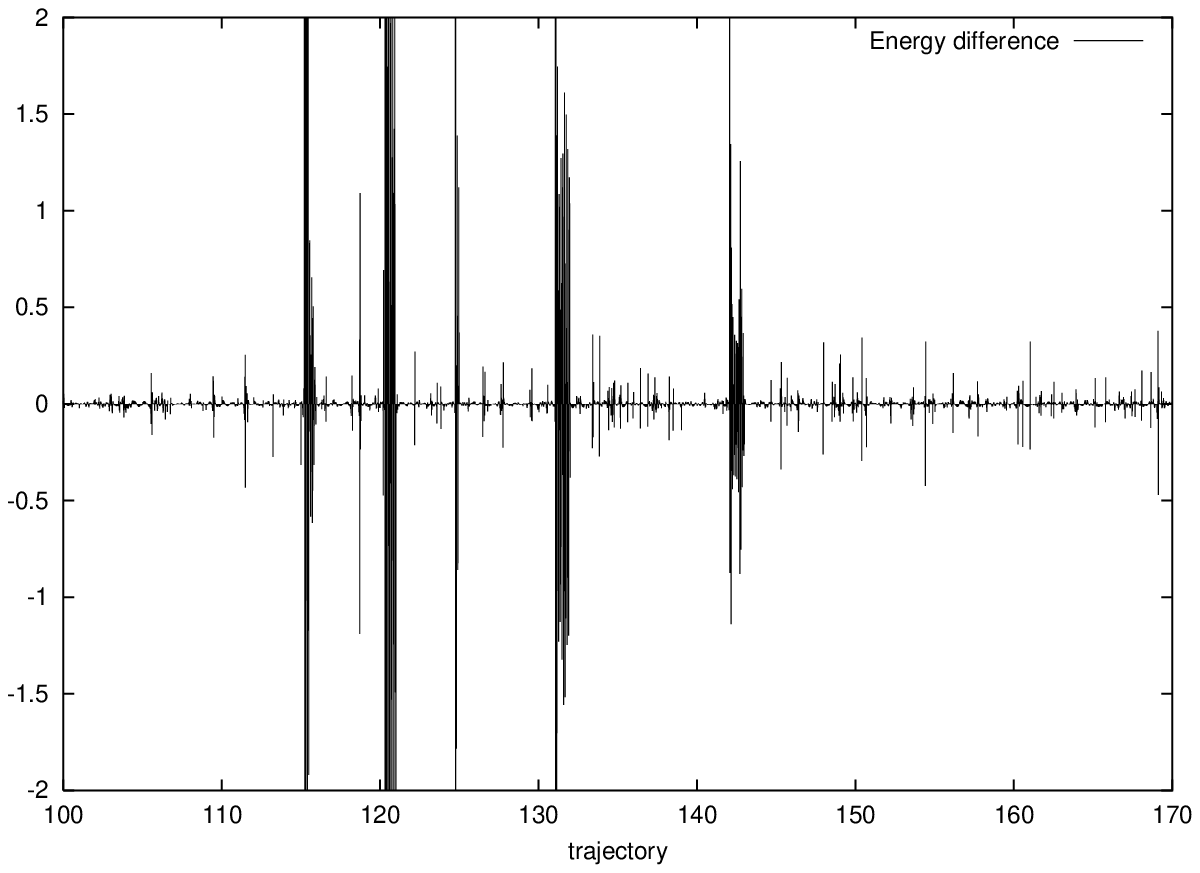}\\
\includegraphics[width = 14cm,height =
  9cm]{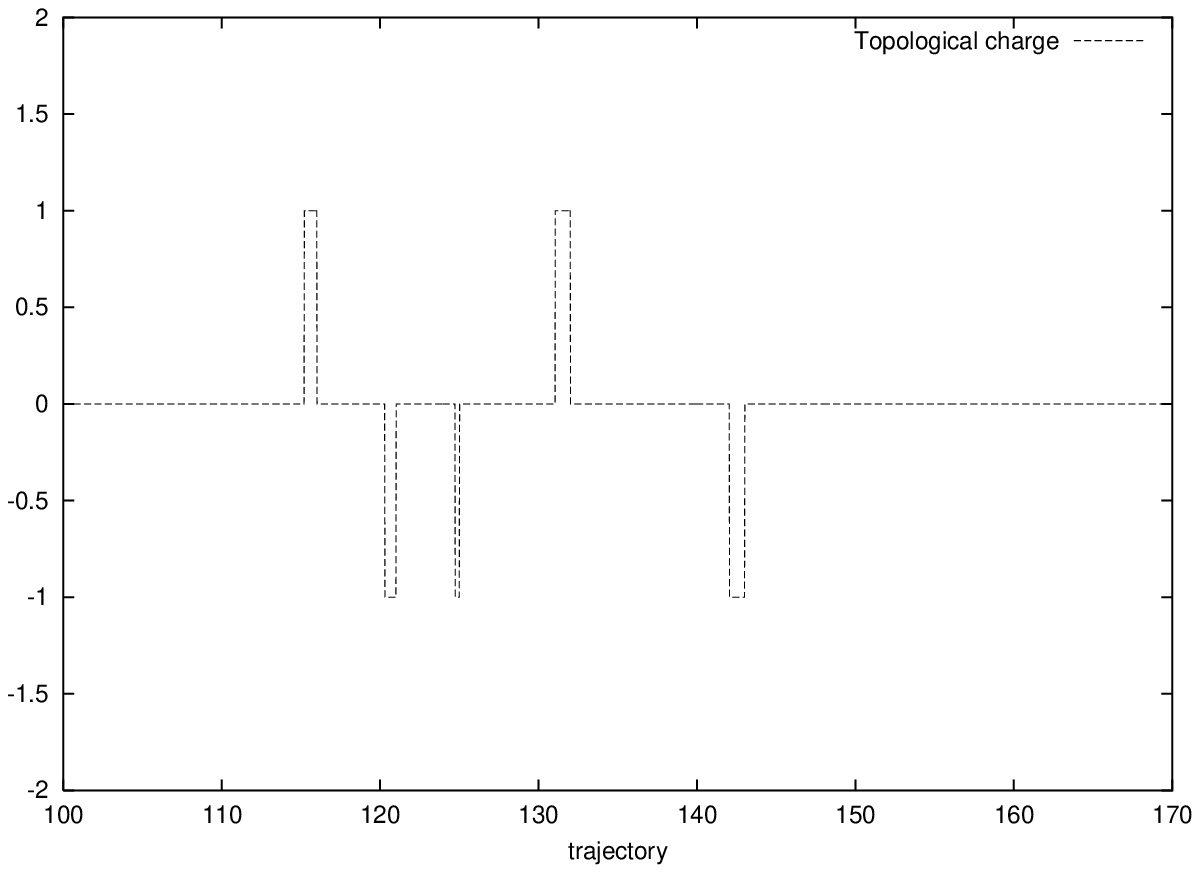}
\end{tabular}
\end{center}
\caption{The energy difference between two micro-canonical steps
  (top), and the topological index (bottom) plotted against the
  trajectory number for 70 trajectories of the $\beta = 5.4$ $\mu =
  0.05$ $\kappa = 0.225$ ensemble. We used the correction step to
  generate this ensemble.}\label{fig:fig2a}
\end{figure}

\begin{table}
\begin{center}
\begin{tabular}{|l l ||l l||l l||l l|}
\hline$4^4$ &$\mu = 0.5$&$4^4$& $\kappa = 0.225$&$6^4 $&$\kappa =
0.2$&$8^4 $&$\kappa = 0.2$\\$\kappa$&$n_{\text{top}}$&
$\mu$&$n_{\text{top}}$& $\mu$&$n_{\text{top}}$&
$\mu$&$n_{\text{top}}$\\ \hline 0.18&0.000(0)& 0.05&0.045(12)&
0.05&0.040(14) && \\ 0.19&0.006(6)& 0.1&0.103(20)& 0.1&0.074(18)&
0.1&0.153(42) \\ 0.2&0.122(31)& 0.2&0.393(39)& & &&
\\ 0.21&0.579(30)& 0.3&0.567(46)& 0.3&0.547(74)& 0.3&0.969(175)
\\ 0.22&0.901(104)& 0.4&0.827(52)& & && \\ 0.23&1.21(83)&
0.5&1.12(73)& & && \\ \hline \end{tabular}
\end{center}
\caption{The number of topological index changes per trajectory
  ($n_{\text{top}}$) for various masses on our $4^4$, $\kappa =
  0.225$ ensembles (left), the $4^4$, $\mu = 0.5$ ensembles
  (middle), and the $6^4$, $\kappa = 0.2$ ensembles (right). Note
  that the $4^4$ and $6^4$ ensembles were generated at different
  values of $\kappa$ (and $\beta$), so the bare quark mass at
  constant $\mu$ is $20\%$ lower for the $6^4$
  ensembles.}\label{tab:topchargechangemasskappa}
\end{table}
When we go down to smaller masses, a slightly different picture
emerges (see figure \ref{fig:fig2a} --- note that this figure includes
the correction step).  We notice that the topological index, $Q_f$, changes
considerably less frequently at a lower mass (see table
\ref{tab:topchargechangemasskappa}, $2^{\mbox{nd}}$-$4^{\mbox{th}}$
columns), as we would expect. The $Q_f\neq 0$ configurations are
suppressed as we move to lower masses, and as the mass decreases,
$d$ increases,\footnote{More precisely $d\propto \mu^{-2}$.} leading to a
lower probability of transmission.\footnote{The probability of
transmission is proportional to $\min(1,\exp(-2d))$.} The
energy difference depends strongly on the topological index: on the
$4^4$, $\mu = 0.05$ ensembles, the mass is about the same size as
the lowest non-zero eigenvalue, implying that the inverse of the
overlap operator, and therefore in general the fermionic force, will
be larger for a $Q_f\neq 0$ configuration. We have only observed this effect of the large fermionic forces with non-trivial topology on our smallest lattices, although it is possible that it will return in large volumes in the $\epsilon$-regime. It can be avoided by working in the chiral sector with no zero modes~\cite{BODE,Cundy:2005mr}.

\begin{figure}
\begin{center}
\begin{tabular}{c}
\includegraphics[width = 14cm,height =
  9cm]{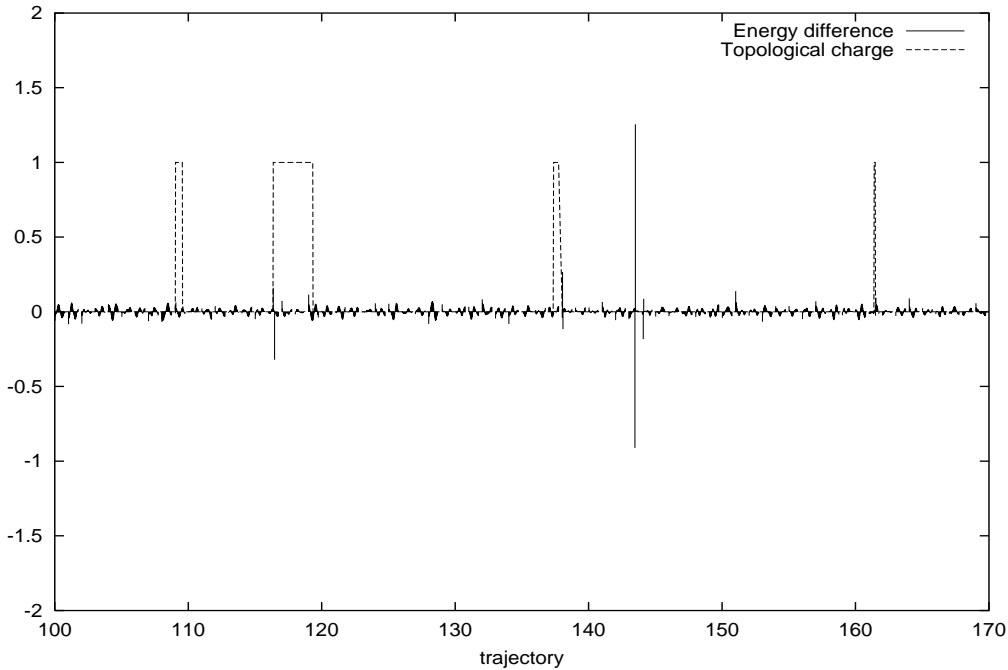}
\end{tabular}
\end{center}
\caption{The energy difference between two micro-canonical steps,
  and the topological index plotted against the trajectory number
  for a section of the $\beta = 5.4$ $\mu = 0.5$ $\kappa =
  0.2$ ensemble.}\label{fig:fig3}
\end{figure}

Finally, it can be seen that reducing $\kappa$ (and therefore
reducing the Wilson mass $m_0$ and as a consequence the bare fermion
mass) generally reduces the number of crossings (see figure
\ref{fig:fig3} and table \ref{tab:topchargechangemasskappa} (first
column)). As expected, see~\cite{Edwards:1999yi,Edwards:1998yw},
there are no changes in the topological index below the critical
value of $\kappa$, which is around 0.2 on the $4^4$ lattices. We
confirm that the overlap operator has very few small eigenvalues
below the critical value of $\kappa$~\cite{FODORCOMMUNICATION}.

\subsubsection{Complexity of correction step}\label{sec:energycons2}
To investigate the $\Delta\tau$ dependence of the energy
conservation violation for the correction step, we took 100 $\,$
$6^4$, $\mu = 0.3$ configurations, containing 35 transmission
correction steps and 123 reflection correction steps. We ran the
molecular dynamics as normal (with $\Delta\tau = 1/30$) up to the
eigenvalue crossing. We then reduced the time step for the molecular
dynamics to $\Delta\tau/n_t$, running $n_t-1$ normal leapfrog
updates and one modified leapfrog update. We calculated the absolute
value of the error in the energy for the modified leapfrog step only
and averaged over the configurations. This procedure ensures that on
each configuration the gauge field and momentum at the crossing have
only a small dependence on $n_t$. The energy differences for our
procedures (algorithm \ref{alg:algmod1}) and the algorithm proposed
in~\cite{FODOR,FODOR2} are given in tables \ref{tab:correctionerror}
for the transmission step and \ref{tab:correctionerror2} for the
reflection step. We expect (see appendix \ref{app:corrsteperror}) an
$O(\tau_c)$ error for the algorithm proposed in~\cite{FODOR,FODOR2}
and an $O(\tau_c^2)$ error for the algorithm presented in this
paper, where $-\Delta\tau/(2n_t)<\tau_c<\Delta\tau/(2n_t)$. Fitting
the results to the form $a_0{n_t}^{a_1}$ gave the results as
presented in table \ref{tab:correctionerrorfit}.

\begin{table}
\begin{center}
{\scriptsize
\begin{tabular}{|l l l | l l l| l l l|}
\hline&&&&~\cite{FODOR}&&&section \ref{sec:prescription}&
\\
$n_t$& $1/n_t$&$1/n_t^2$&$|\Delta E_{0}|$&$\frac{\Delta E_{0}}{\Delta
  E_{0}(n_t = 1)}$&$\Delta\tau\frac{\Delta E_0}{\tau_c}$&$|\Delta E|$&
  $\frac{\Delta E}{\Delta E(n_t = 1)}$&$\Delta\tau\frac{\Delta E}{\tau_c}$\\
\hline
1&1.000&1.000&
0.120(21)&1.000(172)&0.50(9)&
0.059(6)&1.000(105)&0.83(33)
\\
2&0.500&0.250&
0.050(9)&0.414(76)&0.37(5)&
0.009(1)&0.154(25)&0.14(4)
\\
3&0.333&0.111&
0.031(7)&0.261(57)&0.32(5)&
0.0030(7)&0.051(12)&0.16(8)
\\
4&0.250&0.063&
0.019(4)&0.154(32)&0.31(4)&
0.0016(4)&0.027(7)&0.21(14)
\\
5&0.200&0.040&
0.015(3)&0.122(27)&0.31(5)&
0.0009(2)&0.015(3)&0.032(8)
\\
6&0.167&0.028&
0.013(3)&0.107(27)&0.35(6)&
0.0005(1)&0.008(2)&0.018(4)
\\
7&0.143&0.020&
0.011(2)&0.090(19)&0.34(5)&
0.0004(1)&0.007(2)&0.019(4)
\\ \hline \end{tabular}}
\end{center}
\caption{The energy difference $\Delta E$ for the transmission
  correction algorithm \ref{alg:algmod1} compared with the molecular
  dynamics time step, $\Delta\tau = 1/30 n_t$. For comparison,
  results for the algorithm given in~\cite{FODOR}, here denoted as
  $\Delta E_{0}$, are also included.} \label{tab:correctionerror}
\end{table}

\begin{table}
\begin{center}
{\scriptsize
\begin{tabular}{|l l l | l l l| l l l|}
\hline&&&&~\cite{FODOR}&&&section \ref{sec:prescription}&
\\
$n_t$& $1/n_t$&$1/n_t^2$&$|\Delta E_{0}|$&$\frac{\Delta E_{0}}{\Delta
  E_{0}(n_t = 1)}$&$\Delta\tau\frac{\Delta E_0}{\tau_c}$&$|\Delta E|$&
  $\frac{\Delta E}{\Delta E(n_t = 1)}$&$\Delta\tau\frac{\Delta E}{\tau_c}$\\
\hline
1&1.000&1.000&
0.296(30)&1.000(101)&1.35(12)&
0.105(23)&1.000(215)&0.94(23)
\\
2&0.500&0.250&
0.133(13)&0.451(44)&1.16(10)&
0.0114(30)&0.109(28)&0.16(3)
\\
3&0.333&0.111&
0.099(11)&0.335(37)&1.12(9)&
0.0030(5)&0.028(4)&0.08(2)
\\
4&0.250&0.063&
0.064(7)&0.217(25)&1.16(10)&
0.0027(16)&0.026(2)&0.08(2)
\\
5&0.200&0.040&
0.059(6)&0.198(19)&1.16(9)&
0.0011(4)&0.011(4)&0.07(3)
\\
6&0.167&0.028&
0.045(5)&0.151(16)&1.12(9)&
0.00049(9)&0.005(1)&0.03(1)
\\
7&0.143&0.020&
0.036(4)&0.123(14)&1.12(9)&
0.00037(8)&0.004(1)&0.06(2)
\\ \hline \end{tabular}}
\end{center}
\caption{The energy difference $\Delta E$ for the reflection
  correction algorithm \ref{alg:algmod1} compared with the molecular
  dynamics time step, $\Delta\tau = 1/30 n_t$. For comparison,
  results for the algorithm given in~\cite{FODOR}, here denoted as
  $\Delta E_{0}$, are also included.} \label{tab:correctionerror2}
\end{table}

\begin{table}
\begin{center}
\begin{tabular}{|l|l l l|}
\hline
&$a_0$&$a_1$&$\chi^2$\\
\hline
$\Delta E_0$ (\cite{FODOR})&$0.119(^{+0.027}_{-0.026})$&$-1.271(^{+0.183}_{-0.173})$&0.52
\\
$\Delta E$ (section ~\ref{sec:prescription})&$0.058(^{+0.009}_{-0.009})$&$-2.642(^{+0.137}_{-0.139})$&1.11
\\
\hline
$\Delta E_0$ (\cite{FODOR})&$0.291(^{+0.037}_{-0.036})$&$-1.044(^{+0.094}_{-0.092})$&0.92
\\
$\Delta E$ (section ~\ref{sec:prescription})&$0.083(^{+0.029}_{-0.026})$&$-2.879(^{+0.266}_{-0.245})$&4.01
\\ \hline \end{tabular}
\end{center}
\caption{Fits and $\chi^2$ values for the data presented in tables
  \ref{tab:correctionerror} and \ref{tab:correctionerror2} for the
  functional form $\Delta E = a_0 {n_t}^{a_1}$ for transmission
  (top) and reflection (bottom). The errors are the $68\%$
  confidence levels for the fit.}\label{tab:correctionerrorfit}
\end{table}

First, consider the results for the algorithm presented
in~\cite{FODOR,FODOR2}. Comparing the raw data with $\Delta\tau$ and
$\Delta\tau^2$ in tables \ref{tab:correctionerror} and
\ref{tab:correctionerror2} suggests that the energy conservation is
dominated by $O(\Delta\tau)$ terms. This is also indicated by the
results of the fits in table
\ref{tab:correctionerrorfit}. Furthermore, $\Delta E_0/\tau_c$ is
large and constant (there are still some large contributions from
the standard leapfrog part of our algorithm at $n_t = 1$),
suggesting that the violation in energy conservation is dominated by
a large $O(\tau_c)$ term. For the energy differences given by our
algorithm, a different picture emerges: the violation in energy
conservation is dominated by the $O(\Delta\tau^3)$ terms from the
normal leapfrog part (steps 1, 2, 6 and 7) of the algorithm, while
the $O(\tau_c^2)$ error which we expect from the correction steps
(steps 3-5 in algorithm \ref{alg:algmod1}) is small. This is an
encouraging result that deserves to be investigated on larger
lattices and smaller masses.

\subsection{Polyakov loop and plaquette}

Next we calculate several observables to check that we get sensible
numbers.  Here we look at the value of the plaquette---or,
equivalently, the average gauge energy per lattice site, $S_{g}=
\big\langle1-\frac{1}{2 N_C} (U_{\mu\nu} +
U^{\dagger}_{\mu\nu})\big\rangle$ --- and the real part of the
Polyakov loop.
\begin{figure}
\begin{center}
\begin{tabular}{c}
\includegraphics[width = 14cm,height =
  9cm]{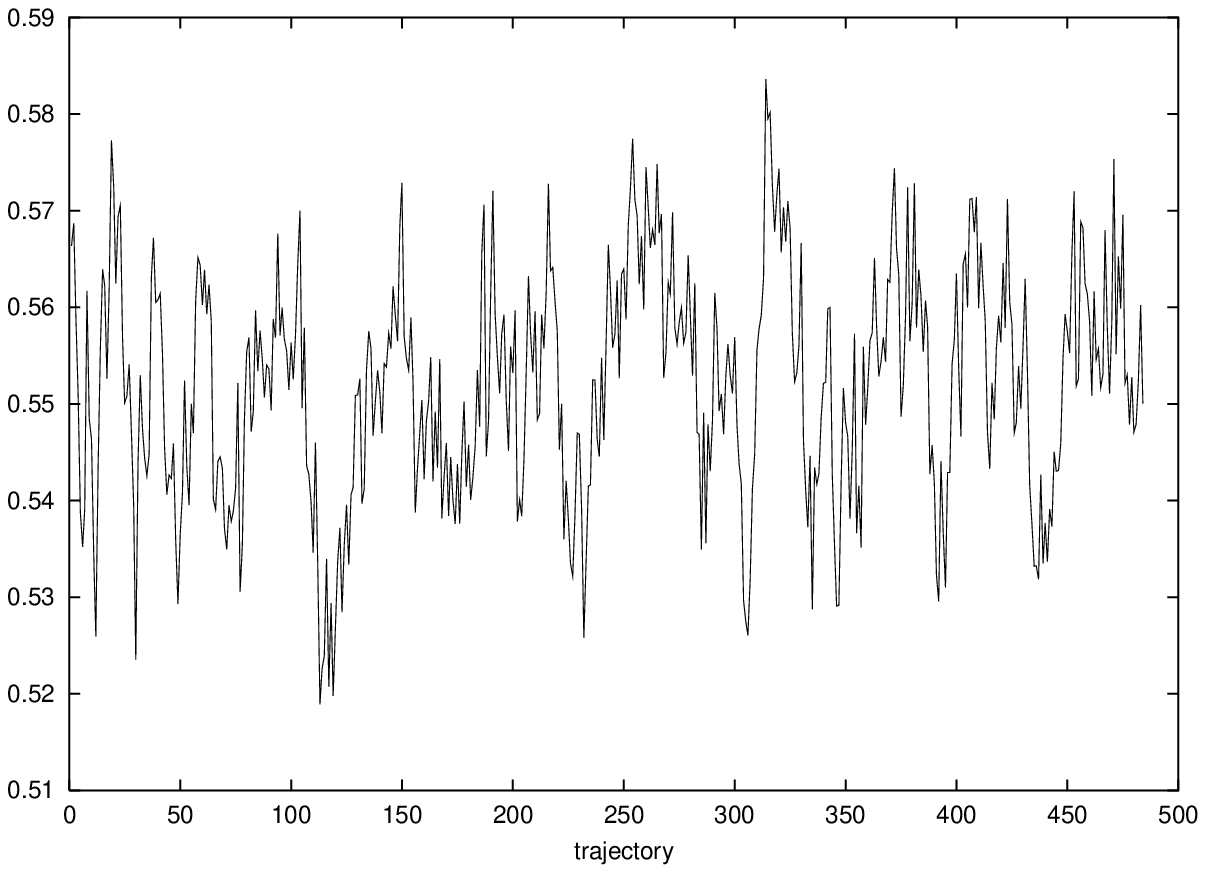}\\
\includegraphics[width = 14cm,height =
  9cm]{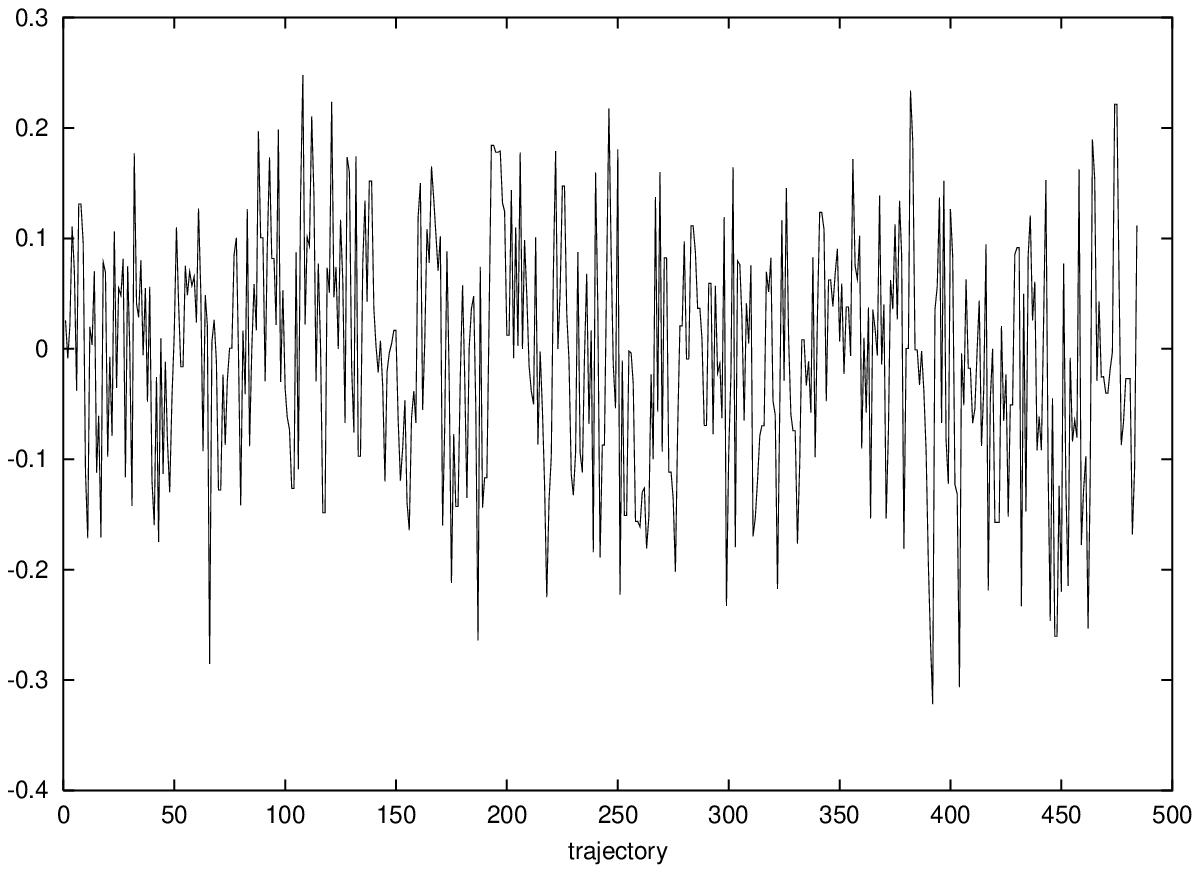}
\end{tabular}
\end{center}
\caption{The gauge energy per lattice site (top) and Polyakov loop
  (bottom) for each trajectory on the $4^4$ ,$\mu = 0.5$, $\kappa =
  0.225$ ensemble.}\label{fig:fig4}
\end{figure}

The values of the gauge energy per lattice site and the Polyakov
loop for one of our ensembles are plotted against computer time in
figure \ref{fig:fig4} (other ensembles are similar). Both the
plaquette and the Polyakov loop are stable over computer time. The
average Polyakov loop, given in tables \ref{tab:plaqmasskappa} and
\ref{tab:plaq64} is small for all our configurations, suggesting
that the configurations are all confined. The average value of
$S_{g}$ per lattice site is given in tables \ref{tab:plaqmasskappa}
and \ref{tab:plaq64}. The statistical errors on $S_{g}$ shown in
these tables were obtained by calculating the jackknife error
$\sigma_j^{(n _j)}$ with $n_j$ consecutive configurations in each
bin. The jackknife method should give a stable value when $n_j \gg
\tau_i$, the auto-correlation length. By fitting the jackknife errors
to the curve $\sigma_j^{(n _j)} = a + b(1-e^{-n_j/\tau_j})$ (the
errors on $\sigma_j^{(n _j)}$ were calculated using the bootstrap
method) we can get an estimate of the value of the plateau
($\sigma_j = a+b$), and a measure, $\tau_j$, of the exponential
auto-correlation length. On the $4^4$ lattices, the auto-correlation
length is small at the larger masses, but increases roughly as
$1/\mu$ as we decrease the mass. On the larger lattices, the
auto-correlation time is more stable with the mass. We were not able
to obtain a reliable estimate of the integrated auto-correlation
times on such small lattices.

\begin{table}
\begin{center}
{\scriptsize
\begin{center}
\begin{tabular}{|l|l l l|| l|l l l|}
\hline$\mu$&$<S_g>$&$\tau_j$&$P_l$&$\kappa$&$<S_g>$&$\tau_j$&$P_l$\\
\hline
0.05&0.5754(10)&25.05($^{+32.57}_{-12.63}$)&-0.0716(40)&
0.18&0.5485(14)&6.31($^{+1.66}_{-1.30}$)&-0.0137(31)
\\
0.1&0.5733(7)&12.76($^{+4.07}_{-4.71}$)&-0.0778(51)&
0.19&0.5522(8)&3.77($^{+0.71}_{-0.59}$)&-0.0153(58)
\\
0.2&0.5678(14)&11.90($^{+4.36}_{-3.21}$)&-0.0506(61)&
0.2&0.5555(10)&6.71($^{+1.55}_{-1.21}$)&-0.0054(50)
\\
0.3&0.5638(8)&4.60($^{+3.19}_{-1.94}$)&-0.0376(55)&
0.21&0.5573(8)&1.82($^{+0.71}_{-0.54}$)&-0.0241(80)
\\
0.4&0.5570(9)&3.98($^{+0.74}_{-0.62}$)&-0.0223(65)&
0.22&0.5560(11)&1.88($^{+2.14}_{-1.88}$)&-0.0266(94)
\\ 
0.5&0.5523(13)&4.46($^{+0.77}_{-0.94}$)&-0.0050(80)&
0.23&0.5528(14)&5.26($^{+1.36}_{1.08}$)&-0.0015(69)
\\ \hline \end{tabular}
\end{center}
}
\end{center}
\caption{The average gauge energy per lattice site, $<S_g>$, the
  real part of the Polyakov loop $P_l$ and jackknife auto-correlation
  lengths for the plaquette, for the $4^4$, $\kappa = 0.225$
  ensembles (left) and the $\mu = 0.5$ ensembles
  (right).}\label{tab:plaqmasskappa}
\end{table}

\begin{table}
\begin{center}
\begin{tabular}{|l|l l l|}
\hline$\mu$&$<S_g>$&$\tau_j$&$P_l$\\
\hline
0.05&0.5353(10)&9.97($^{+1.71}_{-1.48}$)&-0.0082(48)
\\
0.1&0.5350(6)&8.59($^{+0.79}_{-0.62}$)&-0.0049(33)
\\
0.3&0.5230(19)&10.93($^{+2.73}_{-2.18}$)&-0.0018(28)
\\ \hline \end{tabular}
\end{center}
\caption{The average gauge energy per lattice site, $<S_g>$, the
  real part of the Polyakov loop $P_l$ and jackknife auto-correlation
  lengths for the $6^4$ ensembles.}\label{tab:plaq64}
\end{table}

\subsection{Topological susceptibility}

\begin{figure}
\begin{center}
\begin{tabular}{c}
\includegraphics[width = 14cm,height =
  9cm]{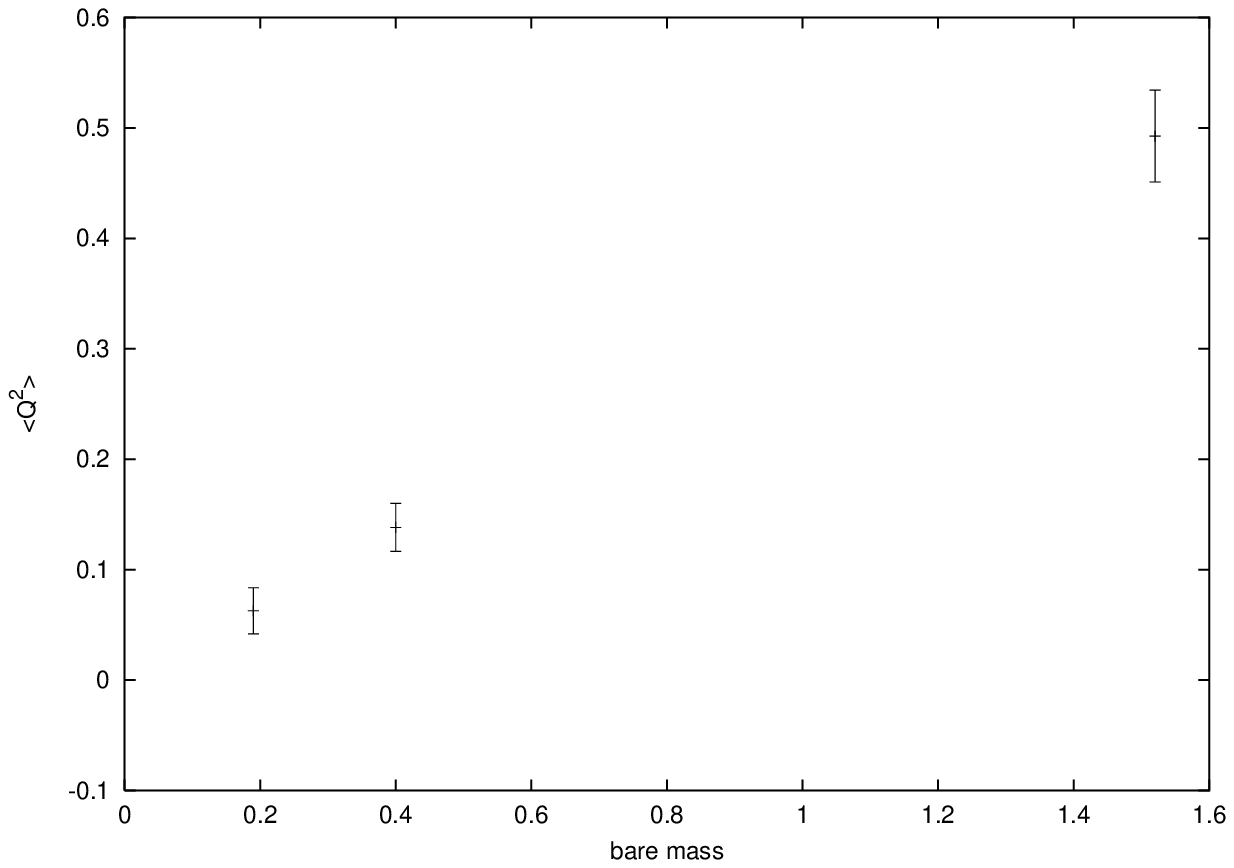}\\
\includegraphics[width = 14cm,height =
  9cm]{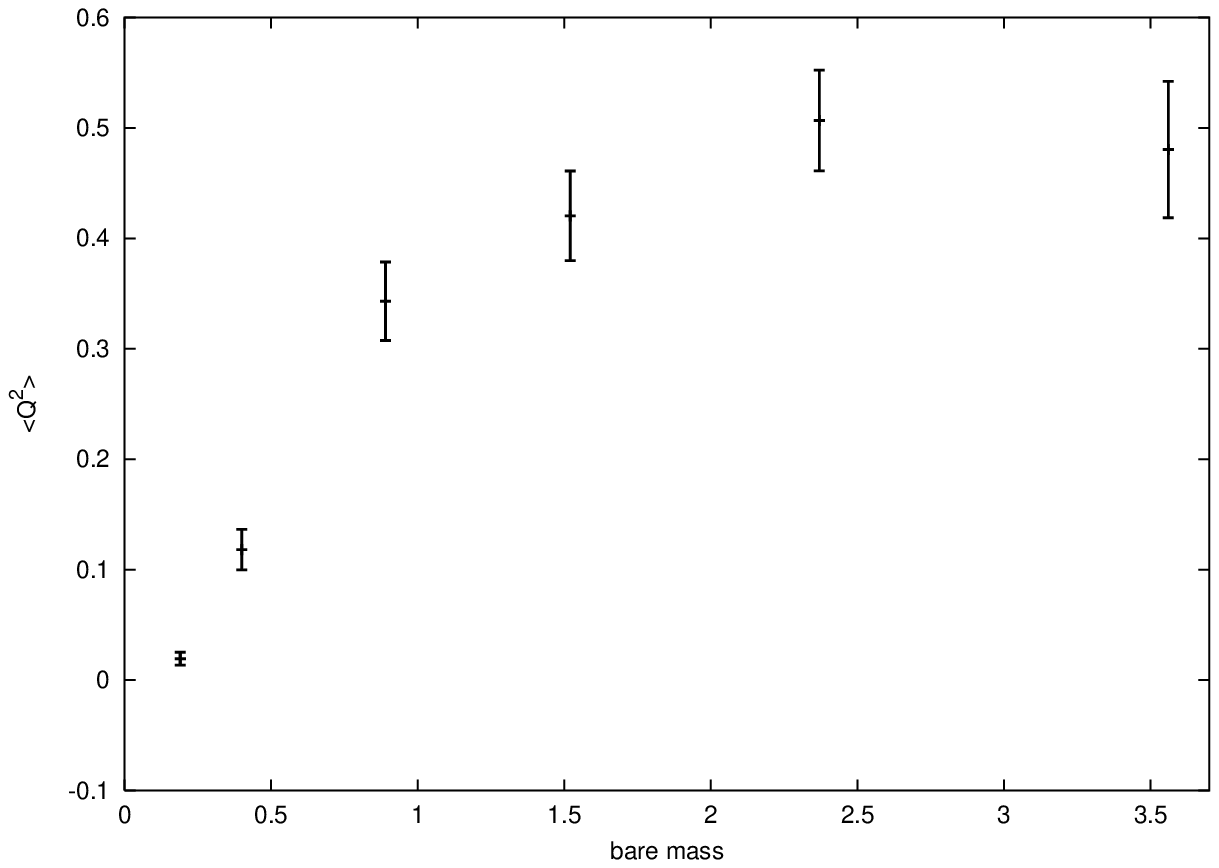}
\end{tabular}
\end{center}
\caption{The ensemble average of $Q^2_f$ plotted against bare quark
  mass on the $6^4$ (top) and $4^4$ (bottom)
  ensembles. }\label{fig:sus}
\end{figure}

The average values of the topological index, $Q_f=n_- - n_+$
($n_{\pm}$ is the number of positive/negative chirality zero modes),
and of $Q_f^2$ are given in tables \ref{tab:susmass} and
\ref{tab:sus64} and are plotted in figure \ref{fig:sus}. The tables
show that the average values of $Q_f$ are generally close to 0,
suggesting that there is no bias towards either a positive or negative topological index.

The topological susceptibility, which is proportional to $<Q_f^2>$, can
be related to the quark mass using chiral perturbation
theory~\cite{Crewther,Vecchia,Leutwyler}. It is expected that at low
quark masses on large enough volumes the
topological susceptibility should be proportional to the quark mass,
the square of the pion mass: 
\begin{gather}
\chi = \frac{<Q_f^2>}{V} \sim \frac{f_{\pi}^2 m_{\pi}^2}{2N_F} + O(m_{\pi}^4).
\end{gather}
There have been several attempts to verify this relation in recent
years (for recent examples, see~\cite{Bernard:2003gq,Bali:2001gk}),
with mixed success. At larger quark masses, the topological
susceptibility should tend asymptotically towards its quenched
value. The transition between the two forms, in a large volume, is
expected to be around 50-90 MeV~\cite{Durr:2001ty}. Our bare quark
masses, which we estimate (based on our early calculations of the
lattice spacing on our $6^4$ ensembles) range from about 100 MeV to
a few GeV and should be in the transitional region between the two
known limits: we cannot expect to see a linear decrease with the
quark mass. However, we recognize a decrease in the topological
susceptibility, and our results are not inconsistent with the
expected functional form, although again we emphasise that our
lattices are far too small to draw any meaningful conclusions.

\begin{table}
\begin{center}
\begin{tabular}{|l l|l l l ||l|l l l|}
\hline$\mu$&$m_b$&$<Q_f^2>$ &$<Q_f>$&$n_{\text{conf}}$&$\kappa$&$<Q_f^2>$ &$<Q_f>$&$n_{\text{conf}}$\\
\hline
0.05&0.19&0.019(6)&$-0.011(5)$&734&
0.18&0.000(0)&0.000(0)&459
\\
0.1&0.40&0.118(18)&0.021(33)&959&
0.19&0.003(4)&0.004(4)&733
\\
0.2&0.89&0.343(36)&0.044(34)&868&
0.2&0.044(16)&$-0.001(14)$&760
\\
0.3&1.52&0.421(41)&$-0.085(35)$&967&
0.21&0.255(35)&0.014(41)&442
\\
0.4&2.37&0.507(46)&0.034(39)&905&
0.22&0.461(44)&$-0.047(62)$&706
\\
0.5&3.56&0.480(62)&0.057(47)&523&
0.23&0.667(89)&0.088(66)&433
\\ \hline \end{tabular}
\end{center}
\caption{The average values of the topological index $Q_f$,
  $Q_f^2$, and the number of configurations for the $4^4$, $\kappa =
  0.225$ ensembles(left) and the $4^4$, $\mu = 0.5$ ensembles
  (right). $m_b$ is the bare fermion mass, given by equation
  (\ref{eq:barefermionmass}).}\label{tab:susmass}
\end{table}

\begin{table}
\begin{center}
\begin{tabular}{|l l|l l l|}\hline
$\mu$&$m_b$&$<Q_f^2>$ &$<Q_f>$&$n_{\text{conf}}$\\
\hline
0.05&0.19&0.063(21)&$-0.008(21)$&377
\\
0.1&0.40&0.138(22)&0.026(20)&350
\\
0.3&1.52&0.493(62)&0.086(61)&420
\\ \hline \end{tabular}
\end{center}
\caption{The average values of the topological index $Q_f$,
  $Q^2_f$, and the number of configurations for the $6^4$
  ensembles. $m_b$ is the bare fermion mass.}\label{tab:sus64}
\end{table}

\subsection{Testing the area conservation of the reflection condition.}\label{sec:tuningdmax}

\begin{table}
\begin{center}
{\scriptsize
\begin{tabular}{|l |l l l| l l  l|}
\hline condition&$n_{\text{conf}}$&AR&$n_{\text{top}}$&$<Q_f^2>$ &$<S_{g}>$&$<P_l>$\\
\hline
$|d|> 0.5$&709&81.9\%&1.048(68)&0.354(35)&0.5625&$-0.0325$
\\
$|d|> 0.45$&855&84.1\%&0.978(57)&0.326(26)&0.5653&$-0.0286$
\\
$|d|> 0.4$&864&86.2\%&0.920(60)&0.317(33)&0.5640&$-0.0377$
\\
$|d|> 0.35$&460&86.5\%&0.843(74)&0.350(33)&0.5645&$-0.0270$
\\
$|d|> 0.3$&705&87.8\%&0.688(59)&0.311(35)&0.5646&$-0.0300$
\\
$|d|>0.25$&698&92.0\%&0.598(50)&0.317(34)&0.5657&$-0.0431$
\\
$|d|>0.2$&604&96.0\%&0.439(55)&0.285(31)&0.5653&$-0.0353$
\\
\hline
$1+\frac{d_2}{\sum_k(\eta^2_k,\Pi_-)^2}>0$&768&97.8\%&1.047(62)&0.357(32)&0.5639&$-0.0357$
\\ \hline \end{tabular}
}
\end{center}
\caption{The acceptance rate, number of topological index changes
  per trajectory, $n_{\text{top}}$, and average values of $Q_f^2$,
  $S_g$ and $P_l$ for $4^4$ ensembles generated at $\mu = 0.3$,
  $\kappa = 0.225$ and $\beta = 5.4$, but with different reflection
  conditions.}\label{tab:suscond}
\end{table}

As remarked earlier, the reflection condition applied is to be
tested numerically in order to verify that area conservation is
fulfilled and no systematic errors are spoiling the results.  We compared the results from our algorithm with the
results from a second algorithm which explicitly satisfies detailed
balance, but where the molecular dynamics does not always conserve
energy. This algorithm reflected whenever $d$ was smaller than a
constant $d_{max}$ and used no correction step for
transmission. When $d$ was larger than this constant, it violated
energy conservation.  A high $d_{\text{max}}$ leads to many
topological index changes and therefore a short topological
auto-correlation length but has a low acceptance rate. A low
$d_{\text{max}}$ would have a higher acceptance rate but fewer
topological index changes.

In table \ref{tab:suscond}, we show how a varying $d_{\text{max}}$
affects the acceptance rate and the number of topological index
changes per trajectory. Using the exact area-conserving algorithm
has a cost in either the acceptance rate or rate of topological
charge changes. Setting $d_{max}=0.5$ has a similar number of
topological index changes per trajectory, but at a cost of 15\% in
the acceptance rate. Lowering $d_{max}$ so that the acceptance rate
is similar to the momentum-dependent algorithm reduces the number of
topological index changes by over a factor of 2, and increases the
auto-correlation correspondingly.

The table shows that the average values of plaquette, topological
susceptibility and Polyakov loop are independent of the reflection
condition. We therefore feel confirmed in assuming that a potential
 systematic error is extremely small. However, this issue will
need to be checked again when we move to larger lattices at smaller
masses.

\subsubsection{Volume scaling of the overlap HMC}

The authors of Ref.~\cite{Degrandschaefer2} remark that the
algorithm might scale with the square of the lattice
volume. Their argument goes as follows:
\begin{enumerate}
\item The density of small eigenvalues should be independent of the volume;
  therefore the number of small eigenvalues scales with the
  volume.
\item\label{point2} The number of Wilson eigenvalues attempting
  to cross the zero line will be proportional to the density of
  small eigenvalues. Therefore, there will be O($V$) correction
  steps.\footnote{See the results of~\cite{Schaefer:2006bk}; we are intending to test these results.}
\item In total, the algorithm might scale as $V^2$.
\end{enumerate}
There are, however, arguments against this reasoning:
The density of small eigenvalues is not just a function of the lattice volume as point (\ref{point2}) suggests, but also the lattice spacing and the fermion mass. Thus,
concluding that the number of correction steps will be
proportional to the volume is too simplistic, as on larger volumes we are also likely to change the other parameters in the action. One could use the same argument to suggest that the algorithm will accelerate as the lattice spacing decreases, since there will be fewer attempted crossings. More importantly, this argument does not take into account the effect of the auto-correlation. Our numerical experience shows that the overlap eigenvalues are remarkably stable under the molecular dynamics, except when there is a topological index change. Our results~\cite{2006slft.confE..49C} demonstrated that small non-zero eigenvalues of the overlap operator are only generated significantly faster through the mechanism of two topological index changes rather than the normal evolution of the molecular dynamics. Thus the auto-correlation time depends strongly on the rate of topological index changes, which in turn, assuming that the transmission rate is large enough, increases proportional to the rate of attempted eigenvalue crossings. Thus the more attempted topological index changes, the shorter the auto-correlation times should be.

Nonetheless, we expect problems with the condition number of both the overlap and Wilson operators increasing as the volume increases, leading to a slowing down in both the eigenvalue and inversion routines.

Results from test runs on lattices ranging from size $4^4$ to
$16^3 32$ show that the number of required correction steps is
certainly not proportional to the volume; still it is increasing
slightly.

\subsection{Computer performance}

The timings to generate a $4^4$ trajectory are shown in table
\ref{tab:timemass}. They are based on running on four nodes of
the cluster computer ALiCE at Wuppertal University. The $6^4$
ensembles (table \ref{tab:timelarger}) were generated on eight
nodes and the $8^4$ ensembles on 16 nodes.\footnote{In parallel
  computations, we estimated---by explicitly calculating the
  number of floating point operations for one trajectory---that
  ALiCE runs at around 0.4 GFlops/node for the $4^4$ ensembles
  and 0.7 GFlops/node for the $6^4$ and $8^4$ ensembles.}  We
used CG with relaxation and preconditioning~\cite{cundy2} to
perform the inversion, which has approximately a factor of 4
performance gain over the standard CG.  Our timings are quite
independent of the mass, although we expect this fact to change
on larger and less well conditioned systems. However, the scaling
of the computer time with mass will be better than the $\mu^{-3}$
behaviour which we assume for Wilson or staggered fermions
because with overlap fermions there are fewer crossings for small
masses. It can be seen by comparing the timings on the $\mu =
0.5$ ensembles (table \ref{tab:timemass}) below and above the
critical $\kappa$ that at this large mass the time needed for the
treatment of the crossings was around $40\%$ of the total
computer time (there are usually 3-4 reflection or transmission
steps per trajectory at $\mu = 0.5$ and $\kappa$ well above the
critical $\kappa$). The acceptance rate was stable for all our
$4^4$ ensembles.  We increased the number of molecular dynamics
steps ($n_{md}$) to 50 (keeping $\Delta\tau\, n_{md} = 1$) for
the $4^4$ ensemble generated at $\mu = 0.05$ and $\kappa = 0.225$
in order to counteract the large fermionic force in non-trivial
topological sectors (see section \ref{sec:energycons}), but
otherwise we did not have to change the trajectory length as we
changed the mass to maintain the acceptance rate.  On our larger
lattices, we had to increase $n_{md}$ at lower masses.

\begin{table}
\begin{center}
\begin{tabular}{|l|l l l||l|l l l|}
\hline$\mu$&Time &$n_{md}$&AR&$\kappa$&Time &$n_{md}$&AR \\
\hline
0.05&2322(130)&50&94\%&
0.18&1150(40)&30&92\%
\\
0.1&1713(200)&30&87\%&
0.19&1339(80)&30&95\%
\\
0.2&1782(130)&30&87\%&
0.2&1275(70)&30&92\%
\\
0.3&1779(110)&30&90\%&
0.21&2192(100)&30&91\%
\\
0.4&1836(180)&30&87\%&
0.22&2485(130)&30&90\%
\\
0.5&1972(220)&30&90\%&
0.23&1937(190)&30&90\%
\\ \hline \end{tabular}
\end{center}
\caption{The average timings for one trajectory (in seconds), the
  number of molecular dynamics steps and the acceptance rate for
  the $\kappa = 0.225$ ensembles (left), and the $\mu = 0.5$
  ensembles (right). The total time of the trajectory,
  $\Delta\tau n_{md}$, is equal to $1$.}\label{tab:timemass}
\end{table}

\begin{table}
\begin{center}
\begin{tabular}{|l|l l l|}
\hline Configuration&Time &$n_{md}$&AR \\
\hline
$6^4$ $\mu = 0.05$&6714(250)&40&84\%
\\
$6^4$ $\mu = 0.1$&6674(240)&40&86\%
\\
$6^4$ $\mu = 0.3$&5535(110)&30&79\%
\\
$8^4$ $\mu = 0.1$&30049(1087)&50&79\%
\\
$8^4$ $\mu = 0.3$&26456(540)&40&80\%
\\ \hline
\end{tabular}
\end{center}
\caption{The average timings for one trajectory (in seconds), the
  number of molecular dynamics steps, and the acceptance rate on
  the larger lattices, with $\kappa = 0.2$.  The total time of
  the trajectory, is equal to $1$ .}\label{tab:timelarger}
\end{table}

\section{Conclusions and Outlook}

In this paper, we have presented the basics of an improved
simulation scheme for full QCD including the chirally symmetric
overlap fermion discretization. We have constructed an exact
Hybrid Monte Carlo algorithm being capable of generating gauge
field configurations with two flavours of dynamical overlap
fermions.  The HMC algorithm proposed aims at the exploration of
the low quark mass regime of lattice QCD with a lattice action
observing exact chiral symmetry.  The method can be extended to
solve the problem of low modes that are mixing as well as to
improve topological tunnelling rates.

The main conceptual difficulty constructing a dynamical overlap
algorithm is the appearance of a Dirac delta function within the
fermionic force. A naive straightforward integration leads to action
differences from the discrete leap frog integration which turn out
to be much too large to allow acceptance in the global Monte
Carlo decision step at the end of a trajectory. These large action
differences are caused by zero-mode crossings of the low eigenvalues
of the Wilson kernel used within the overlap operator. They force
the algorithm to slow down and to get stuck eventually.

In order to avoid the action violations as caused by eigenvalue
crossings, we have developed an explicit integration procedure for
evolving the low crossing modes within the molecular dynamics part
of the HMC. In case of reflection on the action wall the change in
energy is zero by definition and also in case of transmission a
quite smooth integration over the singularity is achieved.  The new
scheme can cope with multiple eigenvalue crossings within the same
MD time step.

The proposed HMC for overlap fermions has several advantages:
\begin{itemize}
\item[--] The action differences are small enough to allow for
  reasonable acceptance rates of the HMC.
\item[--] The scheme is constructed to allow for a systematic
  improvement of the complexity of the action violation as arising
  from a finite integration time-step.
\item[--] The action violation of the specific realization
  constructed in this paper is comparable to the $O(\Delta\tau^2)$
  error of the standard HMC leap-frog integration over the entire
  trajectory.
\item[--] The algorithm is shown to satisfy reversibility and area
  conservation.
\item[--] The universal ansatz in this paper for the correction
  step in terms of orthonormal vectors $\eta_k$ has the potential
  to construct a {\em non-area-conserving} higher-order
  integration method. This approach has the potential to solve
  the problem of the mixing of low modes that occasionally leads
  to large action violations on larger lattices, spoiling the
  acceptance rate even after correction of single mode zero
  crossings. Such an improved scheme, based on the present
  approach, will be described in a subsequent paper
  \cite{Cundy:2007df}.
\item[--] In addition, the universal ansatz of this paper for the
  correction step is part of the improvement of the tunnelling of the
  topological index to be described in an upcoming paper
  \cite{Cundy:2008zc}.
\end{itemize}

We have tested the algorithm on Teraflop-Class computers with
lattice sizes ranging from $4^4$ to $8^4$, calculating a set of
basic physical observables.  Despite of large finite-size effects
appearing on such small lattices, we do get results for, {\em
  e.g.}, the topological susceptibility, which behave qualitatively
as expected. This demonstrates that intermediate lattices are in
reach of Teraflop-class computers.

With the help of the new algorithm and improvements to be
presented in three forthcoming papers~\cite{Cundy:2007df,Cundy:2008zc,cundyforthcoming08}, larger lattice sizes, required to realize
physically meaningful systems, will become accessible on upcoming
Petaflop-class systems.

\section*{Acknowledgements}

We would like to thank W. Bietenholz, A. Borici, S. Katz, T. Kennedy,
S. Schaefer, K. Szabo,  V. Weinberg and Z. Fodor for many
important discussions and comparisons of results. We would also
like to thank the anonymous referee for his many useful
comments. NC was supported by the EU Marie-Curie fellowship
Number MC-EIF-CT-2003-501467, by the Deutsche
Forschungsgemeinschaft under grant Li 701/5-1 and the support of grant 930183 from the EU RP-6 "Hadron Physics" project, from the DFG "Gitter-Hadronen Ph\"anomenologie" project, number 458/14-4. Furthermore, this
research was part of the EU integrated infrastructure initiative
HADRONPHYSICS project under contract number RII3-CT-2004-506078
and later the EU integrated infrastructure project I3HP
``Computational Hadron Physics'' contract
No. RII3-CT-2004-506078.  SK and GA were supported by the
Deutsche Forschungsgemeinschaft under grants Li 701/4-1 and Li
701/5-2. The numerical work was carried out on the cluster
computers ALiCE and ALiCEnext at Wuppertal University and in a
final stage on the Blue Gene /L supercomputer, a cray XD-1 and PC cluster at the J\"ulich
Supercomputing Centre. We thank the John von Neumann Institute
for Computing for granting a substantial amount of computer time
on the BG/L for this project.

\appendix
\section*{Appendix}

\section{Detailed Balance and the Size of the Correction Step Error}\label{app:detailedbalance}

In this section we prove that the general correction update
(\ref{eq:mostgensolution}) satisfies detailed balance. Extending this
proof to the slightly modified update in equations (\ref{eq:40}) and
(\ref{eq:41}) is straightforward.

\subsection{Detailed balance and reversibility}

The initial and final leap frog steps (steps 1 and 2, and 6 and 7)
in algorithm \ref{alg:algmod1} are known to be reversible. Thus, it just has to
be shown that the correction steps (steps 3, 4 and 5) are
reversible.  Here, the proof of reversibility for the correction
step for the case of transmission in algorithm \ref{alg:algmod1} is
given. The proof for the reflection update is similar.

\subsubsection{The momentum update}
It is assumed that when reversing the sign of $\Delta\tau$,
$d_k\rightarrow -d_k\;\forall k$. This condition is satisfied for our algorithms, because $\tau_c$ and $\eta^j_k$ remain constant,
$F^+\rightarrow F^-$ and $d\rightarrow -d$, since the sign of
the smallest eigenvalue is opposite for the reverse update. Our
general transmission momentum update (\ref{eq:mostgensolution})
reads:
\begin{align}
\Pi^+ =& \Pi^- - \tau_c(F^- - F^+) + \tau_c\eta(\eta,F^- - F^+) +
  \frac{1}{3}\Tr(F^- - F^+) + \nonumber\\
&\eta  (\Pi^-,\eta)\left(\sqrt{1+ \frac{d_1
  }{ \left(
  \Pi^-,\eta\right)^2}} - 1\right)\nonumber\\
& + \sum_k\left(\eta^k_1 (\eta^k_1, \Pi^- - \frac{\tau_c}{2}(F^- + F^+)) +\eta^k_1 (\eta^k_2, \Pi^-
- \frac{\tau_c}{2}(F^- + F^+))\right)\times\nonumber\\
&\left(\sqrt{1 +
  \frac{d_k}{(\eta^k_1,
    \Pi^- - \frac{\tau_c}{2}(F^- + F^+))^2 +  (\eta^k_2, \Pi^- -
    \frac{\tau_c}{2}(F^- + F^+))^2}} - 1\right),\label{eq:momupdate}
\end{align}
so that
\begin{align}
&(\Pi^+,\eta) = (\Pi^-,\eta)\sqrt{1+ \frac{ d_1
  }{ \left(\Pi^-,\eta\right)^2}};\nonumber\\
&(\Pi^+ + \frac{\tau_c}{2}(F^- + F^+),\eta_{k_i}) =  (\eta_{k_i}, \Pi^- -
  \frac{\tau_c}{2}(F^- + F^+))\times\nonumber\\
&\phantom{space}\left(\sqrt{1 +
  \frac{d_k}{(\eta^k_1,
    \Pi^- - \frac{\tau_c}{2}(F^- + F^+))^2 +  (\eta^k_2, \Pi^- -
    \frac{\tau_c}{2}(F^- + F^+))^2}} - 1\right).
\end{align}
The time-reverse update, with $\Delta\tau\rightarrow -
\Delta\tau$, is given by
\begin{align}
\Pi'_- =& \Pi^+ +\tau_c(F^+ - F^-) - \tau_c\eta(\eta,F^+ - F^-)
 -\frac{1}{3}\Tr(F^- - F^+) +\nonumber\\
&    \eta  (\Pi^+,\eta)\left(\sqrt{1- \frac{d_1
 }{ \left(
  \Pi^+,\eta\right)^2}} - 1\right)+\nonumber\\
& \sum_k\left(\eta^k_1 (\eta^k_1, \Pi^+ - \frac{\tau_c}{2}(F^- + F^+)) +\eta^k_2 (\eta^k_2, \Pi^+
- \frac{\tau_c}{2}(F^- + F^+))\right)\times\nonumber\\
&\left(\sqrt{1 -
  \frac{d_k}{(\eta^k_1,
    \Pi^+ - \frac{\tau_c}{2}(F^- + F^+))^2 +  (\eta^k_2, \Pi^+ -
    \frac{\tau_c}{2}(F^- + F^+))^2}} - 1\right).\label{eq:rev1}
\end{align}
We note that
\begin{align}
(\Pi^+,\eta)&\left(\sqrt{1- \frac{d_1
  }{\left(
  \Pi^+,\eta\right)^2}} - 1\right)\nonumber\\ =& (\Pi^-,\eta)\sqrt{1+ \frac{d_1
  }{ \left(
  \Pi^-,\eta\right)^2}}\left(\sqrt{1- \frac{d_1}{ \left(
  \Pi^-,\eta\right)^2\left(1+ \frac{d_1}{ \left(\Pi^-,\eta\right)^2}\right)}} - 1\right)\nonumber\\
=&(\Pi^-,\eta)\left(1-\sqrt{1+ \frac{d_1}{ \left(
  \Pi^-,\eta\right)^2}}\right);\label{eq:rev2}\\
(\Pi^+ -& \frac{\tau_c}{2}(F^- + F^+),\eta^k_1)\times\nonumber\\
&\left(\sqrt{1 -
  \frac{d_k}{(\eta^k_1,
    \Pi^+ + \frac{\tau_c}{2}(F^- + F^+))^2 +  (\eta^k_2, \Pi^+ -
    \frac{\tau_c}{2}(F^- + F^+))^2}} - 1\right)\nonumber\\
=&  -(\Pi^- - \frac{\tau_c}{2}(F^- + F^+),\eta^k_1)\times\nonumber\\
&\left(\sqrt{1 +
  \frac{d_k}{(\eta^k_1,
    \Pi^- - \frac{\tau_c}{2}(F^- + F^+))^2 +  (\eta^k_2, \Pi^- -
    \frac{\tau_c}{2}(F^- + F^+))^2}} - 1\right).\label{eq:rev3}
\end{align}
Combining (\ref{eq:momupdate}), (\ref{eq:rev1}), (\ref{eq:rev2}) and
(\ref{eq:rev3}) gives $\Pi'_- = \Pi^-$; therefore, the momentum
update is reversible.


\subsubsection{The gauge field update}

The gauge field update is given by:
\begin{gather}
U^+ = e^{-i\tau_c\Pi^+}e^{i\tau_c\Pi^-}U^-\label{eq:gfrev}.
\end{gather}
Switching $\Delta\tau\rightarrow - \Delta\tau$ leads to
\begin{gather}
U'_- = e^{-i\tau_c\Pi^-}e^{i\tau_c\Pi^+}U^+,
\end{gather}
which gives $U'_- = U^-$; therefore, the gauge
field update is reversible.

\subsection{Detailed Balance: Area Conservation}\label{sec:areaconstrans}

\subsubsection{Transmission}
The transmission update, given by algorithm \ref{alg:algmod1} and
using equation (\ref{eq:mostgensolution}) can be shown to be area
conserving. For simplicity, it is assumed that $d_1$ is a function
of the gauge field at the crossing only and has no additional
dependence on $\Pi$, and all the other $d_k$s are functions of the
gauge field of the crossing and $(\eta,\Pi)$ only. The matrices
$\eta^i_j$ and the fermionic forces $F^{\pm}$ are just functions of
the gauge field at the crossing. These conditions can be relaxed,
but not necessarily for the purpose of this paper.

$D_i$ is a derivative of the gauge fields $U$, defined as
\begin{gather}
D_i f(U) = \lim_{h\rightarrow 0}\frac{\partial}{\partial h} f\left(e^{ih T_i} U\right),
\end{gather}
where $T_i$ are some suitable generators of the SU($N_C$) gauge
group ({\em e.g.} in SU(3), $T_i= \lambda_i/2$, where $\lambda_i$
are the Gell-Mann matrices). One can write the gauge fields in
terms of $N_C^2 - 1$ parameters $r_i$ so that
\begin{gather}
D_i f(U) = \frac{\partial}{\partial r_i} f(U).\label{eq:A1.9}
\end{gather}
Similarly, one can define coordinates for the momentum $\Pi$ and the
matrix $\eta$ in terms of the generators $T$:
\begin{align}
 \Pi_{\mu}(x)=&\pi_i^{x\mu} T_i, &\eta_{\mu}(x) =& \eta_i^{x\mu}
 T_i,&F_{\pm}-\frac{1}{3}\Tr(F_{\pm}) = &f_{\pm i}^{x\mu}T_i,\nonumber\\
\eta^k_{j,\mu}(x) = &n_{j, i}^{k,x\mu} T_i. 
\end{align}
$x$ refers to the lattice site, and $\mu$ is a direction
index. The normalization condition $(\eta,\eta) = 1$ implies that
$n^k_j$ is normalized according to $n_{j,i}^{k,x\mu}
n_{j,i}^{k,x\mu} = 2$. For notational convenience, we shall
define:
\begin{align}
N_i^{x\mu}(U,\Pi) =& \frac{1}{2}\eta^{x\mu}_i(U)
\eta_j^{y\nu}(U)\pi_j^{y\nu}\left(\sqrt{1+\frac{4d_1}{(\eta_k^{z\rho}(U)\pi_k^{z\rho})^2}}-1\right)
- \nonumber\\
& \tau_c\left(\delta_{ij}\delta^{xy}\delta^{\mu\nu} -
\eta^{x\mu}_i(U)\eta^{y\nu}_j(U)\right)\left(f_{-j}^{y\nu}(U) -
f_{+j}^{y\nu}(U)\right)+\nonumber\\
&\sum_k\left(n_{1,i}^{k,x\mu}(U) N_{e1}^{k}(U,\Pi) +n_{2,i}^{k,x\mu}(U) N_{e2}^{k}(U,\Pi)\right)\nonumber\\&  \left(\sqrt{1 +
  \frac{d_k}{(N_{e1}^{k}(U,\Pi))^2 + (N_{e2}^{k}(U,\Pi))^2}} - 1\right).\\
N_{e1}^{k} = &\frac{1}{2}\left(n_{1,j}^{k,z\rho}\left(\pi_j^{z\rho} -
\frac{\tau_c}{2}(f_{+j}^{z\rho} -
f_{-j}^{z\rho})\right)\right),\nonumber\\
N_{e2}^{k} = &\frac{1}{2}\left(n_{2,j}^{k,z\rho}\left(\pi_j^{z\rho} -
\frac{\tau_c}{2}(f_{+j}^{z\rho} -
f_{-j}^{z\rho})\right)\right).
\end{align}
The generator of a time update $\Delta\tau$ is
\begin{align}
e^{\Delta\tau \frac{\partial}{\partial\tau}} = & e^{\Delta\tau \left(\frac{\partial
    \pi_i}{\partial\tau}\frac{\partial}{\partial \pi_i} + \frac{\partial
    r_i}{\partial\tau}\frac{\partial}{\partial r_i}\right)}\nonumber\\
=& e^{\Delta\tau \left(-\frac{\partial
    E}{\partial r_i}\frac{\partial}{\partial \pi_i} + \frac{\partial
    E}{\partial \pi_i}\frac{\partial}{\partial r_i}\right)}\nonumber\\
=&e^{\Delta\tau \hat{H}}.
\end{align}
For the standard leapfrog update, we define the operators
\begin{align}
\hat{P} =&e^{\frac{\Delta\tau}{2} F'_i\frac{\partial}{\partial \pi_i}}
  \nonumber\\
\hat{Q}=&e^{\frac{\Delta\tau}{2} \pi_i \frac{\partial}{\partial r_i}},
\end{align}
then we have
\begin{align}
\hat{P}f(\pi,r) =& f(\pi+F'\frac{\Delta\tau}{2},r)&\hat{Q}f(\pi,r) =& f(\pi,r +
\frac{\Delta\tau}{2} \pi).
\end{align} 
$F'$ is the total force in the momentum update (the sum of the
gauge field and fermionic forces). The standard leapfrog update
is given by
\begin{gather}
\hat{P}\hat{Q}\hat{Q}\hat{P} = 
e^{\Delta\tau(\hat{H}
    + O(\Delta\tau^2))}.
\end{gather}
$\hat{P}\hat{Q}\hat{Q}\hat{P}$ is manifestly area conserving,
because the operators $\hat{P}$ and $\hat{Q}$ are area
conserving.

For the correction step, we need to define two new update operators,
$\hat{P}_c$ and $\hat{Q}_c$, so that $\hat{Q}_c f(\pi,r) =
f(\pi,r - \tau_c\pi)$ and $\hat{P}_cf(\pi,r) = f(\pi
+ N_i,r)$.  
$\hat{P}_c$ and $\hat{Q}_c$ are
\begin{align}
\hat{P}_c = & 1 + N_i \frac{\partial}{\partial \pi_i} + \frac{1}{2}
N_i N_j \frac{\partial^2}{\partial \pi_i\pi_j} + \frac{1}{6}
N_i N_jN_k \frac{\partial^3}{\partial \pi_i\pi_j\pi_k} +
\ldots\nonumber\\
\hat{Q}_c = & 1- \tau_c \pi_i\frac{\partial}{\partial r_i} 
+ \frac{1}{2}\tau_c^2 \pi_i\pi_j\frac{\partial^2}{\partial
  r_ir_j} + \frac{1}{6}\tau_c^3 \pi_i\pi_j\pi_k\frac{\partial^3}{\partial
  r_ir_jr_k} +\ldots.
\end{align}
Because
$[\partial/\partial r_i,\tau_c] \neq 0$ and
$[\partial/\partial r_i,N_j] \neq 0$, we cannot factorise
$\hat{P}_c$ and $\hat{Q}_c$ into a simple exponential form, although
using equation (\ref{eq:A1.9}) we can write
\begin{gather}
\hat{Q}_c U=  e^{-i\tau_c\pi_i T_i} U = e^{-i\tau_c\Pi^-} U.
\end{gather}
 These operators
by themselves are not area conserving.
The full correction update is
$\hat{Q}_c^{\dagger}\hat{P}_c\hat{Q}_c$:
\begin{align}
 \hat{Q}_c^{\dagger}\hat{P}_c\hat{Q}_c \Pi^- = & \Pi^- + \frac{1}{2}\eta (\eta_j^{x\mu}\pi_j^{x\mu})\left(\sqrt{1+\frac{4d_1}{(\eta_m^{y\nu}(\tau_c)\pi_m^{y\nu})^2}} -
  1\right)+ \nonumber\\
&(F^+ - F^-) - \frac{1}{2}\eta \eta^{x\mu}_{j}(f_+ - f_-)^{x\mu}_j
  +\nonumber\\
&\sum_k(\eta^{k}_{1} N_{e1}^{k} + \eta^{k}_{2} N_{e2}^{k})\left(\sqrt{1+\frac{d_k}{(N_{e1}^{k})^2 + (N_{e2}^{k})^2}} - 1\right)= \Pi^+\nonumber\\
\hat{Q}_c^{\dagger}\hat{P}_c\hat{Q}_c U^-
  =&\hat{Q}_c^{\dagger}\hat{P}_c e^{-i \tau_c \Pi^-} U^-\nonumber\\
=&\hat{Q}_c^{\dagger}e^{-i \tau_c (\Pi^- + N(U^-,\Pi^-))}
  U^-\nonumber\\
=&e^{-i \tau_c ({\Pi}_- + N(U_c,{\Pi}_-))} e^{i{\tau}_c\Pi^-} U^- = U^+,
\end{align}
where $d$, $f$, $\eta$ etc. are all calculated at the crossing gauge
field $U_c$.
Therefore, the update in algorithm \ref{alg:algmod1} can be
written as $\hat{P}\hat{Q}\hat{Q_c}^{\dagger}\hat{P}_c\hat{Q}_c
QP$. We know that $\hat{P}$ and $\hat{Q}$ are area conserving, so we
just need to show that $\hat{Q_c}^{\dagger}\hat{P}_c\hat{Q}_c$ is also
area conserving. We write the updated gauge field and momentum as
$q$ and $p$ respectively (recalling that the initial fields were $r$
and $\pi$ in our notation). This gives us
\begin{align}
&p^{x\mu}_i = \hat{Q_c}^{\dagger}\hat{P}_c\hat{Q}_c\pi^{x\mu}_i =  \pi^{x\mu}_i
+ \frac{1}{2}\eta_i^{x\mu} \pi^{y\nu}_j \eta^{y\nu}_j \left(\sqrt{1+\frac{4
    d_1}{(\pi^{z\rho}_m\eta^{z\rho}_m)^2}} - 1\right) - \nonumber\\
&\phantom{p^{x\mu}_i =} \tau_c\left(\delta_{ij}\delta^{xy}\delta^{\mu\nu} -
\frac{1}{2}\eta^{x\mu}_i\eta^{y\nu}_j\right)\left(f_{-j}^{y\nu} -
f_{+j}^{y\nu}\right)+\nonumber\\
&\phantom{p^{x\mu}_i =}\sum_k\left(n_{1,i}^{k,x\mu} N_{e1}^{k} +n_{2,i}^{k,x\mu} N_{e2}^{k}\right)\left(\sqrt{1 +
  \frac{d_k}{(N_{e1}^{k})^2 +(N_{e2}^{k})^2 }} - 1\right)\label{eq:alphaeq};\\
&q^{x\mu}_i =\hat{Q_c}^{\dagger}\hat{P}_c\hat{Q}_c r =
 r^{x\mu}_i + \tau_c(\pi^{x\mu}_i - p^{x\mu}_i)\label{eq:epsiloneq}.
\end{align} 
It has to be shown that the determinant
\begin{gather}
J = \left|\begin{array}{c c}
\frac{\partial p_i}{\partial\pi_m}&\frac{\partial q_i}{\partial\pi_m}\\
\frac{\partial p_i}{\partial r_m}&\frac{\partial q_i}{\partial r_m}\end{array}\right|
= 1.
\end{gather} 
$ r_c$ is the gauge field at the eigenvalue crossing:
\begin{gather}
 r^{x\mu}_{ci} = \hat{Q}_c r_i^{x\mu} =  r^{x\mu}_i + \pi^{x\mu}_i \tau_c.
\end{gather}
Thus, we can write
\begin{align}
\tau_c =& \frac{( r_c^{y\nu} -  r^{y\nu})_i\eta^{y\nu}_i}{\pi^{z\rho}_j\eta^{z\rho}_j},\\
\frac{\partial{\tau_c}}{\partial\pi^{x\mu}_k} = & \frac{1}{\pi_j^{z\rho}\eta_j^{z\rho}}\left(\frac{\partial
   r^{y\nu}_{ci}}{\partial\pi^{x\mu}_k}\eta^{y\nu}_i - \tau_c \eta^{x\mu}_k +
\frac{\partial{\eta_j^{y\nu}}}{\pi^{x\mu}_k}\left( r_c^{y\nu} -  r^{y\nu} - \tau_c
\pi^{y\nu}\right)_j\right),\\
 \frac{\partial{\tau_c}}{\partial r^{x\mu}_k} = & \frac{1}{\pi_j^{z\rho}\eta_j^{z\rho}}\left(\frac{\partial
   r^{y\nu}_{ci}}{\partial r^{x\mu}_k}\eta^{y\nu}_i  - \eta^{x\mu}_k +
\frac{\partial{\eta^{y\nu}_j}}{\partial r^{x\mu}_k}\left( r^{y\nu}_c -  r^{y\nu} - \tau_c
\pi^{y\nu}\right)_j\right).
 \end{align}
We choose $\eta$ so that
\begin{gather}
\frac{\partial
   r^{y\nu}_{ci}}{\partial r_k^{x\mu}}\eta^{y\nu}_i=\frac{\partial
   r^{y\nu}_{ci}}{\partial\pi_k^{x\mu}}\eta^{y\nu}_i = 0.\label{eq:etaeqn}
\end{gather}
Hence
\begin{gather}
\frac{\partial \tau_c}{\partial \pi_k^{x\mu}}  = \tau_c
\frac{\partial \tau_c}{\partial  r_k^{x\mu}} = -\tau_c \frac{\eta^{x\mu}_k}{\pi_j^{z\rho}\eta_j^{z\rho}}.
\end{gather}
Since $ r_c$ is a point on the surface of constant $\lambda = 0$, one solution to
(\ref{eq:etaeqn}) is
that $\eta_i$ is normal to this surface, i.e. proportional to the vector
\begin{align}
\frac{\partial \lambda}{\partial  r^{x\mu}_{ci}} =& -\kappa
\bra{\psi(x)}\gamma_5\left((1-\gamma_{\mu})\frac{\partial U}{\partial
   r^{x\mu}_{ci}}\delta_{x,y+\mu} +(1+\gamma_{\mu})\frac{\partial U^{\dagger}}{\partial
   r^{x\mu}_{ci}}\delta_{x,y-\mu}\right)  \ket{\psi(y)}\label{eq:etadef}.
\end{align}
We also have for any function $g$ of $ r + \pi \tau_c$ (such as
$\eta$ or $d$),
\begin{gather}
\frac{\partial g}{\partial \pi^{z\rho}_k} = \tau_c \frac{\partial
  g}{\partial  r^{z\rho}_k}.
\end{gather}
We are now in a position to write down the components of $J$ by
differentiating (\ref{eq:alphaeq}) and (\ref{eq:epsiloneq}). By
writing $p$ and $q$ as functions of $r_c$ and $\pi$, we have 
\begin{align}
\left(\frac{\partial p_i^{x\mu}}{\partial\pi_m^{y\nu}}\right)_r =
&\left(\frac{\partial p_i^{x\mu}}{\partial\pi_m^{y\nu}}\right)_{r_c} +
\left(\frac{\partial p_i^{x\mu}}{\partial
  r_{cj}^{z\rho}}\right)_{\pi}\left(\frac{\partial
  r_{cj}^{z\rho}}{\partial \pi_{m}^{y\nu}}\right)_{r}\nonumber\\
=&\left(\frac{\partial p_i^{x\mu}}{\partial\pi_m^{y\nu}}\right)_{r_c}
+ \tau_c \left(\frac{\partial p_i^{x\mu}}{\partial
  r_{m}^{y\nu}}\right)_{\pi}\nonumber\\
=&\delta_{im}\delta_{xy}\delta_{\mu\nu} + \frac{1}{2}\eta^{x\mu}_i
\eta^{ y\nu}_m
\left(\frac{1}{\sqrt{1+\frac{4d_1}{(\pi^{z\rho}_n\eta^{z\rho}_n)^2}}} - 1\right) +
\tau_c\frac{\partial  p_i^{x\mu}}{\partial
   r_m^{y\nu}}+\nonumber\\
&\frac{1}{2}\sum_kn_{1,i}^{k,x\mu}n_{1,m}^{k, y\nu}\left(-1 + \frac{1 + (N^{k}_{e2})^2\frac{d_k}{((N^{k}_{e1})^2 + (N^{k}_{e2})^2)^2}}{\sqrt{1+\frac{d_k}{(N^{k}_{e1})^2 + (N^{k}_{e2})^2}}}\right) + \nonumber\\
&\frac{1}{2}\sum_kn_{2,i}^{k,x\mu}n_{2,m}^{k, y\nu}\left(-1 + \frac{1 + (N^{k}_{e1})^2\frac{d_k}{((N^{k}_{e1})^2 + (N^{k}_{e2})^2)^2}}{\sqrt{1+\frac{d_k}{(N^{k}_{e1})^2 + (N^{k}_{e2})^2}}}\right) - \nonumber\\
&\frac{1}{2}\sum_k\left(n_{1,i}^{k,x\mu}n_{2,m}^{k, y\nu}+n_{2,i}^{k,x\mu}n_{1,m}^{
k,  y\nu}\right)\left(
\frac{N^{k}_{e1} N^{k}_{e2}\frac{d_k}{((N^{k}_{e1})^2 + (N^{k}_{e2})^2)^2}}{\sqrt{1+\frac{d_k}{(N^{k}_{e1})^2 + (N^{k}_{e2})^2}}}\right)
+ \nonumber\\
&\frac{1}{2}\sum_k\left(n_{1,i}^{k,x\mu}N_{e1}^{k} +
n_{2,i}^{k,x\mu}N_{e2}^{k}\right)\eta_m^{k, y\nu}\frac{\left(\frac{\partial d_k}{\partial(\eta^{z\rho}_n\pi^{z\rho}_n)}\right)_{r_c}}{\sqrt{1+\frac{d_k}{(N^{k}_{e1})^2 + (N^{k}_{e2})^2}}}.
\\
\frac{\partial q_i^{x\mu}}{\partial\pi_m^{y\nu}} = &-\tau_c\frac{\partial p_i^{x\mu}}{\partial\pi_m^{y\nu}} +
\tau_c\left(\frac{\partial q_i^{x\mu}}{\partial r_m^{y\nu}} +
\tau_c\frac{\partial p_i^{x\mu}}{\partial r_m^{y\nu}}\right). \\
\frac{\partial q_i^{x\mu}}{\partial r_m^{y\nu}} = & \delta_{im}
\delta_{xy}\delta_{\mu\nu} -\frac{\eta^{y\nu}_m}{\eta^{z\rho}_j
  \pi^{z\rho_j}}(\pi^{x\mu}_i - p^{x\mu}_i) - \tau_c \frac{\partial
  p^{x\mu}_i}{\partial r_m^{y\nu}}.
\end{align}
Therefore, the Jacobian is
\begin{align}
J = &\left|\begin{array}{l l}
\frac{\partial p_i}{\partial\pi_k}&\frac{\partial q_i}{\partial\pi_k}\\
\frac{\partial p_i}{\partial r_k}&\frac{\partial q_i}{\partial r_k}\end{array}\right|
= \left|\begin{array}{c c}
\frac{\partial p_i}{\partial\pi_i}&\frac{\partial q_i}{\partial\pi_k}
+ \tau_c\frac{\partial p_i}{\partial\pi_k}\\
\frac{\partial p_i}{\partial r_k}&\frac{\partial q_i}{\partial r_k}+
\tau_c\frac{\partial p_i}{\partial r_k}\end{array}\right|\nonumber\\
=&\left|\begin{array}{l c l}
\frac{\partial p_i}{\partial\pi_k} - \tau_c\frac{\partial p_i}{\partial r_k}&\phantom{s}&\frac{\partial q_i}{\partial\pi_k}
+ \tau_c\frac{\partial p_i}{\partial\pi_k} - \tau_c\left(\frac{\partial q_i}{\partial r_k}+
\tau_c\frac{\partial p_i}{\partial r_k}\right)=0\\
\frac{\partial p_i}{\partial r_k}&\phantom{s}&\frac{\partial q_i}{\partial r_k}+
\tau_c\frac{\partial p_i}{\partial r_k}\end{array}\right|.
\end{align}
So $J$ is the product of the two determinants 
\begin{gather}
\left|\frac{\partial p_i}{\partial\pi_m} - \tau_c\frac{\partial
  p_i}{\partial r_m}\right| =\frac{1}{\sqrt{1+\frac{4d_1}{(\pi^{z\rho}_n\eta^{z\rho}_n)^2}}},\label{eq:dett1} 
\end{gather}
and
\begin{align}
\left| \frac{\partial q_i}{\partial r_m}+
\tau_c\frac{\partial p_i}{\partial r_m}\right| = \sqrt{1+\frac{4 d_1}{(\pi^{z\rho}_n\eta^{z\rho}_n)^2}}.\label{eq:dett2} 
\end{align}

The two determinants (equations (\ref{eq:dett1}) and
(\ref{eq:dett2})) can easily be calculated by using the relations
$|\delta_{ij} + N_i^a M_{ab} N_j^b| = |\delta_{ab}+M_{ab}|$
(where $N_i^a N_i^b = \delta_{ab}$ and $N_i^a$ are real) and
$|\delta_{ij} + v_i u_j| = 1 + v_i u_i$, and by recalling that
$\eta^{x\mu}_i$ and $n_{1,i}^{k,x\mu}$ are normalised to 2. Note
that equation (\ref{eq:dett2}) would not hold if we included
terms in the fermionic force parallel to $\eta$. Multiplying the
two determinants together gives $J=1$.  Therefore, this procedure
is area conserving.

\subsubsection{Reflection}
The proof that the reflection algorithm is area conserving follows
the same line as that for transmission: instead of equations
(\ref{eq:dett1}) and (\ref{eq:dett2}), we get
\begin{align}
\left|\frac{\partial p_i}{\partial\pi_m} - \tau_c\frac{\partial
  p_i}{\partial r_m}\right| =&-1\\
\left| \frac{\partial q_i}{\partial r_m}+
\tau_c\frac{\partial p_i}{\partial r_m}\right| =& 1,
\end{align}
which gives $J= -1$.
\subsubsection{The combined update step}\label{App:areaconsrefltrans}
There is one further requirement for area conservation to be
satisfied. It is not sufficient for the standard update and the
corrected update to be area conserving separately; the combination
of them which we use in our molecular dynamics procedure must also
conserve the measure. We define $F^{\Pi,U}_{T}$ and $F^{\Pi,U}_{R}$ as
the momentum updates for transmission and reflection
respectively. We write the transmission condition as $x(\Pi^-,U_C)
>0$, so we transmit when this condition holds and reflect when
$x<0$. The question remains what we should do at $x=0$, and that is
what we shall address in this section. $x$ is a function of the
initial momentum and the gauge field at the crossing (which is
itself a function of the initial momentum and the initial gauge
field, as discussed above). We can therefore write the momentum and
gauge updates combining the reflection and transmission updates as
\begin{align}
p_i =& F^{\Pi}_{Ti} \theta\left(x\right) + \left(1-\theta\left(x\right)\right) F^{\Pi}_{Ri},\label{eq:noname}\\
q_i =& F^{U}_{Ti} \theta\left(x\right) + \left(1-\theta\left(x\right)\right) F^{U}_{Ri}.
\end{align}
 $\theta(x)$ is the step
function, i.e. $\theta = 1$ for $x>0$ and $\theta(x) = 0$ for
$x<0$. We can now differentiate these equations to get the Jacobian (using the same determinant manipulation as before in an attempt to bring one of the sub-determinants to zero):
\begin{align}
J=&\left|\begin{array}{c c}
\frac{\partial p_i}{\partial\pi_j}&\frac{\partial q_i}{\partial\pi_j};\\
\frac{\partial p_i}{\partial r_j}&\frac{\partial q_i}{\partial
  r_j}\end{array}\right| =
\left|\begin{array}{c c}
J_{11}&J_{12}\\
J_{21}&J_{22}\end{array}\right|\\
J_{11}=&\theta(x)\left(\frac{\partial F^{\Pi}_{Ti}}{\partial\pi_j} - \tau_c\frac{\partial
F^{\Pi}_{Ti}}{\partial r_j}\right) + (1-\theta(x))\left(\frac{\partial F^{\Pi}_{Ri}}{\partial\pi_j} - \tau_c\frac{\partial
F^{\Pi}_{Ri}}{\partial r_j}\right)+\nonumber\\
&\delta(x)\left(\frac{\partial x}{\partial\pi_j}+\frac{\partial x}{\partial
  r_c}\frac{\partial r_c}{\partial \pi_j}-\tau_c \frac{\partial x}{\partial
  r_c}\frac{\partial r_c}{\partial r_j}\right)(F^{\Pi}_{Ti}-F^{\Pi}_{Ri}) \label{eq:J11b}
\end{align}
\begin{align}
J_{12}=&\theta(x)\left(\frac{\partial F^{U}_{Ti}}{\partial\pi_j} - \tau_c\frac{\partial
F^{U}_{Ti}}{\partial r_j} + \tau_c\frac{\partial F^{\Pi}_{Ti}}{\partial\pi_j} - \tau_c^2\frac{\partial
F^{\Pi}_{Ti}}{\partial r_j}\right) +\nonumber\\
&(1-\theta(x))\left(\frac{\partial F^{U}_{Ri}}{\partial\pi_j} - \tau_c\frac{\partial
F^{U}_{Ri}}{\partial r_j} - \tau_c\frac{\partial F^{\Pi}_{Ri}}{\partial\pi_j} + \tau_c^2\frac{\partial
F^{\Pi}_{Ri}}{\partial r_j}\right)+\nonumber\\
&\theta(x)(1-\theta(x))\left(-\tau_c\frac{\partial
F^{U}_{Ti}}{\partial r_j}+ \tau_c^2\frac{\partial
F^{\Pi}_{Ti}}{\partial r_j}+\tau_c\frac{\partial
F^{U}_{Ri}}{\partial r_j}- \tau_c^2\frac{\partial
F^{\Pi}_{Ri}}{\partial r_j}\right) +
\nonumber\\
&\delta(x)(F^U_{Ti}-F^U_{Ri})\left(\frac{\partial x}{\partial r_c}\frac{\partial
  r_c}{\partial \pi_j} -\frac{\partial x}{\partial r_c}\frac{\partial
  r_c}{\partial r_j}\tau_c + \frac{\partial x}{\partial
  \pi_j}\right) + \nonumber\\
&\delta(x)\tau_c(F^{\Pi}_{Ti}-F^{\Pi}_{Ri})(2\theta(x)-1)\left(\frac{\partial x}{\partial r_c}\frac{\partial
  r_c}{\partial \pi_j} -\frac{\partial x}{\partial r_c}\frac{\partial
  r_c}{\partial r_j}\tau_c + \frac{\partial x}{\partial
  \pi_j}\right);\\
J_{21}=&\theta(x)\frac{\partial F^{\Pi}_{Ti}}{\partial r_j} +
(1-\theta(x))\frac{\partial F^{\Pi}_{Ri}}{\partial r_j} + \delta(x)\left(F^{\Pi}_{Ti} -F^{\Pi}_{Ri}\right)\frac{\partial x}{\partial r_c}\frac{\partial r_c}{\partial r_j}; 
\end{align}
\begin{align}
J_{22}=&\theta(x)\left(\frac{\partial F^{U}_{Ti}}{\partial r_j} + \tau_c\frac{\partial
F^{\Pi}_{Ti}}{\partial r_j}\right) + (1-\theta(x))\left(\frac{\partial
  F^{U}_{Ti}}{\partial r_j} - \tau_c\frac{\partial
F^{\Pi}_{Ti}}{\partial r_j}\right)+\nonumber\\
&\theta(x)(1-\theta(x))\tau_c\left(-\frac{\partial
F^{\Pi}_{Ti}}{\partial r_j} + \frac{\partial
F^{\Pi}_{Ri}}{\partial r_j}\right)+\nonumber\\
&\delta(x)\frac{\partial x}{\partial
  r_c}\left(\frac{\partial r_c}{\partial r_j}(F^{U}_{Ti}-F^{U}_{Ri}) +\tau_c(2\theta(x)-1)\left(\frac{\partial r_c}{\partial r_j}(F^{\Pi}_{Ti}-F^{\Pi}_{Ri})\right) \right).
\end{align}
The $[2\theta(x)-1]$-terms are caused by the different updates of
the gauge field in the algorithms for reflection and
transmission. In one case, we need to add the first column to the
second column in the determinant calculation; in the other case we
need to subtract. These terms, when they are multiplied by the step functions in $p$ and $q$, additionally lead to the factors of $[\theta(x)(1-\theta(x))]$. $[\theta(x)(1-\theta(x))]$-terms will not contribute to the Jacobian unless multiplied by a $\delta$-function (even though they will in practice be multiplied by factors containing $1/\sqrt{x}$): they are finite at $x = 0$ and zero everywhere else: they will not contribute to an integral (we define $\mathbb{D}(x)$ as a function which is 1 at $x = 0$ and zero elsewhere. This can be seen simply by noting that the integration over these terms (either in terms of the gauge fields or the momentum) is zero. Alternatively, this can be shown by an explicit calculation using a particular momentum update. For example,
 using the algorithm presented in ~\cite{FODOR}, which has $d_1 =
 4d$ and $x = 1+4d/(\eta,\Pi^-)^2$ (and a slightly different
 reflection routine) one can demonstrate that the total Jacobian,
 including these terms, is
\begin{gather}
J = 1 + (1-x)^2\frac{d}{dx}\left(\frac{2\theta(x) -
  (1+\sqrt{x})\theta^2(x)}{\sqrt{x} - 1}\right) = 1-2\delta(x).
\label{deltafunction}
\end{gather}

The rest of the determinant can be simplified by using the results of the previous two sections, particularly the result $\tau_c \partial r_c/\partial r_i = \partial r_c/\partial x_i$. The non-$\delta$-function terms in $J_{12}$ cancel immediately as before. Because the terms multiplying the $\delta$-functions are the outer product of two matrices, we can remove all but one of these functions by determinant manipulation. For example, by subtracting $\partial x/\partial \pi_j \mathbb{D}(x)/(\partial x/\partial r_c \partial r_c/\partial r$ multiplied by $J_{11}$ and $J_{12}$ from the $J_{21}$ and $J_{22}$ allows us to remove the $\delta$-functions in those terms, and a similar manipulation allows the removal of the $\delta$-function in $J_{12}$, so that the only $\delta$-function which remains is contained within $J_{11}$. We can use a Schur decomposition to factorise the determinant in a similar manner to before
\begin{align}
J = &\left|J_{11} - J_{12}J_{22}^{-1}J_{21}\right| |J_{22}|.
\label{eq:J11}
\end{align}
$|J_{22}|$ contains no $\delta$-function. Because the Schur compliment is proportional to $\mathbb{D}$ and contains no $\delta$-functions, it will not contribute when we integrate over it. But the single $\delta$-function, multiplied by $\partial x/\partial \pi_j$, in $J_{11}$ remains. We can use a stopping criterion with $\partial x/\partial \pi_j \ll 1$ (setting $\partial x/\partial \pi_j =0 $ renders the above argument invalid, but we can safely set it arbitrarily close to zero then take a limit) will remove the effect of the $\delta$-function in a numerical simulation; and we use this result to justify our numerical tests in section \ref{sec:tuningdmax}. But in a real life simulation we cannot use a fixed $x$, and we have to determine whether the $\delta$-function in the Jacobian will affect the final results. Except at precisely $x = 0$, the Jacobian is 1.

\subsubsection{Discussion of impact of the $\delta$-function Jacobian}\label{App:A.2.4}

We now turn our attention to the question of whether the
$\delta$-function of equation~(\ref{eq:J11}) will affect the
equilibrium ensemble of the Monte Carlo. 

The updating procedure needs to satisfy
\begin{gather}
W_c[U'] = \int d[U]  P[U'\leftarrow U] W_c[U],\label{eq:canprobeq}
\end{gather}
 where the probability of moving from a gauge field $U$ to the gauge
 field $U'$ is $P[U'\leftarrow U]$, and $W_c$ is the canonical
 distribution for the gauge fields. This is satisfied by the
 detailed balance condition
\begin{gather}
\int d[\Phi]d[\Phi^{\dagger}]P[U'\leftarrow U,\Phi] W_c[U,\Phi] = \int d[\Phi]d[\Phi^{\dagger}]P[U\leftarrow U',\Phi] W_c[U',\Phi].\label{eq:detailedbal}
\end{gather}
To prove the detailed balance condition for the
HMC~\cite{MontvayMunster}, we define
\begin{align}
 P_{\Pi}[\Pi] =& e^{-\frac{1}{2}\Pi^2}\nonumber\\ 
W_{c}[U,\Phi]
=& e^{-\Phi^{\dagger}\frac{1}{H^2[U]}\Phi}e^{-S_g[U]}\nonumber\\ 
P_A([\Pi',\Phi,U']\leftarrow[\Pi,\Phi,U]) = &\min\left(1,e^{-E[\Pi',\Phi,U']+
  E[\Pi,\Phi,U]}\right)\nonumber\\
P_{\Pi} W_c[U] = & e^{-E([\Pi,\Phi,U])}\nonumber\\
P_{MD}([\Pi',\Phi,U'] \leftarrow[\Pi,\Phi,U]) =&
\delta([\Pi',\Phi,U'] - T_{MD}[\Pi,\Phi,U]),
\end{align}
where $T_{MD}$ is the molecular dynamics trajectory, and 
\begin{align}
P[U'\leftarrow U,\Phi] = &\int
d[\Pi]d[\Pi']P_{\Pi}[\Pi]\nonumber\\
&P_A([\Pi',\Phi,U']\leftarrow[\Pi,\Phi,U])P_{MD}([\Pi',\Phi,U'] \leftarrow[\Pi,\phi,U]).
\end{align}
If the Jacobian (which is introduced when the variables are changed in the $P_{MD}$ term) is of the form $ J = 1 + J(\Pi)\delta(\Pi - \Pi_c)$, then when we perform the integration over momentum, the detailed balance condition given in equation (\ref{eq:detailedbal}) will pick up an additional term when integrating over momentum fields. 

It is tempting to argue that because the condition $x=0$ will never be encountered in a numerical simulation we can neglect this Jacobian while implementing the algorithm; we can, for example, place the logarithm of the Jacobian in the Metropolis step, and say that we will account for it precisely when $x=0$, which will never happen in a numerical simulation. However, arguments based around saying that the condition will never be encountered are in effect saying that the measure (in terms of the integration over the initial momentum field) is zero; which is not enough to counter the effect of a Dirac $\delta$-function. Therefore another argument is needed to demonstrate that this will not contribute to the final ensemble. One can also argue the Jacobian will not contribute because we can use an approximate overlap operator, which we will gradually transform to the sign function. The algorithm with the approximate overlap operator will resemble the transmission/reflection algorithm, and it is clearly correct. However, this argument fails because as the approximation of the sign function improves so the molecular dynamics time step must decrease to correctly resolve the step. We, however, are working at a far larger time step than which is necessary to resolve the sign function: the two cases are different. If algorithm works at infinitesimal time-step it does not necessarily follow that it is valid at a finite time step. Therefore, we must find another way of proving that this $\delta$-function does not contribute, or, if it does contribute, how we can compensate for it. 

We proceed by explicitly performing the momentum integral in the detailed balance equation over the Dirac $\delta$-functions from the Jacobian; and we shall show that the contribution from the $\delta$-functions cancel out, meaning that detailed balance is maintained. This can be done easily because there are as many $\delta-$functions as there as momentum integration variables. We begin by introducing auxiliary momentum fields $\Pi_i$ ($i = 1\ldots N_{\Pi}$), which correspond to the momentum fields at each kernel eigenvalue crossing during the trajectory. The gauge field at the crossing is $U_i$, leading to a reflection condition depending on a variable $x_i$. These additional momentum fields are, of course, precisely determined from either the initial momenta using the trajectory $T_i[\Pi,\Phi,U,]$, or from the final momenta and the reverse trajectory $T_i^{-1}[\Pi',\Phi,U']$; and these conditions lead to a delta-function in the integral for each auxiliary momentum field introduced. Since the various $x_i$s are not degenerate with respect to the initial gauge field and momentum, we can write $\prod(1+\delta(x_i)) = 1+\sum\delta(x_i)$. Using this notation, the detailed balance equation reads  
\begin{align}
&\int d\Phi^{\dagger}d\Phi P[U'\leftarrow U,\Phi] W_c[U,\Phi] = \int d\Pi' d\Pi d\Phi^{\dagger}d\Phi \prod d\Pi_i P_A \nonumber\\
&\phantom{sp}\bigg(\delta([\Pi',\Phi,U'] - T_{MD}[\Pi,\Phi,U])\prod_{j=1}^{N_{\Pi}}\delta([\Pi_j,\Phi,U_j]-T_i[\Pi,\Phi,U])+\nonumber\\
&\phantom{sp}\sum_{i=1}^{N_{\pi}}\delta(x_i) J(x_i)\prod_{j=1}^i\delta([\Pi_i,\Phi,U_i]-T_i[\Pi,\Phi,U])\prod_{j=i+1}^{N_{\Pi}}\delta([\Pi_i,\Phi,U_i]-T^{-1}_i[\Pi',\Phi,U'])\bigg).\label{eq:a54}
\end{align}
$J(x_i)$ is the coefficient of the $\delta-$function in the Jacobian. Using the notation of equation (\ref{eq:mostgensolution}), we can convert the delta function from a function of $x$ to a function of $(\Pi_i,\eta_k)$ by multiplying by
$\frac{1}{dx/d(\Pi_i,\eta_k)}$, and summing over each possible solution for $(\Pi_i,\eta_k)$ which gives $x = 0$. Since $x$ is quadratic in each $(\Pi_i,\eta_k)$, which corresponds to one component of the momentum field, there will be two possible solutions to the equation $x=0$ for each of the $N_k$ momentum field components; and we will generate a $\delta-$function for each of these solutions.  Using $x = 1+4d/(\sum_k(\eta_k,\Pi_i)^2$, it follows that
\begin{align}
\delta(x_i)J(x_i) =& 2\sum_k\bigg[\delta\left((\Pi_i,\eta_k) - \sqrt{-\left(4d + \sum_{m\neq k} (\Pi_i,\eta_m)^2\right)}\right)\frac{(\Pi_i,\eta_k)\eta_k}{\sum_n (\Pi_i,\eta_n)^2} +\nonumber\\
& \delta\left((\Pi_i,\eta_k) + \sqrt{-\left(4d + \sum_{m\neq k} (\Pi_i,\eta_m)^2\right)}\right)\frac{(\Pi_i,\eta_k)\eta_k}{\sum_n (\Pi_i,\eta_n)^2}\bigg]J(x_i).
\end{align}
It is easily observed that the coefficient of the $\delta-$function for each $k$ is an odd function of $(\Pi_i,\eta_k)$ multiplied by $J(x_i)$. In particular, if $J(x_i)$ is an even function of $(\Pi_i,\eta_n)$, then the coefficients for the positive $(\Pi_i,\eta)$ and negative $(\Pi_i,\eta)$ $\delta$-functions will precisely cancel, removing all the $\delta-$function terms from the detailed balance equation. So the task of demonstrating that this $\delta-$function will not contribute to the final ensemble can be reduced to a demonstration that $J(x)$ is an even function of the momenta.

From equation (\ref{eq:J11b}), we see that $J(x) = \partial x/\partial (\Pi,\eta_j) (F_{Ti}^{\Pi} - F^{\Pi}_{Ri})$. The Jacobian is the determinant of the $N_k\times N_k$ square matrix
\begin{gather}
(\eta_i,F_{T}^{\Pi} - F^{\Pi}_{Ri})\partial x/\partial (\Pi,\eta_j)\delta(x) +\ldots.
\end{gather}
Noting that $(\eta_i,F_{T}^{\Pi}) = 0$ at $x = 0$, this matrix is the outer product of two vectors, both of which are odd functions of $(\eta_j,\Pi_i)$. Therefore, $J(x_i)$ is an even function of  $(\eta_j,\Pi_i)$. As previously argued, this means that the presence of the $\delta-$function will not affect the final ensemble. We demonstrate that $J(x_i)$ is even for an explicit example below.

It might be thought that this argument might be invalid because we cannot sample both positive and negative $\Pi$ with equal probability. If the trajectory starts in one particular topological sector, then it can only approach the topological wall from one direction (unless there is more than topological index change in the trajectory). However, our calculation does not depend on the sign of $\Pi$ but the sign of $(\Pi,\eta)$. The algorithms presented in this article are all even functions of $\eta$; in practice, the sign of $\eta$ will be chosen randomly. In the detailed balance equation, since the both the molecular dynamics trajectory, $T$, and the metropolis acceptance probability, $P_A$, are independent of the sign of $\eta$, both choices for $\eta$ lead to the same final gauge field, and we must include this choice of sign by summing over the two possibilities, weighting them by the appropriate probability; if the algorithm is designed without bias, this probability will be $1/2$. In other words, it is not the integration over the auxiliary momentum fields but this sum over the signs of $\eta$ which causes the $\delta-$functions to cancel. Thus the direction in which the trajectory approaches the topological sector wall, and the limited sampling of the momentum fields are both unimportant.

As an example, we now explicitly calculate the Jacobian for one particular case to demonstrate how this works in practice. We assume that there is just one momentum crossing in the trajectory, and use the update
\begin{align}
F_{Ti}^{\Pi} =& \pi^-_i + \eta_i(\Pi^-,\eta)\left(\sqrt{1+\frac{4d}{(\Pi^-,\eta)^2}} - 1\right) -\nonumber\\& \tau_c ((F^-_i+F^+)-\eta_i(\eta,F^- + F^+))\nonumber\\
F_{Ri}^{\Pi} = & \pi^-_i - 2\eta_i (\eta,\Pi^-) - 2 \tau_c (F^-_i-\eta_i(\eta,F^-))\nonumber\\
F_{Ti}^{U} = & q^-_i + \tau_c(F_{Ti}^{\Pi} - \pi_i^-)\nonumber\\
F_{Ri}^{U} = & q^-_i + \tau_c(F_{Ri}^{\Pi} + \pi_i^-)\nonumber\\
x = &1+\frac{4d}{(\Pi^-,\eta)^2}\nonumber\\
\left.\left(\frac{\partial x}{\partial (\Pi^-,\eta)}\right)\right|_{x=0} =& -\frac{2\eta}{(\Pi^-,\eta)}\nonumber\\
\left.(F_{Ti}^{\Pi} -F_{Ri}^{\Pi})\right|_{x=0}=& (\Pi^-,\eta)\nonumber\\
J(x) = & \left.\left(\frac{\partial x}{\partial (\Pi^-,\eta)}\right)\right|_{x=0} \left.(F_{Ti}^{\Pi} -F_{Ri}^{\Pi})\right|_{x=0}= -2.
\end{align}
Hence $J$ is an even function of $(\Pi,\eta)$. The additional and unwanted term in the detailed balance equation, once we have integrated out all the auxiliary momentum integration variables except $(\Pi_i,\eta)$, reads
\begin{align}
-2\int d\Phi^{\dagger}d\Phi\sum_{\pm\eta}\int(\Pi_i,\eta)  \bigg[\delta((\Pi_i,\eta_i) - \sqrt{-4d}) \frac{1}{(\Pi_i,\eta_i)} &+ \delta((\Pi_i,\eta_i) +\phantom{a}\nonumber\\& \sqrt{-4d}) \frac{1}{(\Pi_i,\eta_i)}\bigg]P_A = 0
\end{align}
Hence the detailed balance condition is, in fact, maintained despite the $\delta$-functions in the Jacobian.

Numerically, in section~\ref{sec:tuningdmax}, we have compared the
algorithm with an approach that explicitly does satisfy detailed
balance (reflecting when $|d|>d_{max}$ and using $\Pi^+=\Pi^-$ when
$|d|<d_{max}$), but does not necessarily conserve energy. In an
extensive simulation we found no difference in the observables
calculated.

\subsection{Correction step error of $O(\tau^2)$}\label{app:corrsteperror}

Suppose that there is a crossing between time $\tau = 0$ (where the
gauge field, momentum, and fermionic force are $U_0$, $\Pi_0$
respectively), and time $\tau = \Delta\tau$ (with gauge field $U_1$,
momentum $\Pi_1$). The gauge fields immediately before and after the
eigenvalue crossing are $U_d$ and $U_u$.\footnote{Note that in
  practice $U_u = U_d = U_c$ - there is no discontinuity in the
  gauge field at the crossing. The notation here is purely for
  clarification.} If $S(U) = E- \frac{1}{2}\Pi^2$ is the action,
then the total force $F'$ acting on the momentum is
\begin{gather}
f^{x\mu}_i(U) = \frac{\partial S(U)}{\partial r_i^{x\mu}}=
-\frac{\partial \pi_i^{x\mu}}{\partial\tau}.
\end{gather}
We shall demonstrate that the energy difference, $\Delta E$, between
time $\tau = \Delta\tau$ and time $\tau = 0$ is of order
$\Delta\tau^2$. The energy change between time $\Delta\tau$ and time
$0$ is
\begin{gather}
\Delta E = \frac{1}{2}(\Pi_1,\Pi_1) - \frac{1}{2}(\Pi_0,\Pi_0) + S(U_1) - S(U_0).
\end{gather}
In this notation, our update for transmission (\ref{eq:40}), with
$s_1^2 = 1$), and neglecting terms of $O(\Delta\tau^2)$ or higher is
\begin{align}
\Pi^- =& \Pi_0 - \frac{\Delta\tau}{2} F_0;\nonumber\\
\Pi^+ = &\Pi^- - \tau_c\left(1 - \ket{\eta} \bra{\eta}\right)(F^- -
F^+)+\tau_c\frac{1}{3}\Tr(F^- - F^+) + \nonumber\\
&\eta(\eta,\Pi^-)\sqrt{1+\frac{4d}{(\eta,\Pi^-)^2}} + \nonumber\\
& \left(\eta_1^2 (\eta_1^2, \Pi^- - \frac{\tau_c}{2}(F^- + F^+)) +\eta_2^2 (\eta_2^2, \Pi^-
- \frac{\tau_c}{2}(F^- + F^+))\right)\times\nonumber\\
&\left(\sqrt{1 -
  2\frac{\tau_c(F^-,\eta)(\Pi^-,\eta) - \tau_c(F^+,\eta)(\Pi^+,\eta)}{(\eta_1^2,
    \Pi^- - \frac{\tau_c}{2}(F^- + F^+))^2 +  (\eta_2^2, \Pi^- -
    \frac{\tau_c}{2}(F^- + F^+))^2}} - 1\right);\nonumber\\
\Pi_1 =& \Pi^+ - \frac{\Delta\tau}{2} F_1\nonumber\\
U^- =& e^{i\frac{\Delta\tau}{2} \Pi^-}U_0\nonumber\\
U^+ = & e^{-i\tau_c \Pi^+}e^{+i\tau_c \Pi^-} U^-\nonumber\\
U_1 = & e^{\frac{i\Delta\tau}{2} \Pi^+} U^+\\
U_d =& e^{i\left(\frac{\Delta\tau}{2} + \tau_c\right) \Pi^-}U_0\nonumber\\
U_1 =& e^{i\left(\frac{\Delta\tau}{2} - \tau_c\right) \Pi^+}U_u.
\end{align}
 There is a jump in the action $S$ at the eigenvalue crossing,
\begin{gather}
S(U_u)- S(U_d) = -2d.
\end{gather}
Using $(\Pi_+-\Pi_-,F_+ - F_-) = O(\Delta\tau^2)$, we can show that
\begin{align}
(\Pi^+,\Pi^+) - (\Pi^-,\Pi^-) =&  4d+ 2\tau_c(\Pi^-,F^+) -
   2\tau_c(\Pi^-,F^-)+   \nonumber\\
&\tau_c (\Pi_+-\Pi_-,F_+ +F_-)+ O(\Delta\tau^2)\nonumber\\
=&4d +2\tau_c(\Pi^-,F^+) -
   2\tau_c(\Pi^-,F^-)+O(\Delta\tau^2),
\end{align}
\begin{align}
(\Pi_1,\Pi_1) - (\Pi_0,\Pi_0) = 4d - &\Delta\tau((\Pi_0, F_0) +
  (\Pi_1,F_1))\nonumber\\
&- 2\tau_c((F_0, \Pi^-) - (F_1, \Pi^+)) + O(\Delta\tau^2),
\end{align}
\begin{align}
S(U_1) - S(U_0) =& S\left(U\left(r_{ci}^+ +
\left(\frac{\Delta\tau}{2} - \tau_c\right) p_i\right)\right)\nonumber\\&
-S\left(U\left(r_{ci}^- + \left(\frac{\Delta\tau}{2} +
\tau_c\right)\pi_i\right)\right)+ O(\Delta\tau^2) \nonumber\\ 
=& S(U_u) - S(U_d) + \left(\frac{\Delta\tau}{2} -
\tau_c\right)\left.\frac{\partial
  S}{\partial r_i}\right|_{r_c^+} p_i +\nonumber\\& \left(\frac{\Delta\tau}{2} +
\tau_c\right)\left.\frac{\partial
  S}{\partial r_i}\right|_{r_c^-}\pi_i+ O(\Delta\tau^2)\nonumber\\
=& -2d + \frac{\Delta\tau}{2}\left((\Pi^+,F_1) + (\Pi^-,F_0)\right) +\nonumber\\&
\tau_c((\Pi^-,F^-) - (\Pi^+, F^+)) + O(\Delta\tau^2).
\end{align}
So,
\begin{gather}
\Delta E =
O(\Delta\tau^2).
\end{gather}
Therefore, our modified correction update for transmission conserves energy up to
$O(\Delta\tau^2)$. It can also be shown that the energy
violating reflection term is also $O(\Delta\tau^2)$. Without the additional
correction term to account for the $O(\Delta\tau)$ error, the energy violation would be 
\begin{gather}
\frac{\tau_c}{2} \left(\left(F^- - F^+,\eta\right)(\Pi^- + \Pi^+,\eta) -
\left(F^- + F^+,\eta\right)(\Pi^- - \Pi^+,\eta)\right) + O(\Delta\tau). 
\end{gather}

\section{Two Crossings within the same Molecular Dynamics Step}\label{sec:updatetwice}

There are several situations when it is difficult to resolve whether
or not there has been an eigenvalue crossing. Firstly, when an
eigenvalue has a minimum close to zero, and secondly when there are
two small eigenvalues. In the first case, one can identify the
minimum by looking at the sign of the differential of the eigenvalue
with respect to computer time. If the eigenvalue is sufficiently
small that there is a danger that the crossing may have occurred,
one can search for the zero eigenvalue. In the second case one has
to search for crossings if there is a large mixing between the two
eigenvectors, or an initial Newton-Raphson extrapolation indicates
that they might cross. In both cases, one moves to the point of the
first crossing, changes the momentum in the usual manner, moves to
the second crossing (if necessary, i.e. the first crossing was
transmission), and changes the momentum again. Care needs to be
taken to ensure that the algorithm will detect the potential
crossing in both directions, otherwise there could be a breakdown in
reversibility.

More serious is the possibility of mixing between a small positive
and small negative eigenvalues of the kernel operator. As an
example, let us consider a system with just two eigenvectors
$|\psi_1\rangle$ and $|\psi_2\rangle$. One can calculate the mixing
between the eigenvectors using a similar method to the one used in
section \ref{sec:overlap}, except in this case we will not expand in
small $\Delta\tau$. One starts as before with the eigenvalue
equations at times $\tau$ and $\tau + \Delta\tau$:
\begin{align}
Q|\psi_1\rangle = & \lambda_1|\psi_1\rangle\nonumber\\
Q|\psi_2\rangle = & \lambda_2|\psi_2\rangle\nonumber\\
|\psi_1\rangle + |\delta_1\rangle = & \cos\theta|\psi_1\rangle + e^{i\phi} \sin\theta |\psi_2\rangle\nonumber\\
|\psi_2\rangle + |\delta_2\rangle = & \cos\theta|\psi_2\rangle + e^{-i\phi} \sin\theta |\psi_1\rangle.
\end{align}
One can now easily solve for the mixing angles $\theta$ and $\phi$:
\begin{align}
e^{i\phi} =& \sqrt{\frac{\langle\psi_2|\delta Q | \psi_1\rangle}{\langle\psi_2|\delta Q | \psi_1\rangle}}\nonumber\\
A = & \frac{(\lambda_2-\lambda_1 + \langle\psi_2|\delta Q | \psi_2\rangle - \langle\psi_1|\delta Q | \psi_1\rangle)^2}{4 \langle\psi_2|\delta Q | \psi_1\rangle \langle\psi_1|\delta Q | \psi_2\rangle}\nonumber\\
\tan^2\theta =& A \left(\sqrt{1+\frac{1}{A}} - 1\right)^2.
\end{align}
An expansion in small $\delta\tau$ (with $\delta Q =
\delta\tau\partial Q/\partial \tau + O(\delta\tau^2)$) gives the
same expression that was derived in
section~\ref{sec:overlap}. However, this expansion is only valid if
\begin{gather}
\left| \frac{\langle\psi_2|\delta Q | \psi_2\rangle - \langle\psi_1|\delta Q | \psi_1\rangle}{\lambda_2-\lambda_1} \right|\ll 1\nonumber
\end{gather}
and
\begin{gather}
\left| \frac{\langle\psi_1|\delta Q | \psi_2\rangle \langle\psi_2|\delta Q | \psi_1\rangle}{(\lambda_2-\lambda_1)^2} \right|\ll 1.\nonumber
\end{gather}
These conditions fail if two eigenvalues are close to each other and
both have maxima or minima in the same time step. One can write the
fermionic force contribution from the small eigenvectors (excluding
the delta function term) as
\begin{gather}
2 \lim_{\delta\tau\rightarrow 0}\frac{1}{\delta\tau}\sin\theta\cos\phi\bra{X}\left\{\gamma_5,\ket{\psi_1}\bra{\psi_2} + \ket{\psi_2}\bra{\psi_1}\right\}\ket{X}(\epsilon(\lambda_1)-\epsilon(\lambda_2)).\nonumber
\end{gather}
If there occurs a small negative eigenvalue and a small positive
eigenvalue---mixing between eigenvectors with the same sign does not
contribute to the force---and they have minima at the same time
step, there could potentially emerge a very large fermionic force.

In our simulations we have indeed observed this effect. Most of the
metropolis rejections on larger lattices can be traced back to
it. In a subsequent paper we are going to describe a solution to this
problem~\cite{Cundy:2007df}.

\bibliographystyle{myelsart-num}
\bibliography{HMCoverlap}

\begin{thebibliography}{10}
\expandafter\ifx\csname url\endcsname\relax
  \def\url#1{\texttt{#1}}\fi
\expandafter\ifx\csname urlprefix\endcsname\relax\def\urlprefix{URL }\fi

\bibitem{Creutz:2008hx}
Michael Creutz, {The saga of rooted staggered quarks} 0804.4307.

\bibitem{Kaplan:1992bt}
D.~B. Kaplan, A Method for simulating chiral fermions on the lattice, Phys.
  Lett. B288 (1992) 342--347, hep-lat/9206013.

\bibitem{Nishy-Ninny}
H.~B. Nielsen and B.~Ninomiya, Nucl. Phys B185 (1981) 20.

\bibitem{Narayanan:1993ss}
R.~Narayanan and H.~Neuberger, Chiral fermions on the lattice, Phys. Rev. Lett.
  71 (1993) 3251--3254, hep-lat/9308011.

\bibitem{Neuberger:1998fp}
H.~Neuberger, Exactly massless quarks on the lattice, Phys. Lett. B417 (1998)
  141--144, hep-lat/9707022.

\bibitem{Hasenfratz}
P.~Hasenfratz, Prospects for perfect actions., Nucl.Phys.Proc.Suppl. 63 (1998)
  53--58, hep-lat/9709110.

\bibitem{Ginsparg:1982bj}
P.~H. Ginsparg and K.~G. Wilson, A remnant of chiral symetry on the lattice,
  Phys. Rev. D25 (1982) 2649.

\bibitem{Luscher:1998du}
M.~L{\"u}scher, Abelian chiral gauge theories on the lattice with exact gauge
  invariance, Nucl. Phys. B549 (1999) 295--334, hep-lat/9811032.

\bibitem{HeJaLe99}
P.~Hern{\'a}ndez, K.~Jansen, and L.~Lellouch, Finite-size scaling of the quark
  condensate in quenched lattice QCD, Phys. Lett. B469 (1999) 198--204,
  hep-lat/9907022.

\bibitem{Hernandez:1999gu}
P.~Hern{\'a}ndez, K.~Jansen, and L.~Lellouch, Chiral symmetry breaking from
  Ginsparg-Wilson fermions, Nucl. Phys. Proc. Suppl. 83 (2000) 633--635,
  hep-lat/9909026.

\bibitem{HeJaLe00}
P.~Hern{\'a}ndez, K.~Jansen, and L.~Lellouch, A Numerical treatment of
  {N}euberger's Lattice {D}irac Operator, in: Frommer et~al.
  \cite{Wuppertal:1999}, pp. 29--39, proceedings of the International Workshop,
  University of Wuppertal, August 22-24, 1999.

\bibitem{Bu98}
B.~Bunk, hep-lat/9805030 (1998).

\bibitem{Hernandez:2000iw}
P.~Hern{\'a}ndez, K.~Jansen, and M.~L{\"u}scher, A note on the practical
  feasibility of domain-wall fermions, hep-lat/0007015 (2000).

\bibitem{Bor99c}
A.~Bori\c{c}i, Fast methods for computing the Neuberger operator, in: Frommer
  et~al.  \cite{Wuppertal:1999}, pp. 40--47, proceedings of the International
  Workshop, University of Wuppertal, August 22-24, 1999.

\bibitem{Bor99b}
A.~Bori\c{c}i, A {L}anczos approach to the inverse square root of a large and
  sparse matrix, J. Comput. Phys. 162 (2000) 123, hep-lat/9910045.

\bibitem{Bor99a}
A.~Bori\c{c}i, On the {N}euberger overlap operator, Phys. Lett. B453 (1999)
  46--53, hep-lat/9810064.

\bibitem{Vor00}
H.~A. {van der Vorst}, Solution of $f({A})x=b$ with Projection Methods, in:
  Frommer et~al.  \cite{Wuppertal:1999}, pp. 18--28, proceedings of the
  International Workshop, University of Wuppertal, August 22-24, 1999.

\bibitem{Neuberger:1998my}
H.~Neuberger, A practical implementation of the overlap-Dirac operator, Phys.
  Rev. Lett. 81 (1998) 4060--4062, hep-lat/9806025.

\bibitem{Neu00}
H.~Neuberger, Overlap Dirac Operator, in: Frommer et~al.
  \cite{Wuppertal:1999}, proceedings of the International Workshop, University
  of Wuppertal, August 22-24, 1999.

\bibitem{EHKN00}
R.~G. Edwards, U.~M. Heller, J.~Kiskis, and R.~Narayanan, Chiral condensate in
  the deconfined phase of quenched gauge theories, Phys. Rev. D61 (2000)
  074504, hep-lat/9910041.

\bibitem{EHN98}
R.~G. Edwards, U.~M. Heller, and R.~Narayanan, A study of practical
  implementations of the overlap-Dirac operator in four dimensions, Nucl. Phys.
  B540 (1999) 457--471, hep-lat/9807017.

\bibitem{EFL02}
J.~van~den Eshof, A.~Frommer, Th. Lippert, K.~Schilling, and H.A. van~de Vorst,
  Numerical Methods for the {QCD} overlap Operator: {I}. Sign-Function and
  Error Bounds, Comput. Phys. Comm. 146 (2002) 203--224.

\bibitem{OVERLAP1}
R.~V. Gavai, S.~Gupta, and R.~Lacaze, Speed and adaptability of overlap fermion
  algorithms, Comput. Phys. Commun. 154 (2003) 143--158, hep-lat/0207005.

\bibitem{Dong:2003im}
S.~J. Dong et~al., Chiral logs in quenched QCD, hep-lat/0304005.

\bibitem{LIUOVERLAP}
T.~Chiu, T.~Hsieh, C.~Huang, and T.~Huang, A note on the Zolotarev optimal
  rational approximation for the overlap Dirac operator, phys. Rev. D66 (2002)
  114502, hep-lat/0206007.

\bibitem{FODOR}
Z.~Fodor, S.~D. Katz, and K.~K. Szabo, Dynamical overlap fermions, results with
  hybrid Monte-Carlo algorithm, JHEP 08 (2004) 003, hep-lat/0311010.

\bibitem{LIU1}
T.~Chiu, Optimal domain-wall fermions, Phys. Rev. Lett. 90 (2003) 071601,
  hep-lat/0209153.

\bibitem{LIU2}
T.~Chiu, Locality of optimal lattice domain-wall fermions, Phys. Lett. B552
  (2003) 97--100, hep-lat/0211032.

\bibitem{LIULATTICE}
T.~Chiu, Optimal lattice domain-wall fermions with finite N(s), hep-lat/0304002
  (2003).

\bibitem{Borici:2001ua}
A.~Borici, A.~D. Kennedy, B.~J. Pendleton, and U.~Wenger, The overlap operator
  as a continued fraction, Nucl. Phys. Proc. Suppl. 106 (2002) 757--759,
  hep-lat/0110070.

\bibitem{Neuberger:1999re}
H.~Neuberger, The overlap lattice Dirac operator and dynamical fermions, Phys.
  Rev. D60 (1999) 065006, hep-lat/9901003.

\bibitem{Wenger:2004gd}
U.~Wenger, The overlap Dirac operator as a continued fraction. Hep-lat/0403003.

\bibitem{Liu:1998hj}
C.~Liu, Hybrid Monte Carlo algorithm for lattice QCD with two flavors of
  dynamical Ginsparg-Wilson quarks, Nucl. Phys. B554 (1999) 313--322,
  hep-lat/9811008.

\bibitem{Jansen1}
P.~Hern{\'a}ndez, K.~Jansen, and L.~Lellouch, A numerical treatment of
  Neuberger's lattice Dirac operator, in: A.~Frommer (Ed.), Wuppertal 1999,
  Numerical challenges in lattice quantum chromodynamics, 2000, pp. 29--39,
  hep-lat/0001008.

\bibitem{Jansen2}
K.~Jansen, Overlap and domain wall fermions: What is the price of chirality?,
  Nucl. Phys. Proc. Suppl. 106 (2002) 191--192, hep-lat/0111062.

\bibitem{Giusti1}
L.~Giusti, C.~Hoelbling, and C.~Rebbi, Light quark masses with overlap fermions
  in quenched QCD, Phys. Rev. D64 (2001) 114508, hep-lat/0108007.

\bibitem{Giusti2}
L.~Giusti, C.~Hoelbling, M.~L{\"u}scher, and H.~Wittig, Numerical techniques
  for lattice QCD in the epsilon-regime, Comput. Phys. Commun. 153 (2003)
  31--51, hep-lat/0212012.

\bibitem{KennedyDuane}
S.~Duane, A.~Kennedy, B.~Pendelton, and D.~Roweth, Phys. Lett. B195 (1987) 216.

\bibitem{BODE}
A.~Bode, U.~M. Heller, R.~G. Edwards, and R.~Narayanan, First experiences with
  HMC for dynamical overlap fermions, in: Dubna 1999, Lattice fermions and
  structure of the vacuum, 1999, pp. 65--68, hep-lat/9912043.

\bibitem{Schwinger1}
S.~D{\"u}rr and Ch. Hoelbling, Staggered versus overlap fermions: A study in
  the Schwinger model with N(f) = 0,1,2, Phys. Rev. D69 (2004) 034503,
  hep-lat/0311002.

\bibitem{Schwinger2}
L.~Giusti, Ch. Hoelbling, and C.~Rebbi, Schwinger model with the overlap-Dirac
  operator: Exact results versus a physics motivated approximation, Phys. Rev.
  D64 (2001) 054501, hep-lat/0101015.

\bibitem{Durr:2004ta}
S.~D{\"u}rr and C.~Hoelbling, Scaling tests with dynamical overlap and rooted
  staggered fermions Hep-lat/0411022.

\bibitem{FODOR2}
Z.~Fodor, S.~D. Katz, and K.~K. Szabo, Dynamical overlap fermions, results with
  HMC algorithm, Nucl. Phys. Proc. Supp. 140C, hep-lat/0409070.

\bibitem{DeGrand:2004nq}
T.~DeGrand and S.~Schaefer, Physics issues in simulations with dynamical
  overlap fermions Hep-lat/0412005.

\bibitem{Hernandez:1998et}
P.~Hernandez, K.~Jansen, and M.~L{\"u}scher, Locality properties of Neuberger's
  lattice Dirac operator, Nucl. Phys. B552 (1999) 363--378, hep-lat/9808010.

\bibitem{vdEetal:03a}
G.~Arnold, N.~Cundy, J.~{van den Eshof}, A.~Frommer, S.~Krieg, Th. Lippert, and
  K.~Sch{\"a}fer, Numerical methods for the QCD overlap operator. {II}: Optimal
  Krylov subspace methods To appear in the proceedings of the Third
  International Workshop on Numerical Analysis and Lattice QCD,
  hep-lat/0311025.

\bibitem{cundy2}
N.~Cundy, J.~{van den Eshof}, A.~Frommer, S.~Krieg, Th. Lippert, and
  K.~Sch{\"a}fer, Numerical methods for the QCD overlap operator. {III}: Nested
  iterations., Comp. Phys. Comm. 165 (2005) 841, hep-lat/0405003.

\bibitem{Krieg:2004xg}
S.~Krieg et~al., Improving inversions of the overlap operator, Nucl. Phys.
  Proc. Supp. 140C, hep-lat/0409030.

\bibitem{Cundy:2004xf}
N.~Cundy, S.~Krieg, A.~Frommer, Th. Lippert, and K.~Schilling, Dynamical
  overlap simulations using HMC, Nucl. Phys. Proc. Supp. 140C (2005) 841,
  hep-lat/0409029.

\bibitem{Omelyan}
I.~P. Omelyan, I.~M. Mrygloda, and R.~Folk, Symplectic analytically integrable
  decomposition algorithms: classification, derivation, and application to
  molecular dynamics, quantum and celestial mechanics simulations, Computer
  Physics Communications 151 (2003) 272.

\bibitem{Takaishi:2005tz}
Tetsuya Takaishi and Philippe de~Forcrand, Testing and tuning new symplectic
  integrators for hybrid Monte Carlo algorithm in lattice QCD, Phys. Rev. E73
  (2006) 036706, hep-lat/0505020.

\bibitem{SW}
J.~C. Sexton and D.~H. Weingarten, Nucl. Phys. B 380 (1992) 665.

\bibitem{Zolotarev}
D.~Ingerman, V.~Druskin, and L.~Knizerhman, Comm. Pure Appl. Math. 53 (2000)
  1039.

\bibitem{zolotarevpaper}
P.~P. Petrushev and V.~A. Popov, Rational approximation of real functions,
  Cambridge University Press, 1987.

\bibitem{Cundy:2007df}
Nigel Cundy, {Small Wilson Dirac operator eigenvector mixing in dynamical
  overlap hybrid Monte-Carlo} 0706.1971.

\bibitem{Cundy:2007dp}
Nigel Cundy, Stefan Krieg, Thomas Lippert, and Andreas Sch{\"a}fer, {Dynamical
  overlap fermions with increased topological tunnelling}, PoS LATTICE (2007)
  030, 0710.1785.

\bibitem{Degrandschaefer2}
Thomas DeGrand and Stefan Schaefer, Chiral properties of two-flavor QCD in
  small volume and at large lattice spacing, Phys. Rev. D72 (2005) 054503,
  hep-lat/0506021.

\bibitem{Neff}
H.~Neff, Efficient computation of low-lying eigenmodes of non-Hermitian
  Wilson-Dirac type matrices, Nucl. Phys. Proc. Suppl. 106 (2002) 1055--1057,
  hep-lat/0110076.

\bibitem{PARPACK}
http://www.caam.rice.edu/software/ARPACK/.

\bibitem{Kalkreuter:1996mm}
T.~Kalkreuter and H.~Simma, An Accelerated conjugate gradient algorithm to
  compute low lying eigenvalues: A Study for the Dirac operator in SU(2)
  lattice QCD, Comput. Phys. Commun. 93 (1996) 33--47, hep-lat/9507023.

\bibitem{Cundy:2005mr}
Nigel Cundy, {Current status of dynamical overlap project}, Nucl. Phys. Proc.
  Suppl. 153 (2006) 54--61, hep-lat/0511047.

\bibitem{DeGrand2006}
Thomas Degrand and Stefan Schaefer, Simulating an arbitrary number of flavors
  of dynamical overlap fermions, JHEP 07 (2006) 020, hep-lat/0604015.

\bibitem{Cundy:2008zc}
Nigel Cundy, Stefan Krieg, Thomas Lippert, and Andreas Sch{\"a}fer,
  {Topological tunneling with Dynamical overlap fermions}
  ArXiv:0803.0294[hep-lat].

\bibitem{cundyforthcoming08}
Nigel Cundy, Stefan Krieg, Thomas Lippert, and Andreas Sch{\"a}fer,
  forthcoming.

\bibitem{Egri:2005cx}
G.~I. Egri, Z.~Fodor, S.~D. Katz, and K.~K. Szabo, Topology with Dynamical
  Overlap Fermions Hep-lat/0510117.

\bibitem{Fukaya:2005cw}
Hidenori Fukaya, Shoji Hashimoto, Takuya Hirohashi, Kenji Ogawa, and Tetsuya
  Onogi, Topology conserving gauge action and the overlap-Dirac operator, Phys.
  Rev. D73 (2006) 014503, hep-lat/0510116.

\bibitem{Brent}
R.~P. Brent, Algorithms for minimization without Derivitives, Englewood Cliffs,
  NJ, 1973.

\bibitem{Brent2}
W.~Press, S.~Teukolsky, W.~Vetterling, and B.~Flannery, Numerical Recipies in
  Fortran, 2nd Ed., Cambridge University Press, 1992.

\bibitem{Garron:2003cb}
N.~Garron, L.~Giusti, Ch. Hoelbling, L.~Lellouch, and C.~Rebbi, B(K) from
  quenched QCD with exact chiral symmetry, Phys. Rev. Lett. 92 (2004) 042001,
  hep-ph/0306295.

\bibitem{Edwards:1999yi}
R.~G. Edwards, U.~M. Heller, and R.~Narayanan, The overlap-Dirac operator:
  Topology and chiral symmetry breaking, Chin. J. Phys. 38 (2000) 594--604,
  hep-lat/0001013.

\bibitem{Edwards:1998yw}
R.~G. Edwards, U.~M. Heller, and R.~Narayanan, A study of practical
  implementations of the overlap-Dirac operator in four dimensions, Nucl. Phys.
  B540 (1999) 457--471, hep-lat/9807017.

\bibitem{FODORCOMMUNICATION}
Z.~Fodor, private communication.

\bibitem{Crewther}
R.~J. Crewther, Phys.Lett. B70 (1977) 349.

\bibitem{Vecchia}
P.~Di~Vecchia and G.~Veneziano, Nucl Phys B171 (1980) 253.

\bibitem{Leutwyler}
H.~Leutwyler and A.~Smilga, Phys. Rev. D45 (1992) 5607.

\bibitem{Bernard:2003gq}
C.~Bernard et~al., Topological susceptibility with the improved Asqtad action,
  Phys. Rev. D68 (2003) 114501, hep-lat/0308019.

\bibitem{Bali:2001gk}
G.~S. Bali et~al., Quark mass effects on the topological susceptibility in QCD,
  Phys. Rev. D64 (2001) 054502, hep-lat/0102002.

\bibitem{Durr:2001ty}
S.~D{\"u}rr, Topological susceptibility in full QCD: Lattice results versus the
  prediction from the QCD partition function with granularity, Nucl. Phys. B611
  (2001) 281--310, hep-lat/0103011.

\bibitem{Schaefer:2006bk}
Stefan Schaefer, {Algorithms for dynamical overlap fermions}, PoS LAT2006
  (2006) 020, hep-lat/0609063.

\bibitem{2006slft.confE..49C}
N.~{Cundy}, {Zero mode topology with dynamical overlap fermions}, in: XXIVth
  International Symposium on Lattice Field Theory, 2006.

\bibitem{MontvayMunster}
I.~Montvay and G.~M{\"u}nster, Quantum Fields on a Lattice, Cambridge
  University Press, 1994.

\bibitem{Wuppertal:1999}
A.~Frommer, Th. Lippert, B.~Medeke, and K.~Schilling (Eds.), Numerical
  Challenges in Lattice Quantum Chromodynamics, Lecture Notes in Computational
  Science and Engineering, Springer Verlag, Heidelberg, 2000, proceedings of
  the International Workshop, University of Wuppertal, August 22-24, 1999.

\end{thebibliography}

\end{document}